\documentclass[11pt]{article}
\usepackage[colorlinks,bookmarksopen,bookmarksnumbered,citecolor=blue,urlcolor=green,hypertexnames=false]{hyperref}
\usepackage{graphicx}
\usepackage{multirow}
\usepackage{natbib}
\usepackage{amsmath, amssymb}
\usepackage{subfigure}
\usepackage{color}
\usepackage{alltt}
\usepackage{inconsolata}
\usepackage[T3,OT1]{fontenc}
\DeclareSymbolFont{tipa}{T3}{cmr}{m}{n}
\DeclareMathAccent{\invbreve}{\mathalpha}{tipa}{16}
\usepackage{dsfont}
\usepackage{geometry}
\renewcommand\appendix{\par
  \setcounter{section}{0}
  \setcounter{subsection}{0}
  \setcounter{figure}{0}
  \setcounter{table}{0}
  \renewcommand\thesection{Appendix \Alph{section}}
  \renewcommand\thefigure{\Alph{section}\arabic{figure}}
  \renewcommand\thetable{\Alph{section}\arabic{table}}
}
\begin{document}
\title{Spatial Models for Field Trials}%
\author{Mar\'ia Xos\'e Rodr\'iguez - \'Alvarez$^{1,2}$, Martin P. Boer$^{3}$, Fred A. van Eeuwijk$^{3}$, Paul H. C. Eilers$^{4}$\\\\
        \small{$^{1}$ Department of Statistics and Operations Research}\\
				\small{and Biomedical Research Centre (CINBIO), University of Vigo}\\
        \small{Campus Lagoas-Marcosende s/n, 36310 Vigo, Spain}.\\
        \small{\texttt{mxrodriguez@uvigo.es}}\\
				\small{$^{2}$ BCAM - Basque Center for Applied Mathematics, Bilbao, Spain}\\
        \small{$^{3}$ Biometris, Wageningen University \& Research, Wageningen, the Netherlands}\\
        \small{$^{4}$ Erasmus University Medical Centre, Rotterdam, the Netherlands}}
\maketitle
\date{}
\sloppy
\begin{abstract}
An important aim of the analysis of agricultural field trials is to obtain good predictions for genotypic performance, by correcting for spatial effects. In practice these corrections turn out to be complicated, since there can be different types of spatial effects; those due to management interventions applied to the field plots and those due to various kinds of erratic spatial trends. This paper presents models for field trials in which the random spatial component consists of tensor product Penalized splines (P-splines). A special ANOVA-type reformulation leads to five smooth additive spatial components, which form the basis of a mixed model with five unknown variance components. On top of this spatial field, effects of genotypes, blocks, replicates, and/or other sources of spatial variation are described by a mixed model in a standard way. We show the relation between several definitions of heritability and the effective dimension or the effective degrees of freedom associated to the genetic component. The approach is illustrated with large-scale field trial experiments. An \texttt{R}-package is provided.   
\end{abstract}
\section{Introduction}
Spatial variation is common in agricultural field trials. Many factors combine to generate micro-environments that differ from plot to plot, strongly influencing yield and other traits. It is necessary to correct for them when estimating genotypic effects.

A part of the spatial variation can be attributed to systematic effects, caused by the way the field was prepared before and during sowing or planting. A familiar example are row and column effects, caused by the movements of machines during ploughing, tilling and other procedures. It is relatively easy to add factors to a statistical model to account for them.

Random spatial variation such as  for example fertility trends is harder to model. There are no covariates (like row numbers) to relate it to, so it is necessary to include a model component for a random field. Roughly speaking, there are two main approaches to model spatial trends: one based on spatial variance-covariance structures; and the other based on smoothing techniques. In the first case, a spatially correlated stochastic component is included into the spatial model. However, this is non-trivial, as correlation in two directions, along the rows and columns of the field has to be modeled. To keep the effort manageable, several assumptions need to be made, and it has become standard to consider separability and stationarity \citep[see e.g. ][]{Zimmerman1991}. Important contributions in this area are the separable autoregressive model proposed by \cite{Cullis1991} and extended in \cite{Gilmour1997}, the separable linear variance model discussed in \cite{Piepho2010}, or the Bayesian first-differencing model in rows and columns given in \cite{Besag1999}. As a complementary approach, smoothing methods model spatial trend variation explicitly. The use of smoothing techniques in the agricultural context dates back to \cite{Green1985}, and it has been further described and extended, among others, by \cite{Durban2003} and \cite{Verbyla1999}. To the best of our knowledge, this modeling technique has been mainly approached in the statistical literature in the one dimensional case, i.e., through separate (or additive) smoothed trend effects along the rows and columns of the field. However, while these approaches have proved useful for modeling large-scale dependence (or global trend), they suffer from the limitation of not always being able to capture small-scale dependence (local trend). As a consequence of this limitation, the inclusion of spatially correlated components might still be necessary \citep{Gilmour1997, Verbyla1999}. 

As an alternative, this paper explores the use of two-dimensional smooth surfaces. We propose the use of tensor product Penalized splines \cite[P-splines,][]{Eilers2003} to explicitly model both sources of spatial dependence. P-splines were introduced by \citet{Eilers1996}, as a simplification of a proposal by \citet{OSullivan1986}. P-splines approach smoothing as penalized regression: a rich B-spline basis is combined with a penalty on (higher order) differences of the B-spline coefficients to avoid overfitting, and estimation is based on penalized ordinary least squares. As it will be seen, the mixed model representation of P-splines \citep{Currie2002, Wand2003} provides us with a general framework for the analysis of field trials. It allows the inclusion of both extra fixed and random components, such as genotypic effects or the correction for rows and columns. Besides, using nested B-spline bases \citep{Lee2013} the computational effort of our approach, which we call SpATS, is moderate, even for large fields trials. Our SpATS model has a number of other attractive properties: (1) an explicit estimate of the spatial trend in the field is obtained; (2) estimation is stable and fast; (3) missing plots, even a large fraction of them, are easily handled; and (4) extension to a non-normal response, along the lines of the generalized linear model, is straightforward. 

We should mention that our approach is not completely new in the agricultural literature. In \cite{Taye2008} and \cite{Robbins2012} the authors discuss similar approaches in the context of field experiments, and in forest research the topic has been covered by \cite{Cappa2008}. This paper goes one step further by proposing a fully anisotropic penalized approach framed within the mixed-effects model context. Use is made of the P-spline ANOVA-type (PS-ANOVA) approach presented in \cite{Lee2013}, which gives rise to a model with five smooth spatial components each having a clear interpretation. We also show the link between the generalized definitions of heritability proposed by \cite{Cullis2006} and \cite{Oakey2006} and the notion of effective dimension of model components, a well-known complexity measure in the smoothing context \citep{Hastie1990}. Finally, we provide software for the practical application of our proposal in a free and easy-to-use \texttt{R}-package \citep{R16}, called \texttt{SpATS}.  

The rest of the paper is structured as follows. We start by motivating our approach in Section \ref{s_motexample}. Section \ref{s_splines} presents background on B- and P-splines in one and two dimensions, including their representation as mixed models. They form the basis for spatial models, which are presented in Section \ref{s_trials}. Simulations comparing our SpATS model and that of \cite{Gilmour1997} can be found in Section \ref{sim_studies} and Section \ref{s_app} presents several applications to large-scale field trials. A Discussion Section closes the paper. Some technical details have been added as Appendices, where we also describe the \texttt{R}-package that accompany this paper.
\section{Motivating example\label{s_motexample}}
Uniformity field trials are trials in which a single genotype or variety is evaluated. In practice, the interest of such field trials is that its statistical analysis can help understanding the different sources of spatial variation present in a field, and thus serve as guidance for the design and subsequent analyses when genetic effects are to be evaluated. In this section we present a series of analyses of a set of barley uniformity data discussed in the paper by \cite{Williams88}. We focus here on presenting the big picture of our approach, leaving the more technical details to subsequent sections. 

\begin{figure}
    \begin{center}
    \subfigure[Raw data]{
		\includegraphics[width=3.5cm]{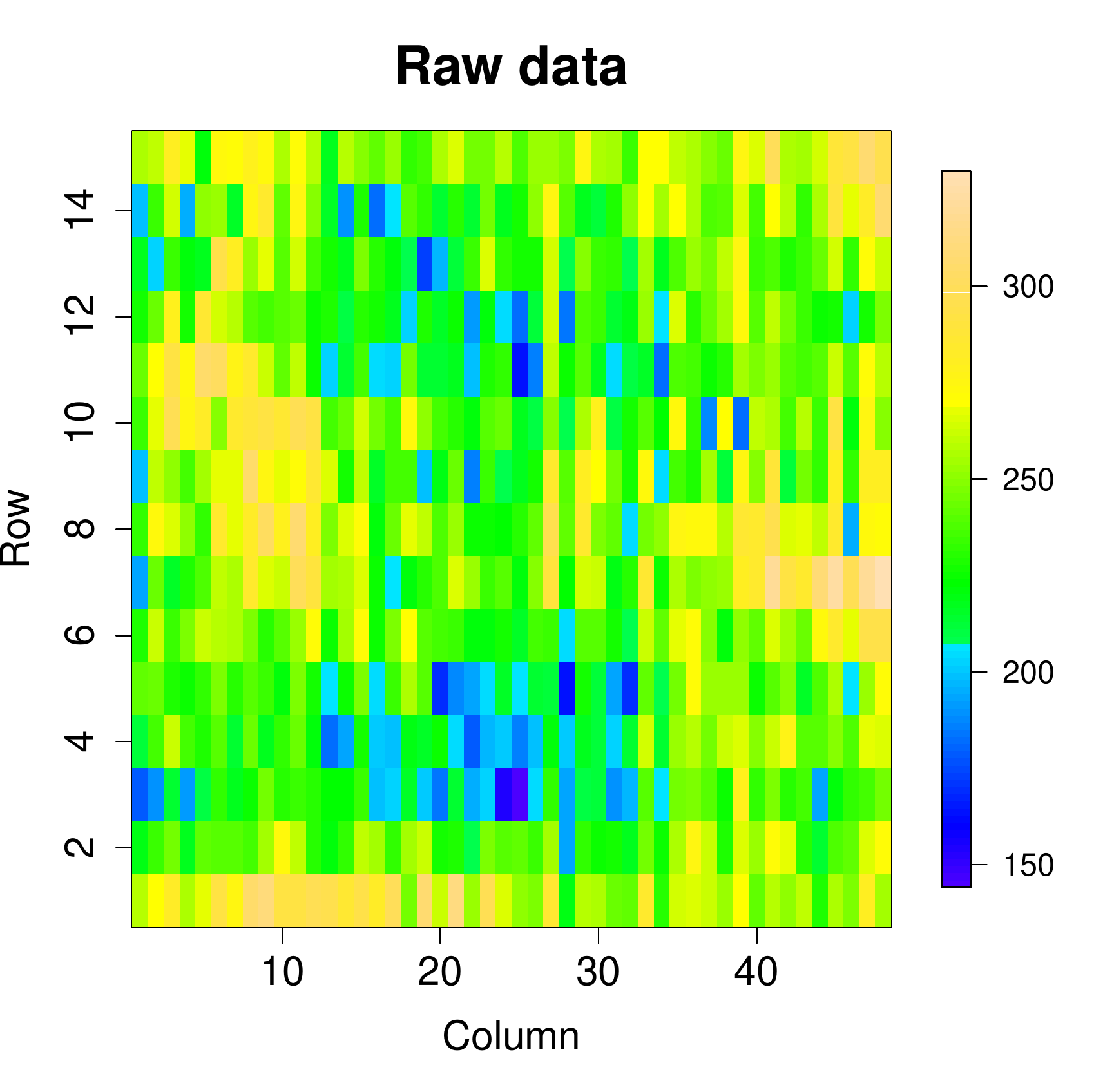}\label{Uniformity_results_raw}}
		\subfigure[Simple model]{
		\includegraphics[width=3.5cm]{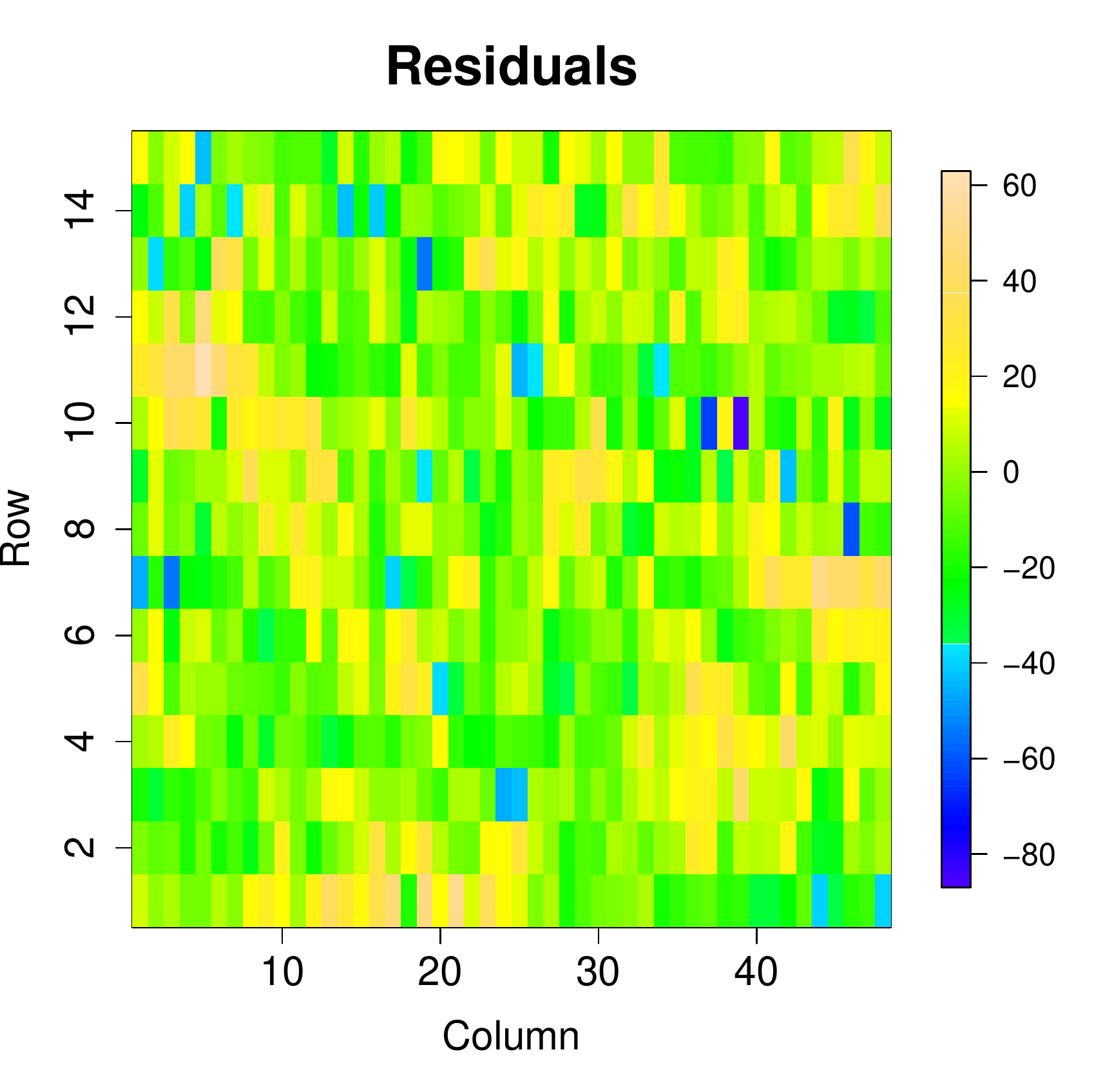}
		\includegraphics[width=3.5cm]{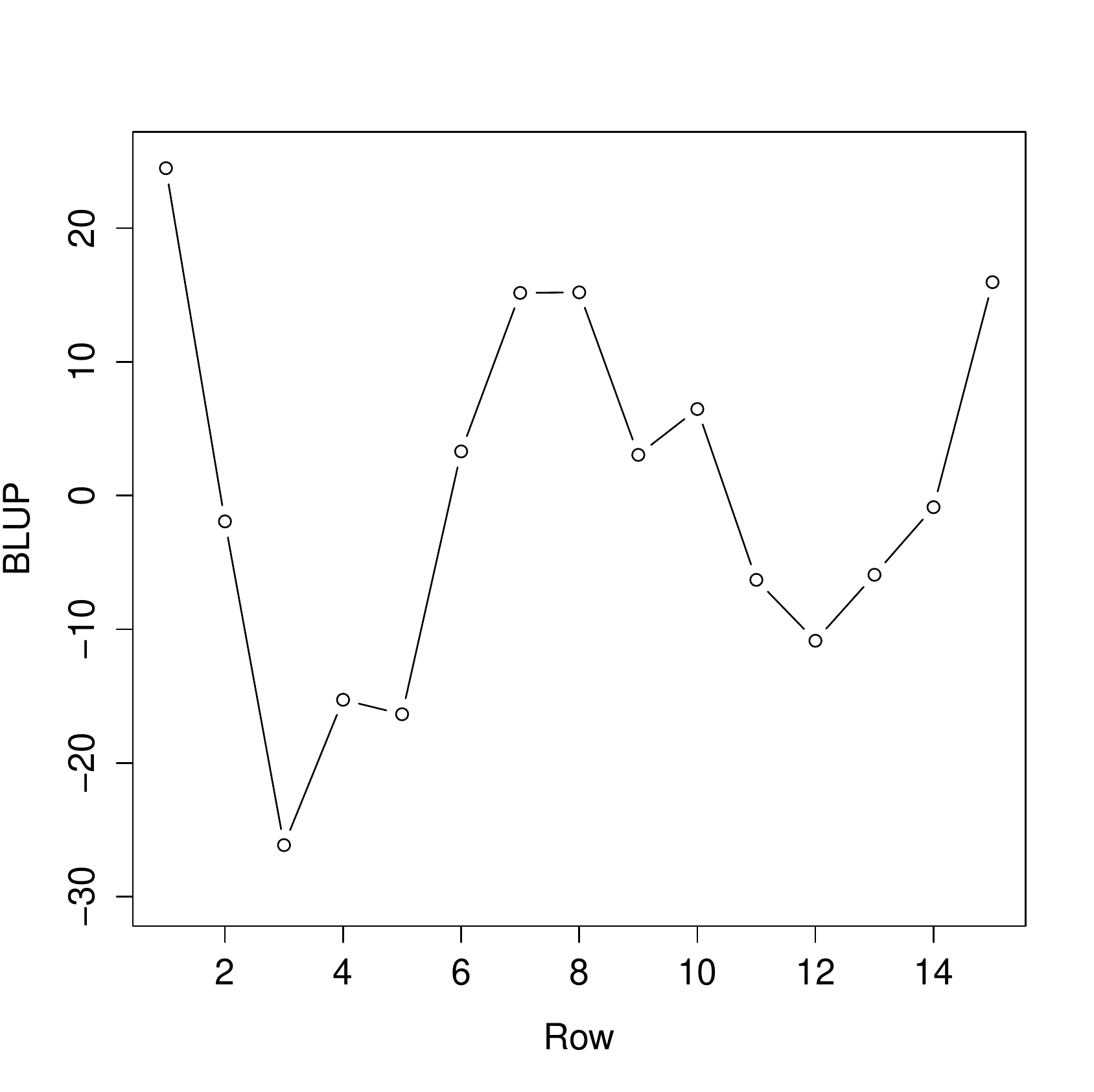}
		\includegraphics[width=3.5cm]{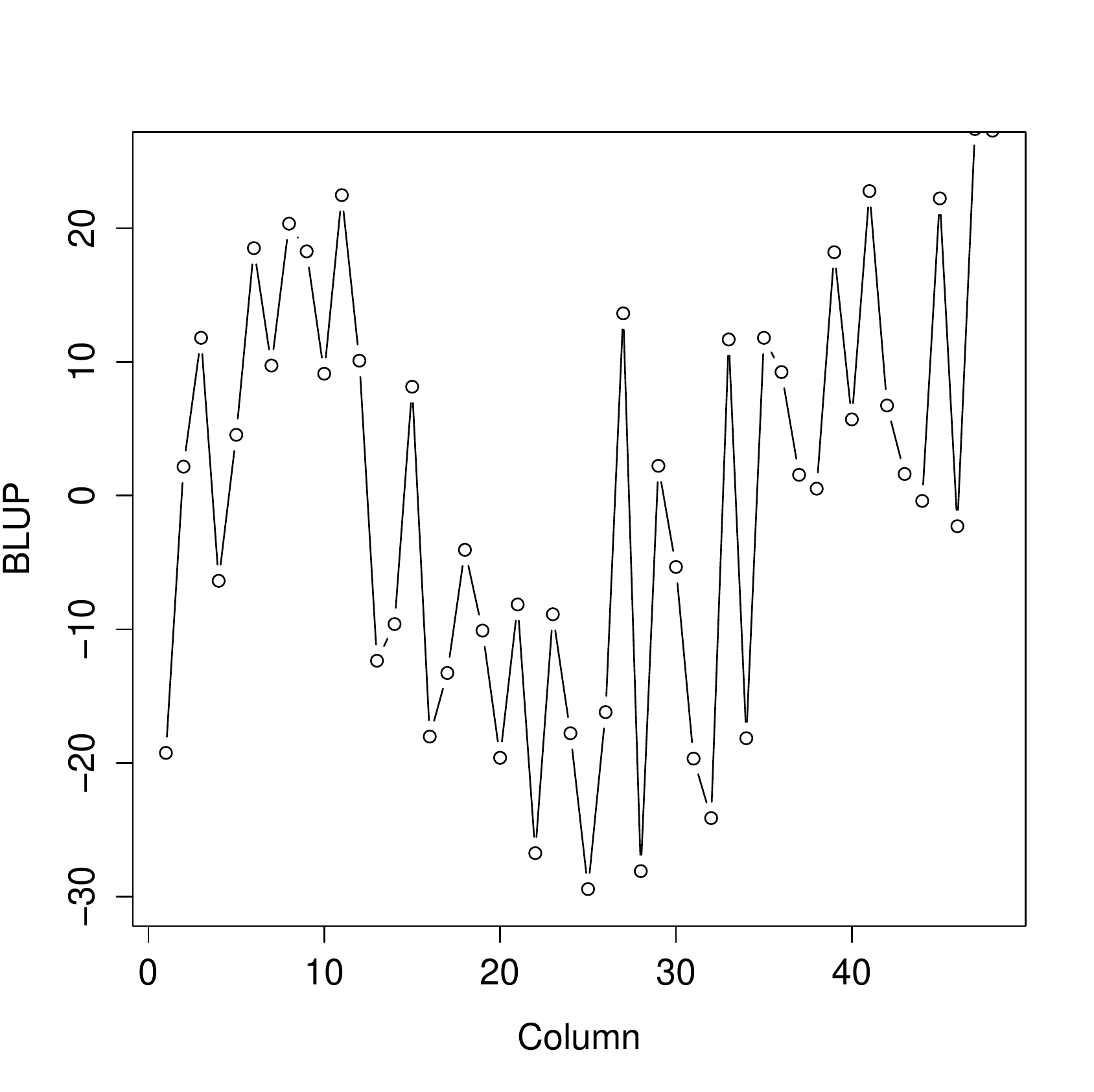}\label{Uniformity_simple_m}}
		\subfigure[Full Model]{
		\includegraphics[width=3.5cm]{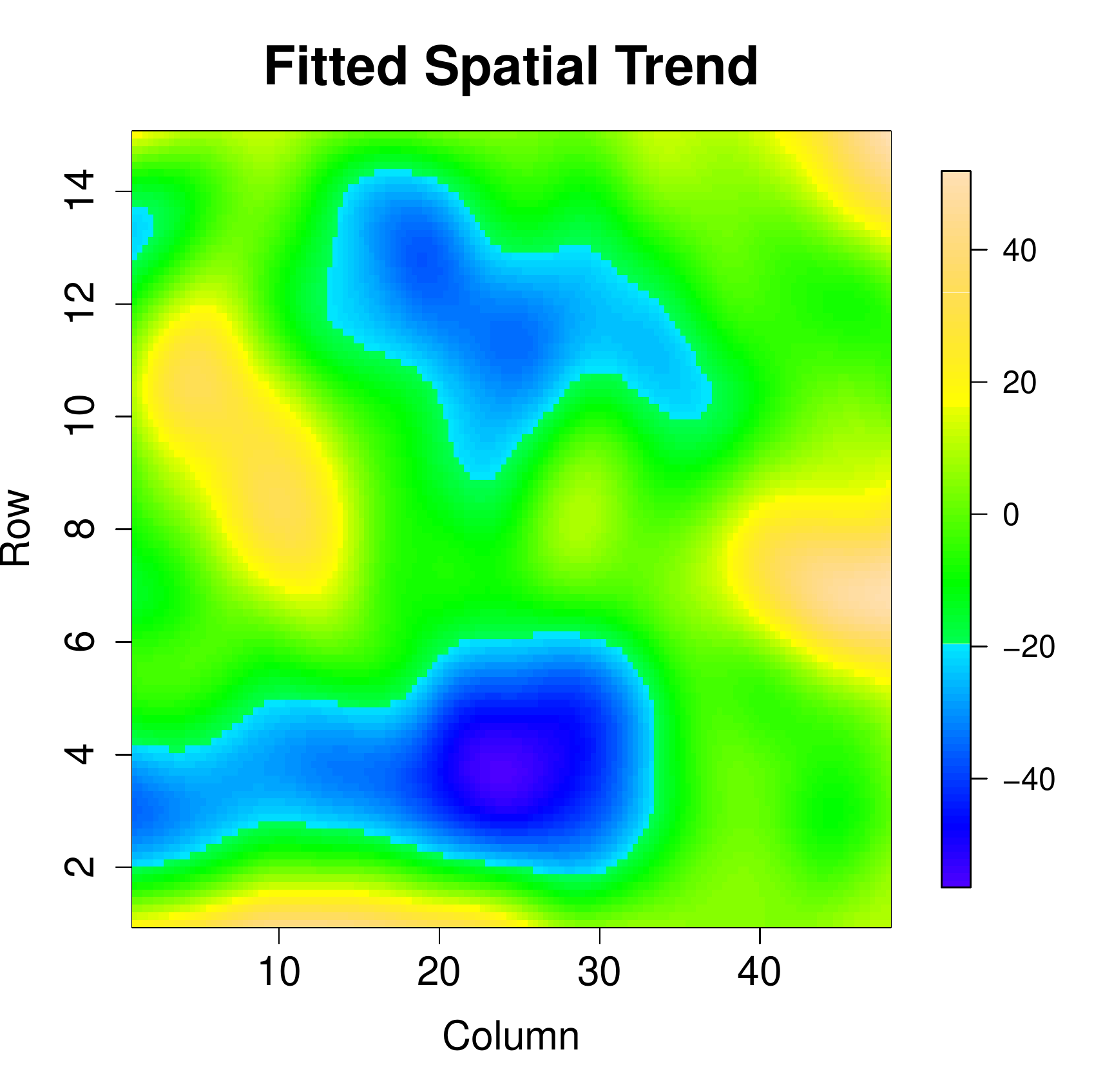}
		\includegraphics[width=3.5cm]{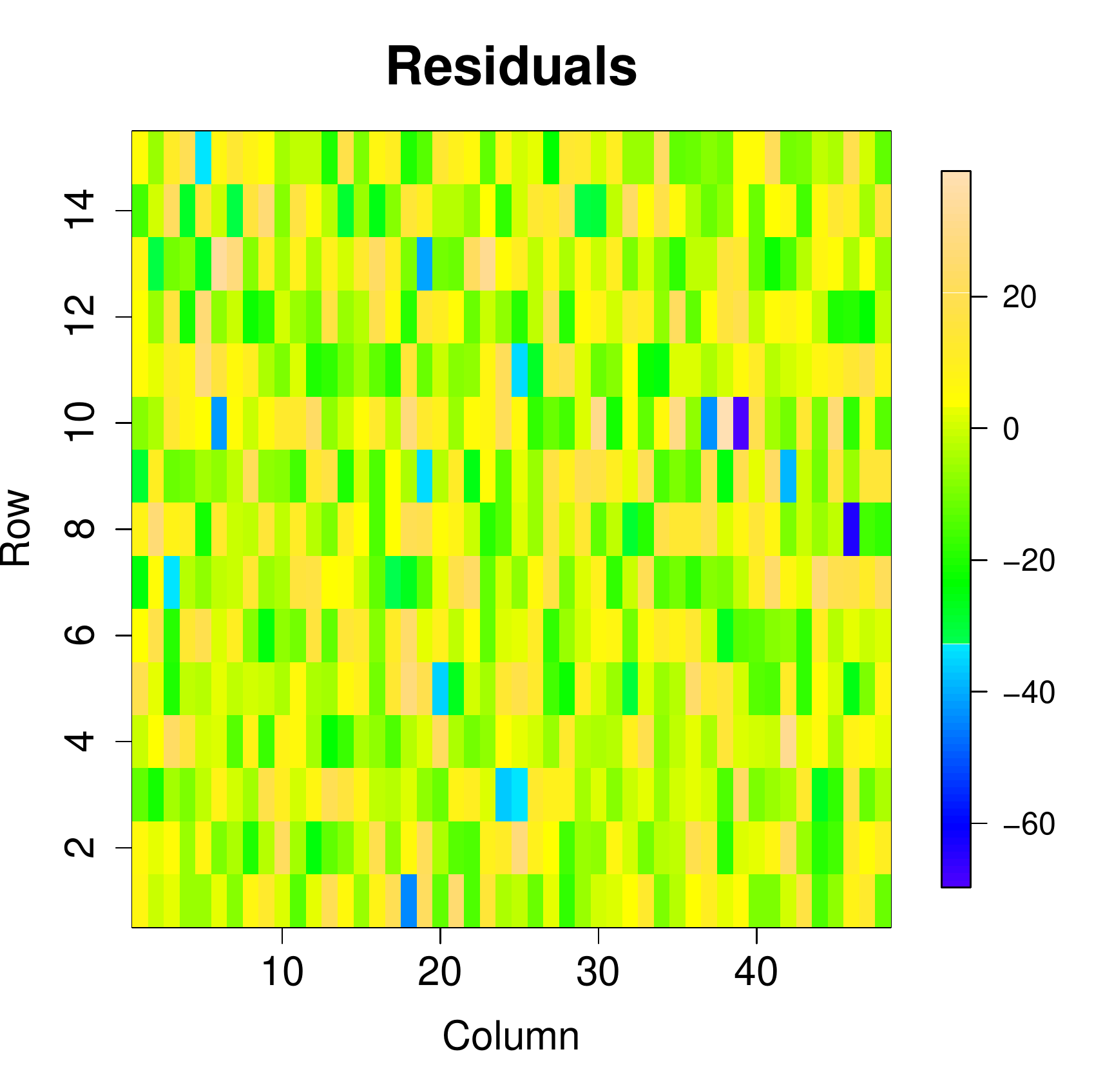}
		\includegraphics[width=3.5cm]{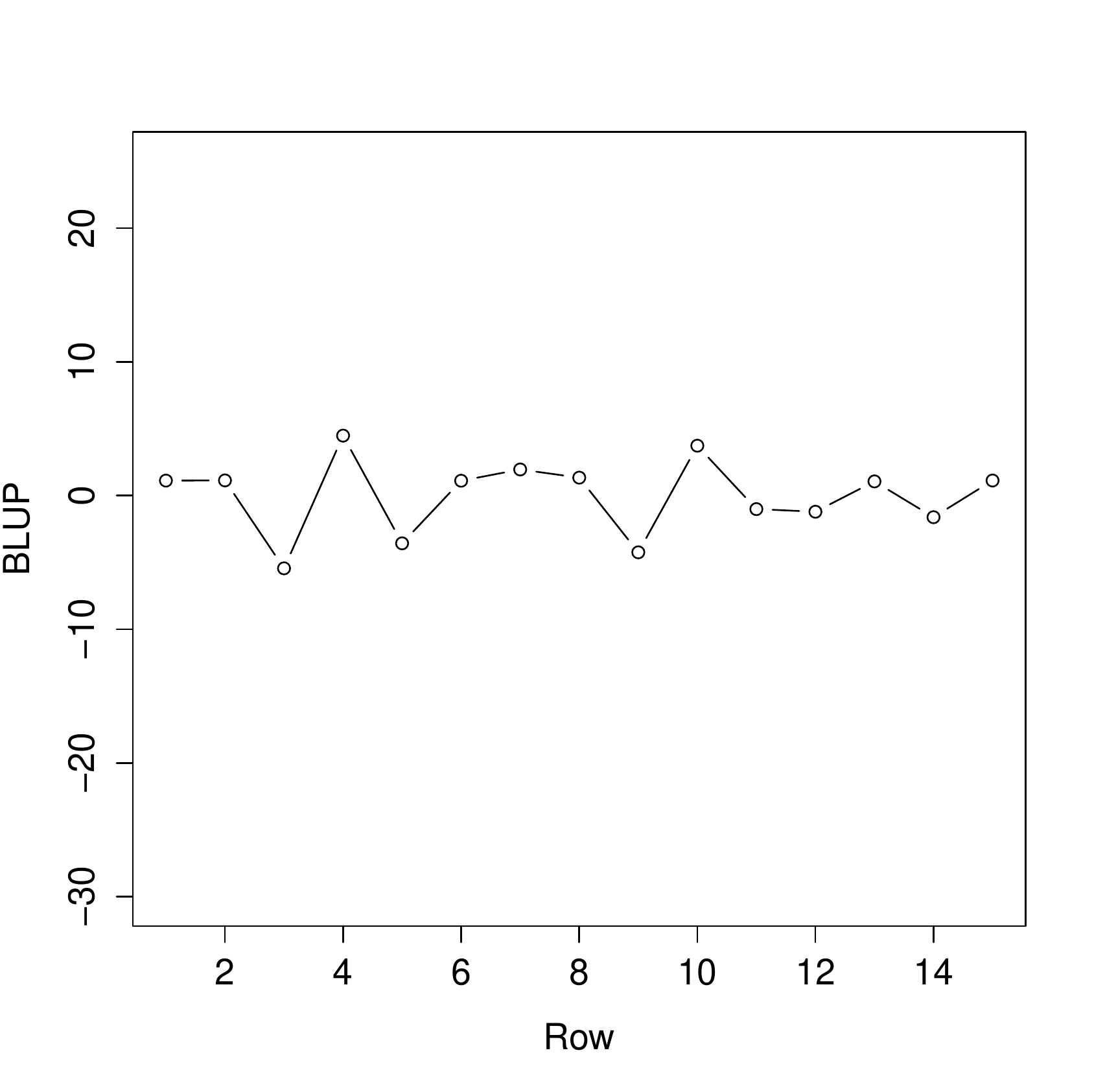}
		\includegraphics[width=3.5cm]{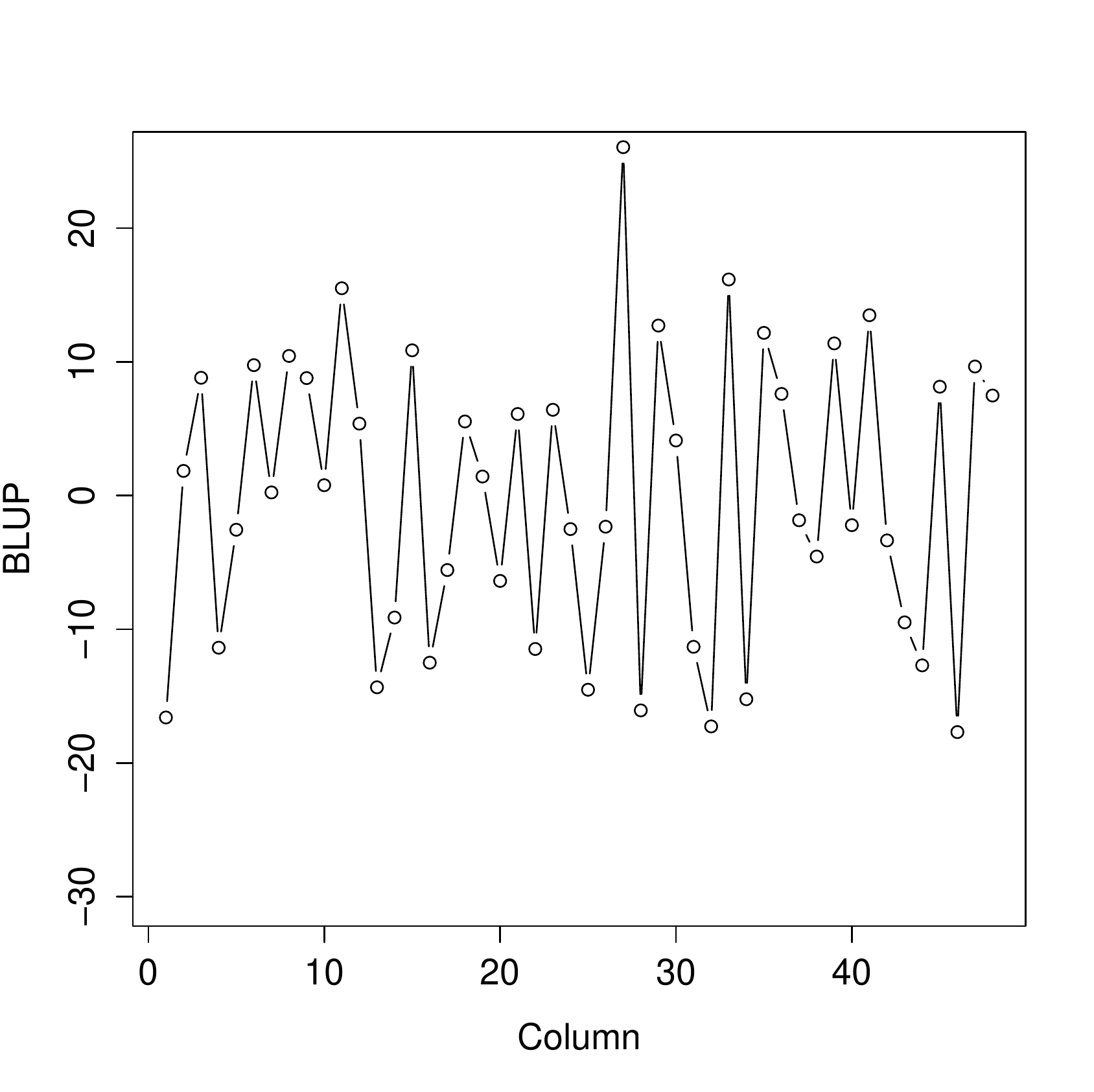}\label{Uniformity_full_m}}
		\end{center}
		 \caption{For the barley uniformity data: Raw data, residuals' spatial plot, best linear unbiased predictions (BLUPs) for the row and column random factors and contour plot of the estimated spatial trend based on the models including (b) only the row and column random factors (Simple model) (c) the smooth spatial trend (Full model)}
		\label{Uniformity_results}
\end{figure}

In this experiment, plots were laid out in a $15$ row by $48$ column grid, and the phenotypic trait of interest was yield. Figure~\ref{Uniformity_results_raw} depicts the raw yield data. Note that there is a rather complex spatial pattern, with patches presenting larger/smaller yield values. Let $y_i$ denotes the yield data (in kg per hectare divided by $10$) obtained at plot $i$ ($i = 1,\ldots, 720$), and $u_i$ and $v_i$ the row and column position respectively, both centered and scaled. A common strategy in the analysis of field trial experiments is to use the following statistical model as starting point
\begin{equation}
\boldsymbol{y} = \boldsymbol{1}_{720}\beta_0 + \boldsymbol{Z}_r\boldsymbol{c}_r + \boldsymbol{Z}_c\boldsymbol{c}_c + \boldsymbol{\varepsilon},
\label{model_uni}
\end{equation}
where $\boldsymbol{y} = \left(y_1, \ldots, y_{720}\right)^{t}$, $\boldsymbol{1}_n$ is a column vector of ones of length $n$, and $\boldsymbol{c}_r = \left(c_{r1},\ldots, c_{r15}\right)^{t}$ and $\boldsymbol{c}_c = \left(c_{c1},\ldots, c_{c48}\right)^{t}$ are, respectively, the random effect coefficients for the rows and columns with $\boldsymbol{c}_r \sim N\left(\boldsymbol{0}, \sigma^2_r\boldsymbol{I}_{15}\right)$ and $\boldsymbol{c}_c \sim N\left(\boldsymbol{0}, \sigma^2_c\boldsymbol{I}_{48}\right)$ and associated design matrices $\boldsymbol{Z}_r$ and $\boldsymbol{Z}_c$. Finally, the random error vector $\boldsymbol{\varepsilon} = \left(\varepsilon_1, \ldots, \varepsilon_{720}\right)^{t} \sim N\left(\boldsymbol{0},\sigma^2\boldsymbol{I}_{720}\right)$.

Figure~\ref{Uniformity_simple_m} depicts the empirical best linear unbiased predictors (BLUPs) for the row and column random factors. As can be observed, and especially for the rows, the BLUPs show a clear evidence of a pattern, indicating that the independent Gaussian distribution assumption does not hold. Besides, the residuals' spatial plot ($\widehat{\boldsymbol{\varepsilon}} = \boldsymbol{y} - \boldsymbol{1}_{720}\widehat{\beta}_0 + \boldsymbol{Z}_r\widehat{\boldsymbol{c}}_r + \boldsymbol{Z}_c\widehat{\boldsymbol{c}}_c$) shown also in Figure~\ref{Uniformity_simple_m} suggests that the complex spatial pattern has not been completely captured, and thus a more complex statistical analysis is required. To that end, we propose a modeling strategy based on incorporating a model (spatial) component that simultaneously accounts for the spatial trend across both directions of the field. Specifically, a smooth bivariate surface, jointly defined over the row and column positions, is assumed  
\begin{equation}
\boldsymbol{y} = f\left(\boldsymbol{u}, \boldsymbol{v}\right) + \boldsymbol{Z}_r\boldsymbol{c}_r + \boldsymbol{Z}_c\boldsymbol{c}_c + \boldsymbol{\varepsilon},
\label{tpspline_model_uni}
\end{equation}
where $\boldsymbol{u} = \left(u_1, \ldots, u_{720}\right)^{t}$, $\boldsymbol{v} = \left(v_1, \ldots, v_{720}\right)^{t}$, and $f\left(\boldsymbol{u}, \boldsymbol{v}\right) = \left(f\left(u_1, v_1\right),\ldots,f\left(u_{720}, v_{720}\right)\right)^{t}$, with $f(\cdot, \cdot)$ representing a smooth bivariate function. Note that the intercept, $\beta_0$, is embedded as part of $f\left(u,v\right)$. To have a better understanding of the interpretation of $f\left(\cdot,\cdot\right)$, we can further decompose it in a nested-type ANOVA structure
\begin{equation*}
f\left(\boldsymbol{u}, \boldsymbol{v}\right) = \underbrace{\boldsymbol{1}_n\beta_0 + \boldsymbol{u}\beta_1 + \boldsymbol{v}\beta_2 + \boldsymbol{u}\odot\boldsymbol{v}\beta_3}_{\mbox{Bilinear polynomial}} + \underbrace{f_u\left(\boldsymbol{u}\right) + f_v\left(\boldsymbol{v}\right) + \boldsymbol{u}\odot h_v\left(\boldsymbol{v}\right) + \boldsymbol{v}\odot h_u\left(\boldsymbol{u}\right) + f_{u,v}\left(\boldsymbol{u}, \boldsymbol{v}\right)}_{\mbox{Smooth part}},
\end{equation*}
where $\odot$ denotes the element-wise vector (matrix) product. There are now two components: the bilinear polynomial and the smooth part. The bilinear (or parametric) component includes the intercept ($\beta_0$), the linear trends along the row ($\beta_1$) and column ($\beta_2$) directions, as well as the linear interaction trend ($\beta_3$). In addition, the smooth component is responsible for modeling the deviation from this compound linear trend. Here, 
\begin{itemize}
\item $f_u(u)$ is a smooth trend along the rows, identical for all columns (i.e., a main smooth effect).
\item $f_v(v)$ is a smooth trend along the columns, identical for all rows.
\item $vh_u(u)$ and $uh_v(v)$ are linear-by-smooth interaction trends. For instance, $uh_v(v)$ is a varying coefficient surface trend, consisting of functions, linear in the rows, for each column, but with slopes that change smoothly along the columns, $h_v(v)$ (the same holds for $vh_u(u)$). 
\item $f_{u,v}(u,v)$ is a smooth-by-smooth interaction trend jointly defined over the row and column directions. 
\end{itemize}

The functions $f_u$, $f_v$, $h_u$ and $h_v$ are constructed with variations on one-dimensional P-splines, while $f_{u,v}$ is based on tensor product P-splines. It may come as a surprise that six components are introduced to model the surface $f$ in (\ref{tpspline_model_uni}) (the bilinear polynomial and the five smooth trends). The reason is that this decomposition translates model (\ref{tpspline_model_uni}) directly to a standard mixed model. In fact, for each of the smooth components the desirable amount of smoothing is computed using restricted maximum likelihood \citep[REML,][]{Patterson71}. This decomposition also allows to reduce the size of the interaction component $f_{u,v}$ for very large fields, to save computation time \citep{Lee2013}. The technical details will be presented in Section \ref{s_splines}.

\begin{figure}
    \begin{center}
		\includegraphics[width=5cm]{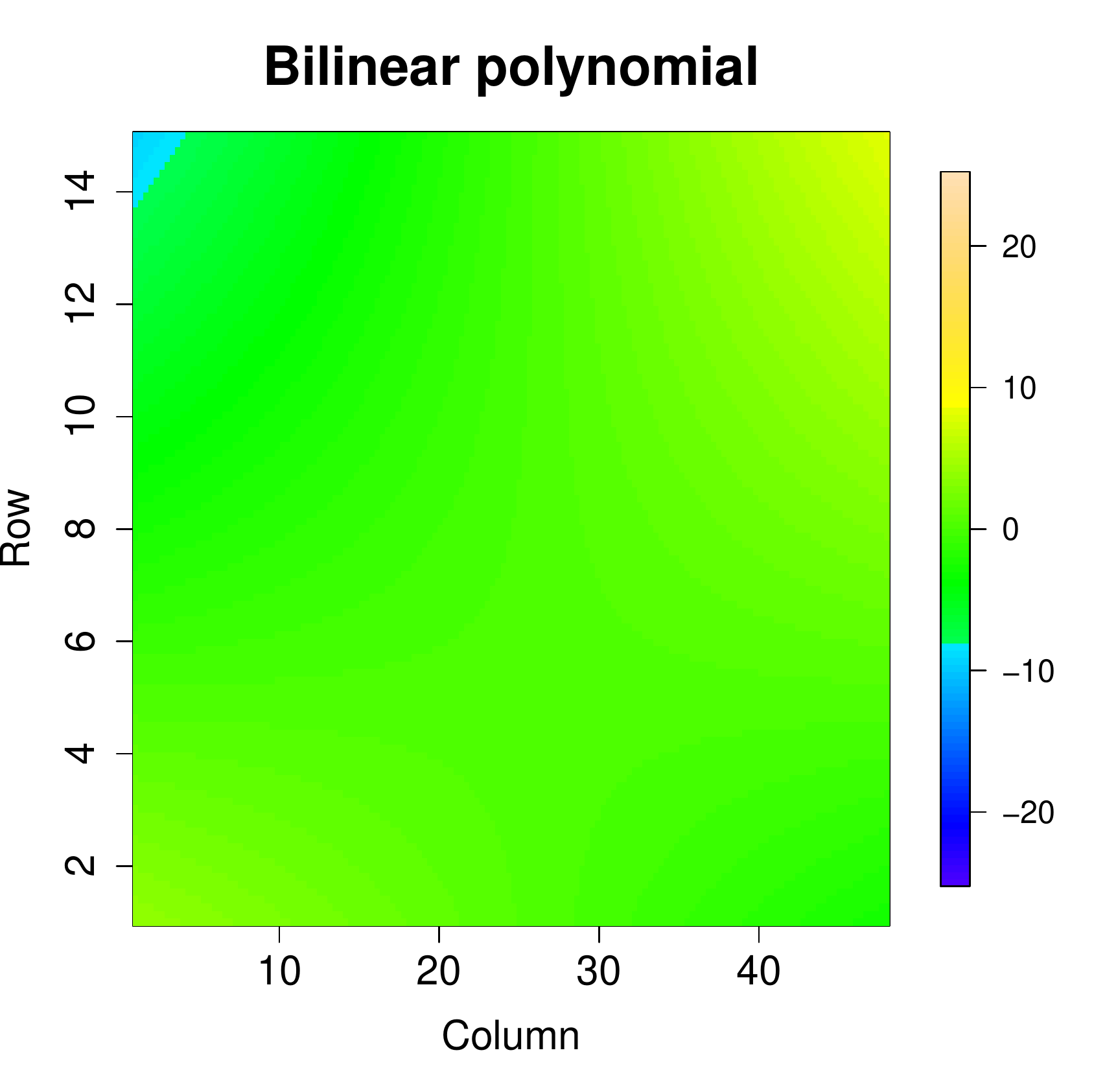}
		\includegraphics[width=5cm]{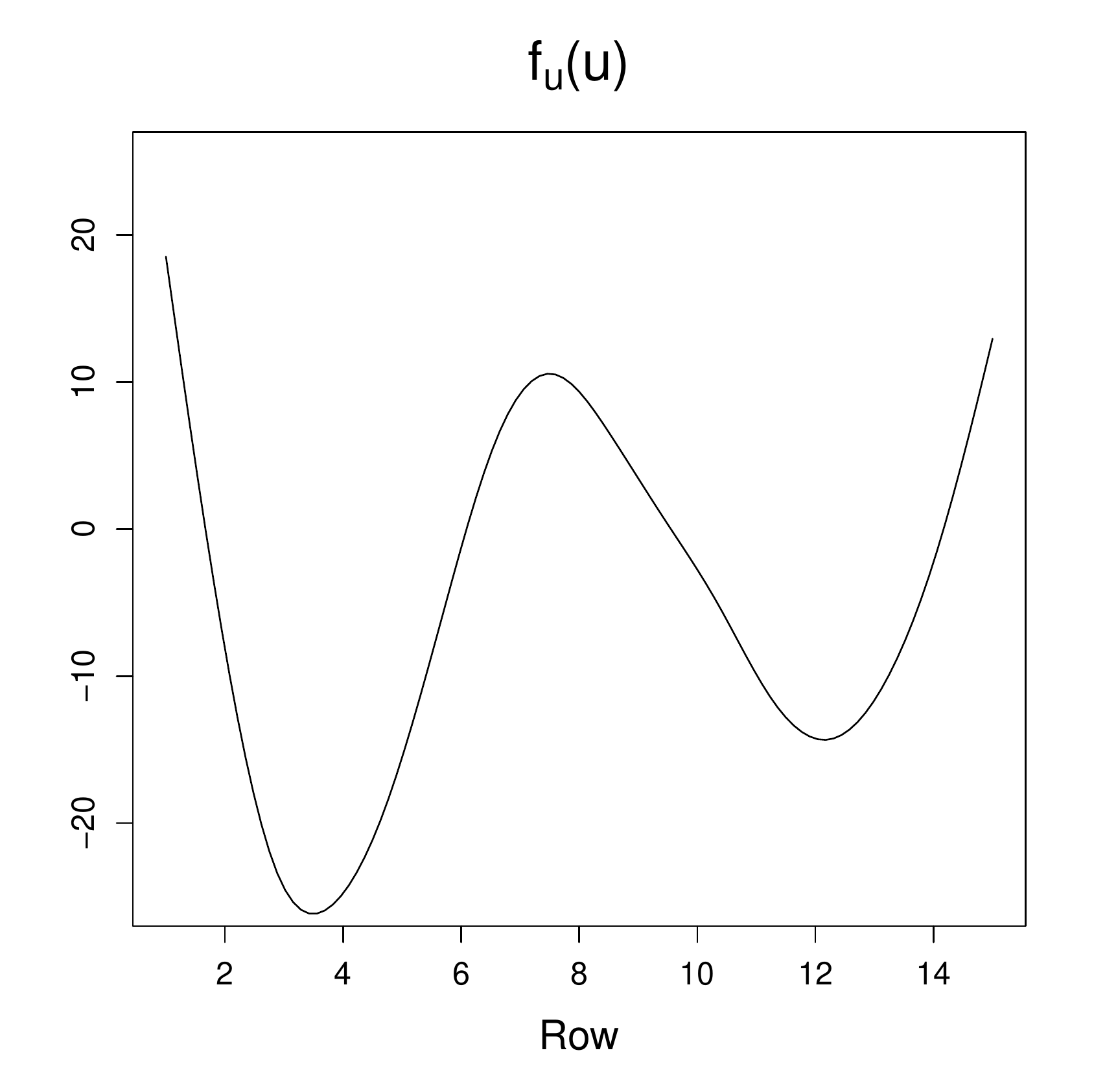}
		\includegraphics[width=5cm]{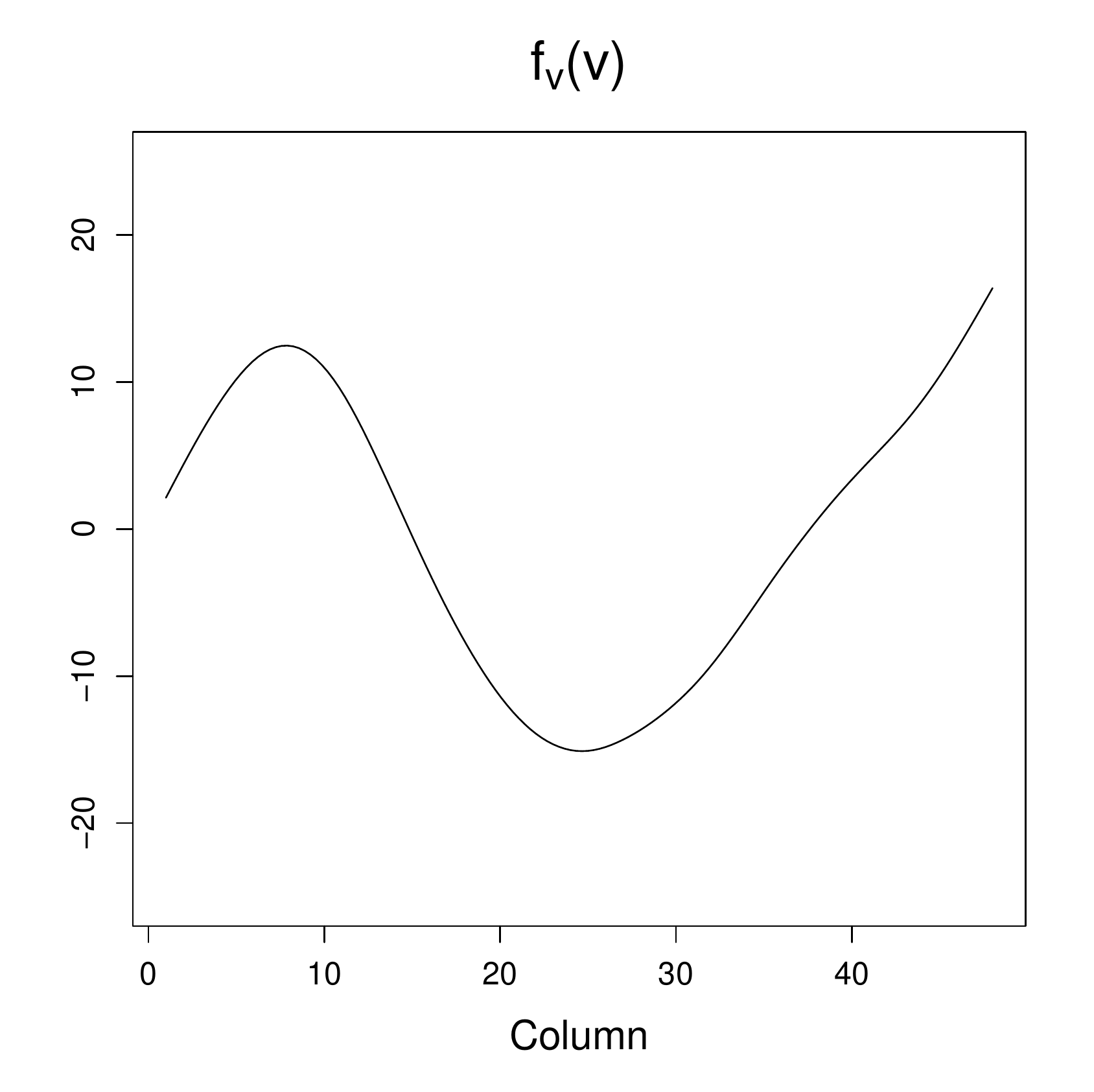}
		\includegraphics[width=5cm]{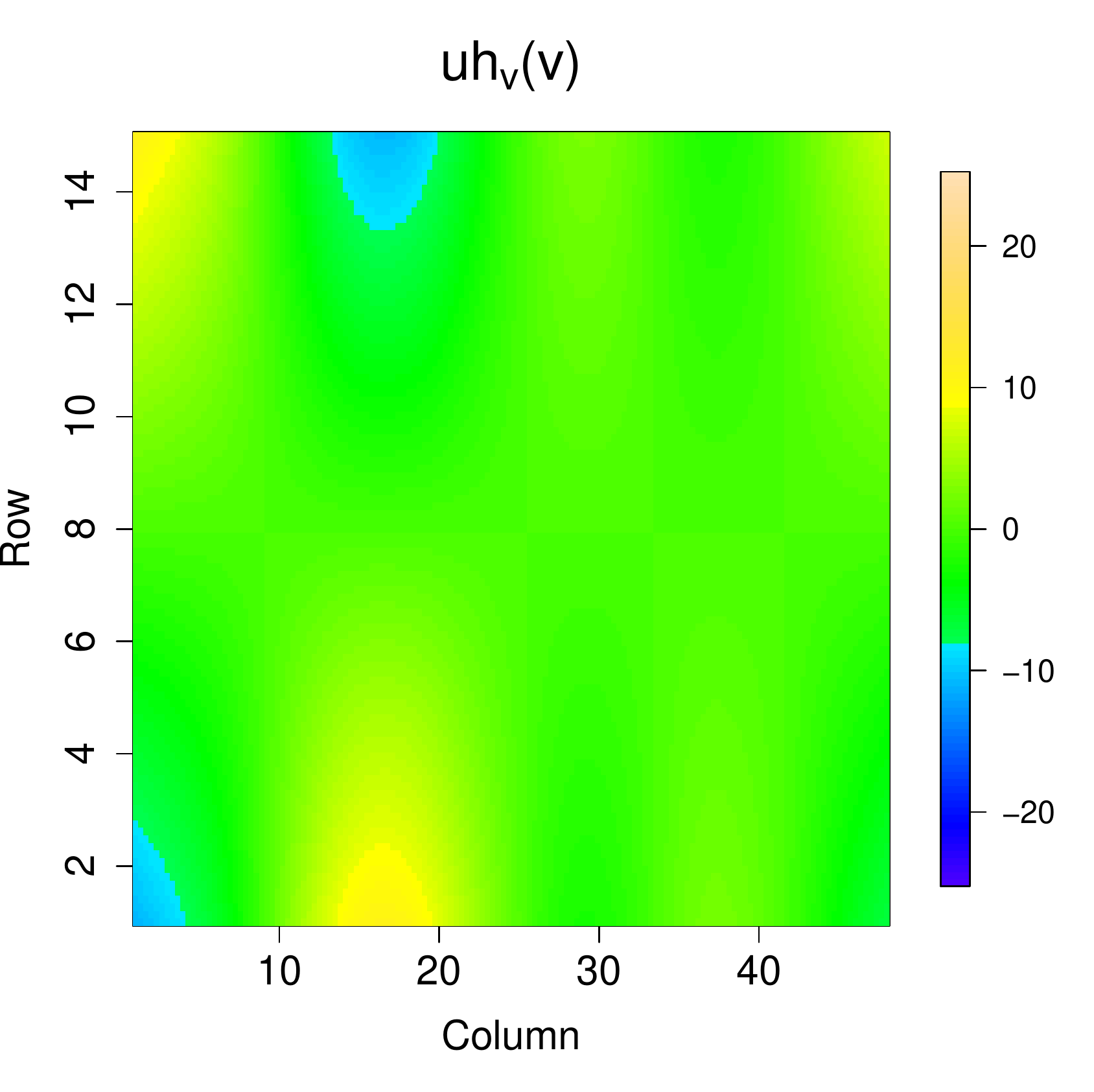}
		\includegraphics[width=5cm]{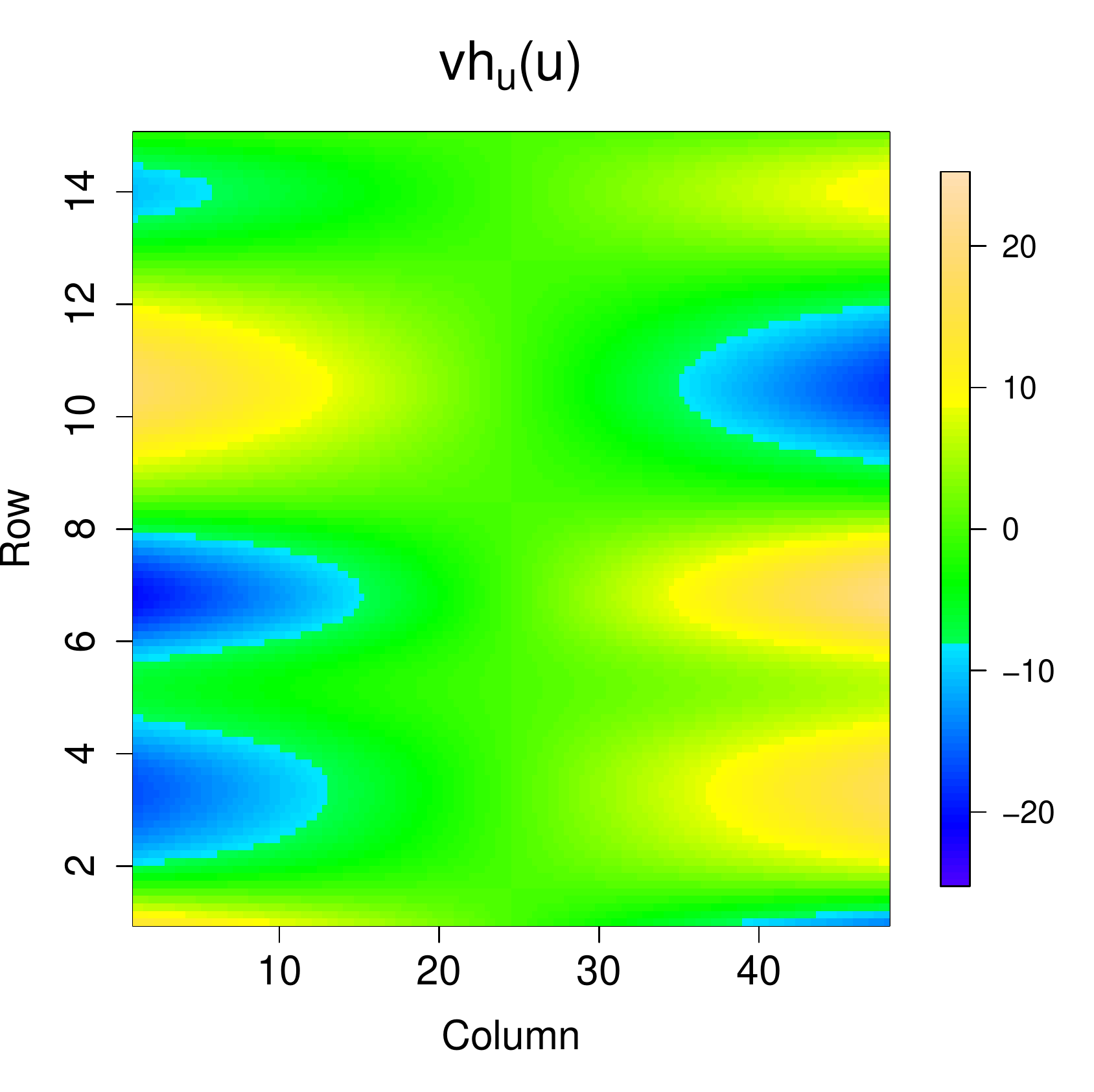}
		\includegraphics[width=5cm]{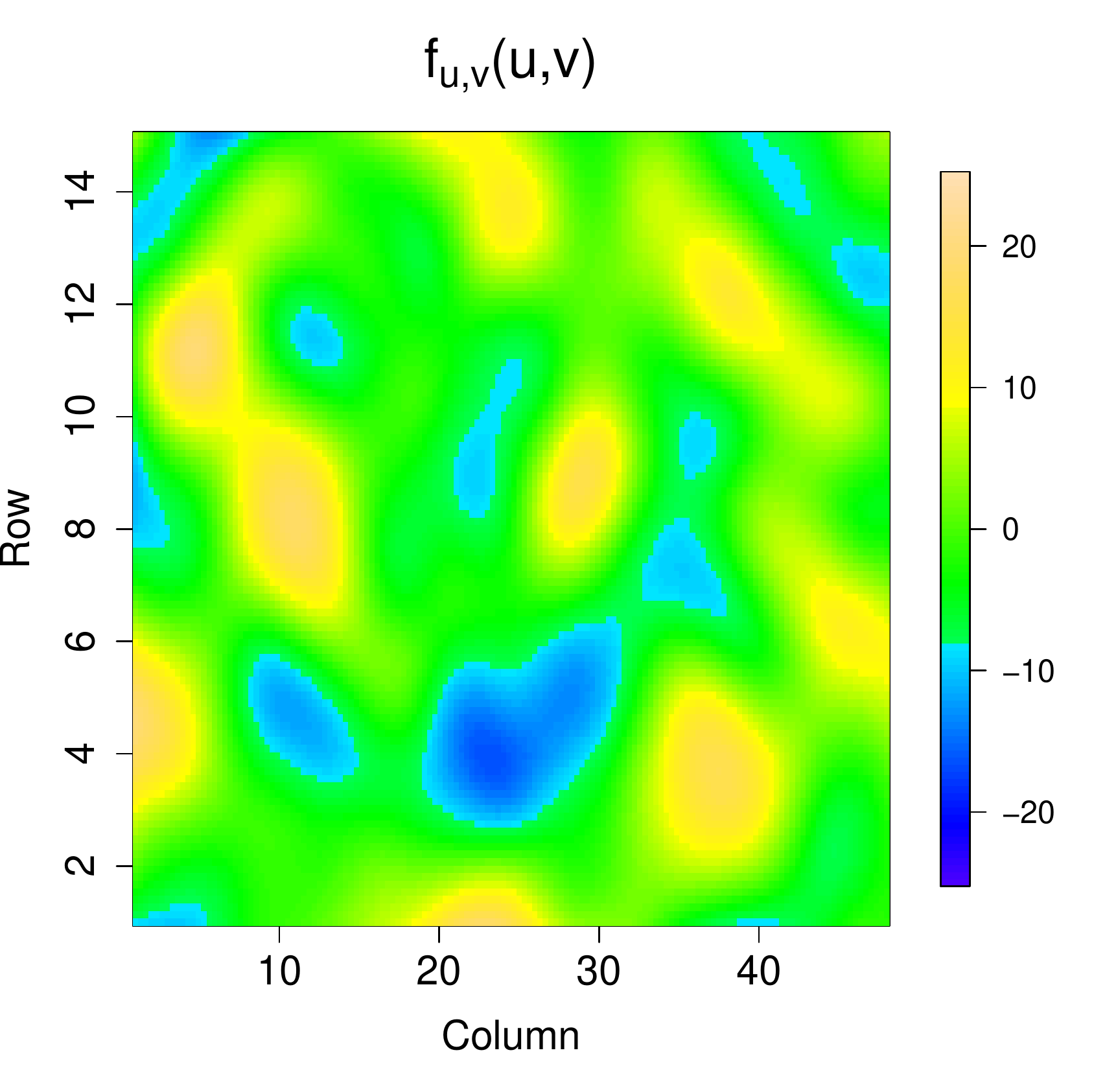}
		\end{center}
		\caption{Bilinear and smooth components of the ANOVA-type decomposition of the estimated spatial trend for the barley uniformity data.}
		\label{ANOVA_decomposition}
\end{figure}

Figure \ref{Uniformity_full_m} shows the estimated spatial trend across the field, i.e. $\widehat{f}(\cdot, \cdot)$, but excluding the intercept. A nice property of our proposal is that it allows depicting the spatial trend in a grid finer than the number of rows and columns, facilitating results interpretation. Note that we recover quite successfully the spatial variation observed in the raw data. The residuals' spatial plot and BLUPs for $\boldsymbol{c}_r$ and $\boldsymbol{c}_c$ shown also in Figure \ref{Uniformity_full_m} suggest that the spatial independence assumption for the error vector $\boldsymbol{\varepsilon}$ might be appropriate, and that no trend is now present in the BLUPs. Figure~\ref{ANOVA_decomposition} shows the bilinear and smooth components of the ANOVA-type decomposition discussed above. Note that the estimated smooth functions defined over the rows ($\hat{f}_u$) and columns ($\hat{f}_v$) capture the trends observed in the BLUPs for the row and column analysis (model (\ref{model_uni}) and Figure~\ref{Uniformity_simple_m}). When we compare the estimated linear-by-smooth interactions trends, we observe that the contribution of $v\hat{h}_u(u)$ to the fitted spatial trend is stronger than that due to $u\hat{h}_v(v)$, but both are mainly responsible for modeling local behaviors at the edges of the field. Finally, the smooth-by-smooth interaction term recovers the local patches observed in the raw data, that the other components would not be able to capture. 

\section{Background on P-splines\label{s_splines}}
This section provides background information on P-splines, their tensor products and equivalent mixed model formulations. We refer the interested reader to \cite{Eilers2010} and \cite{Eilers2015} for an extensive account of many aspects of P-splines. 

\subsection{Modeling surfaces by means of P-splines\label{s_splines_2D}}
Suppose we are given $n$ data triples ($u_i$, $v_i$, $y_i$), with $u_i$ and $v_i$ for positions, e.g, rows and columns, and $y_i$ for a response variable, e.g a phenotype in a trial, and that we are interested in the following model
\begin{equation}
\label{bsp_m}
y_i = f(u_i, v_i) + \varepsilon_i,\;\;\;\; \varepsilon_i \sim N(0,\sigma^2).
\end{equation}  
In the P-spline framework, the smooth bivariate function $f\left(u_i, v_i\right)$ is approximated by the tensor product of B-spline basis \citep{Dierckx1993}. The idea is simple: we form two B-spline bases, $\invbreve{\boldsymbol{B}}$, with $\invbreve{b}_{il} = \invbreve{B}_l(u_i)$ and $\breve{\boldsymbol{B}}$, with $\breve b_{ip} = \breve B_p(v_i)$, where $\invbreve{B}_l(u_i)$ is the $l$-th B-spline evaluated at $u_i$ (and the same holds for $\breve B_p(v_i)$), and take 
\begin{equation*}
f\left(u_i, v_i\right) = \sum_{l = 1}^{L} \sum_{p = 1}^{P} \invbreve B_l(u_i) \breve B_p(v_i)\alpha_{lp},
\end{equation*}
where $\boldsymbol{\alpha} = \left(\alpha_{11},\ldots,\alpha_{1P},\ldots, \alpha_{LP}\right)^{t}$ is a vector of unknown regression coefficients of dimension $(LP \times 1)$. Here $B_{lp}(u,v) = \invbreve B_l(u)\breve B_p(v)$ implicitly defines a bivariate B-spline basis function as the tensor product of two univariate B-splines. Under this representation, model (\ref{bsp_m}) can be expressed in matrix notation as
\begin{equation}
\label{e_bsp2d}
\boldsymbol{y} =  \boldsymbol{B}\boldsymbol{\alpha} + \boldsymbol{\varepsilon},
\end{equation}
with
\[
\boldsymbol{B} = \breve{\boldsymbol{B}}\Box\invbreve{\boldsymbol{B}} = \left(\breve{\boldsymbol{B}}\otimes\boldsymbol{1}^{t}_L\right)\odot \left(\boldsymbol{1}^{t}_P\otimes\invbreve{\boldsymbol{B}}\right),
\] 
where $\otimes$ denotes the Kronecker product and $\Box$ the `row-wise' Kronecker product \citep{Eilers2006}. Note that model (\ref{e_bsp2d}) is purely parametric and can thus be estimated by minimizing the residual sum of squares (with the explicit solution $\widehat{\boldsymbol{\alpha}} = (\boldsymbol{B}^{t}\boldsymbol{B})^{-1}\boldsymbol{B}^{t}\boldsymbol{y}.$). To prevent over-fitting, \cite{Eilers1996} propose to modify the least squares criterion by incorporating a discrete penalty on the coefficient associated to adjacent B-splines. For the two-dimensional case, the vector $\boldsymbol{\alpha}$ can be seen as an $(L \times P)$ matrix of coefficients, $\boldsymbol{A} = [\alpha_{lp}]$. Now the rows and columns of $\boldsymbol{A}$ correspond to the regression coefficients in the $v$ and $u$ direction, respectively. In anisotropic P-splines, a different amount of smoothing is assumed along the $u$ and $v$ directions. It leads to two penalties: one on all rows of $\boldsymbol{A}$, the other on all of its columns; and the penalized least squares objective function becomes \citep{Eilers2003}
\begin{align}\label{e_psp2d}
    S^* = ||\boldsymbol{y} - \boldsymbol{B}\boldsymbol{\alpha}||^2 +
    \invbreve\lambda ||\invbreve{\boldsymbol{D}}\boldsymbol{A}||^2_F + \breve \lambda ||\boldsymbol{A}\breve{\boldsymbol{D}}^{t}||^2_F = ||\boldsymbol{y} - \boldsymbol{B}\boldsymbol{\alpha}||^2 + \boldsymbol{\alpha}^{t}\boldsymbol{P}\boldsymbol{\alpha},
 \end{align}
where $\boldsymbol{P} = \invbreve\lambda(\boldsymbol{I}_P\otimes \invbreve{\boldsymbol{D}}^{t}\invbreve{\boldsymbol{D}}) + \breve \lambda(\breve{\boldsymbol{D}}^{t}\breve{\boldsymbol{D}} \otimes \boldsymbol{I}_L)$ is the penalty matrix, $\invbreve\lambda$ and $\breve \lambda$ are the smoothing parameters acting, respectively, on the columns and rows of $\boldsymbol{A}$, and $\invbreve{\boldsymbol{D}}$ and $\breve{\boldsymbol{D}}$ are matrices that form differences of order $d_u$ and $d_v$ respectively (in this paper we use second order differences, i.e., $d_u = d_v = 2$). Finally, $||\boldsymbol{M}||^2_F$ denotes the Frobenius norm, i.e., the sum of the squares of the elements of the matrix $\boldsymbol{M}$. The minimizer of (\ref{e_psp2d}), given $\invbreve\lambda$ and $\breve\lambda$, is
\begin{equation}\label{e_solvep2}
    \widehat{\boldsymbol{\alpha}} = (\boldsymbol{B}^{t}\boldsymbol{B} + \boldsymbol{P})^{-1}\boldsymbol{B}^{t}\boldsymbol{y}.
 \end{equation}
The only tuning mechanism for smoothness is now the strength of the penalty, i.e the value of the smoothing parameters $\breve{\lambda}$ and $\invbreve{\lambda}$. The number of B-splines will be purposely chosen so large as to get over-fitting without a penalty. Accordingly, a critical issue is setting the right value for the smoothing parameters, which we like to see determined by the data. We discuss this point in detail in Section~\ref{mm_psplines}. The P-spline principle asks for a generous number of B-splines in the bases. (However, we cannot be too generous, as the size of the systems of equations is $LP$, the product of the sizes of $\invbreve{B}$ and $\breve B$). \ref{s_p2mix} contains a proposal to use smaller, nested, B-spline bases \citep{Lee2013} that strongly reduce the computational effort.

\subsection{Mixed model-based smoothing parameter selection\label{mm_psplines}}
As said before, an important aspect when fitting a P-spline model is choosing appropriate values for the smoothing parameters. \citet{Eilers1996} showed how to use several classical criteria, like cross-validation and Akaike's Information Criterion (AIC). For our application, however, we exploit the formal similarity between P-splines and mixed models \citep{Currie2002, Wand2003}, as it provides a general framework for the analysis of field trials. In this approach, the smooth functions are treated as sums of fixed and random components, and the smoothing parameters are replaced by ratios of variances which are estimated by REML. We present here the main ideas, and refer to \cite{Lee2010, Lee2011} for a more detailed description.

Note that, for given $\breve{\lambda}$ and $\invbreve{\lambda}$, the solution to the penalized least squares problem (\ref{e_psp2d}) given in (\ref{e_solvep2}) corresponds to the BLUPs for the $(LP \times 1)$ vector $\boldsymbol{\alpha}$ under the assumption that $\boldsymbol{\alpha} \sim N(\boldsymbol{0}, \sigma^2\boldsymbol{P}^{+})$. Here, $\boldsymbol{P}^{+}$ denotes the Moore-Penrose pseudoinverse of the penalty matrix $\boldsymbol{P}$ given in (\ref{e_psp2d}). Given that both $\invbreve{\boldsymbol{D}}^{t}\invbreve{\boldsymbol{D}}$ and $\breve{\boldsymbol{D}}^{t}\breve{\boldsymbol{D}}$ are rank-deficient, $\boldsymbol{P}$ is also rank-deficient, and thus $N(\boldsymbol{0}, \sigma^2\boldsymbol{P}^{+})$ is a degenerate distribution. This causes numerical instability when applying mixed model estimation techniques. Moreover, it also implies that, for second order differences, the bilinear polynomial discussed in Section \ref{s_motexample} remains unpenalized.  

To obtain a full rank penalty matrix or precision matrix, the key is to write $\boldsymbol{B}\boldsymbol{\alpha} = \boldsymbol{X}_s\boldsymbol{\beta}_s + \boldsymbol{Z}_s\boldsymbol{c}_s$. There are now two bases, $\boldsymbol{X}_s$, with coefficients that are not penalized at all, and $\boldsymbol{Z}_s$, with a size penalty on its coefficients. There are different ways to decompose $\boldsymbol{B}$ \citep[see, e.g.,][]{Ruppert2003,Eilers1999,Currie2006}. In this paper we follow the proposal by \cite{Lee2011}, based on the eigenvalue decomposition (EVD) of $\invbreve{\boldsymbol{D}}^{t}\invbreve{\boldsymbol{D}}$ and $\breve{\boldsymbol{D}}^{t}\breve{\boldsymbol{D}}$. This approach gives rise to a diagonal penalty matrix, making it very appealing from a computational point of view.

Let $\invbreve{\boldsymbol{D}}^{t}\invbreve{\boldsymbol{D}} = \boldsymbol{U}_u\boldsymbol{E}_u\boldsymbol{U}_u^{t}$ and $\breve{\boldsymbol{D}}^{t}\breve{\boldsymbol{D}} = \boldsymbol{U}_v\boldsymbol{E}_v\boldsymbol{U}_{v}^{t}$ be the EVD of $\invbreve{\boldsymbol{D}}^{t}\invbreve{\boldsymbol{D}}$ and $\breve{\boldsymbol{D}}^{t}\breve{\boldsymbol{D}}$ respectively. Here $\boldsymbol{U}_j$ denotes the matrix of eigenvectors and $\boldsymbol{E}_j$ the diagonal matrix of eigenvalues ($j = u,v$). If we use second order differences, $\boldsymbol{E}_u$ and $\boldsymbol{E}_v$ contain $L - 2$ and $P - 2$ non-zero values respectively. Let us also denote by $\widetilde{\boldsymbol{U}}_j$ and $\widetilde{\boldsymbol{E}}_j$ the sub-matrices corresponding to the non-zero eigenvalues. Setting
\begin{equation}
\boldsymbol{X}_s = \left[\boldsymbol{1}_n,\boldsymbol{u},\boldsymbol{v},\boldsymbol{u}\odot\boldsymbol{v}\right] \;\;\;\mbox{and}\;\;\;\boldsymbol{Z}_s = \left[\boldsymbol{Z}_v,\boldsymbol{Z}_u,\boldsymbol{Z}_v\Box\boldsymbol{u},\boldsymbol{v}\Box\boldsymbol{Z}_u,\boldsymbol{Z}_v\Box\boldsymbol{Z}_u\right],
\label{mm_XZ_psp2d}
\end{equation}
where $\boldsymbol{Z}_u = \breve{\boldsymbol{B}}\widetilde{\boldsymbol{U}}_u$ and $\boldsymbol{Z}_v = \invbreve{\boldsymbol{B}}\widetilde{\boldsymbol{U}}_v$, the penalized least squares problem (\ref{e_psp2d}) becomes
\begin{equation}\label{e_psp2d_mm}
S^{*} = ||\boldsymbol{y} - \boldsymbol{X}_s\boldsymbol{\beta}_s - \boldsymbol{Z}_s\boldsymbol{c}_s||^2 + \boldsymbol{c}_s^{t}\widetilde{\boldsymbol{P}}\boldsymbol{c}_s,
\end{equation}
with
\begin{equation*}
\widetilde{\boldsymbol{P}} = \mbox{blockdiag}\left(\breve{\lambda}\widetilde{\boldsymbol{E}}_v, \invbreve{\lambda}\widetilde{\boldsymbol{E}}_u, \breve{\lambda}\widetilde{\boldsymbol{E}}_v, \invbreve{\lambda}\widetilde{\boldsymbol{E}}_u, \breve{\lambda}\widetilde{\boldsymbol{E}}_v \otimes \boldsymbol{I}_{L - 2} + \invbreve{\lambda}\boldsymbol{I}_{P - 2}\otimes \widetilde{\boldsymbol{E}}_u\right).
\end{equation*}
Each block in $\widetilde{\boldsymbol{P}}$ corresponds to each block in $\boldsymbol{Z}_s$ (see (\ref{mm_XZ_psp2d})). The solution to (\ref{e_psp2d_mm}), for given $\breve{\lambda}$ and $\invbreve{\lambda}$, corresponds to the empirical best linear unbiased estimator (BLUE) for the $(4 \times 1)$ vector $\boldsymbol{\beta}_s$, and the BLUPs for the $((LP - 4) \times 1)$ vector $\boldsymbol{c}_s$ under the linear mixed model
\begin{equation*}
\boldsymbol{y} = \boldsymbol{X}_s\boldsymbol{\beta}_s + \boldsymbol{Z}_s\boldsymbol{c}_s + \boldsymbol{\varepsilon},\;\mbox{with}\;\boldsymbol{\varepsilon}\sim N\left(\boldsymbol{0}, \sigma^2\boldsymbol{I}_{n}\right)\;\mbox{and}\;\boldsymbol{c}_s\sim N\left(\boldsymbol{0}, \boldsymbol{G}_s\right), 
\label{mm_psp}
\end{equation*}
where $\boldsymbol{G}_s = \sigma^2\widetilde{\boldsymbol{P}}^{-1}$. Denoting as $\breve{\sigma}^2 = \sigma^2/\breve{\lambda}$ and $\invbreve{\sigma}^2 = \sigma^2/\invbreve{\lambda}$ the variance parameters involved in $\boldsymbol{G}_s$, it follows  
\begin{equation}
\boldsymbol{G}_s^{-1} = \mbox{blockdiag}\left(\frac{1}{\breve{\sigma}^2}\widetilde{\boldsymbol{E}}_v, \frac{1}{\invbreve{\sigma}^2}\widetilde{\boldsymbol{E}}_u, \frac{1}{\breve{\sigma}^2}\widetilde{\boldsymbol{E}}_v, \frac{1}{\invbreve{\sigma}^2}\widetilde{\boldsymbol{E}}_u, \frac{1}{\breve{\sigma}^2}\widetilde{\boldsymbol{E}}_v \otimes \boldsymbol{I}_{L - 2} + \frac{1}{\invbreve{\sigma}^2}\boldsymbol{I}_{P - 2}\otimes \widetilde{\boldsymbol{E}}_u\right).
\label{g_inv_psp2d}
\end{equation}
Besides the possibility of selecting the smoothing parameters ($\breve{\lambda} = \sigma^2/\breve{\sigma}^2$ and $\invbreve{\lambda} = \sigma^2/\invbreve{\sigma}^2$) by (RE)ML, the mixed model representation of the tensor-product P-spline directly provides the interesting ANOVA-type decomposition discussed in Section \ref{s_motexample} \citep{Lee2011, Lee2013}. The block-structure of both $\boldsymbol{X}_s$ and $\boldsymbol{Z}_s$ (see (\ref{mm_XZ_psp2d})) implies 
\begin{align*}
f(\boldsymbol{u},\boldsymbol{v}) & = \boldsymbol{X}_s\boldsymbol{\beta}_s + \boldsymbol{Z}_s\boldsymbol{c}_s\\
& = \boldsymbol{1}_n\beta_0 + \boldsymbol{u}\beta_1 + \boldsymbol{v}\beta_2 + \boldsymbol{u}\odot\boldsymbol{v}\beta_3\\
& + \underbrace{f_v(\boldsymbol{v})}_{\boldsymbol{Z}_v\boldsymbol{c}_{s1}} + \underbrace{f_u(\boldsymbol{u})}_{\boldsymbol{Z}_u\boldsymbol{c}_{s2}} + \underbrace{\boldsymbol{u}\odot h_{v}(\boldsymbol{v})}_{\left[\boldsymbol{Z}_v\Box\boldsymbol{u}\right]\boldsymbol{c}_{s3}} + \underbrace{\boldsymbol{v}\odot h_{u}(\boldsymbol{u})}_{\left[\boldsymbol{v}\Box\boldsymbol{Z}_u\right]\boldsymbol{c}_{s4}}+ \underbrace{f_{u,v}(\boldsymbol{u}, \boldsymbol{v})}_{\left[\boldsymbol{Z}_v\Box\boldsymbol{Z}_u\right]\boldsymbol{c}_{s5}},
\end{align*}
where $\boldsymbol{c}_{sk}$ ($k = 1,\ldots,5$) contains the elements of $\boldsymbol{c}_s$ that correspond to the $k$-th block of $\boldsymbol{Z}_s$, i.e., $\boldsymbol{c}_s = \left(\boldsymbol{c}_{s1}^{t},\ldots,\boldsymbol{c}_{s5}^{t}\right)^{t}$. The dimension of each $\boldsymbol{c}_{sk}$ will depend on the basis dimensions, $L$ and $P$, used for the tensor product (eqn. (\ref{e_bsp2d})). It is easy to show that $\boldsymbol{c}_{s1}$ and $\boldsymbol{c}_{s3}$ are $\left((P-2) \times 1\right)$ vectors, $\boldsymbol{c}_{s2}$ and $\boldsymbol{c}_{s4}$ are $\left((L - 2) \times 1\right)$ vectors, and $\boldsymbol{c}_{s5}$ is a $\left((L-2)(P-2) \times 1\right)$ vector. The nested bases presented in~\ref{s_p2mix} allow reducing the dimension associated to the smooth-by-smooth interaction term, and, as a consequence, the computational effort.

A close look at (\ref{g_inv_psp2d}) shows that, despite the five smooth components, only two variance parameters (or smoothing parameters) control their smoothness: $\breve{\sigma}^2$ and $\invbreve{\sigma}^2$. In fact, the same variance parameter apply to both the main effects and the interaction terms. In \cite{Lee2013} the ANOVA-type decomposition is further exploited, and the authors propose to use a different variance component for each smooth component, i.e., each block in (\ref{g_inv_psp2d}) will have its own variance component. For ease of notation, let $\boldsymbol{\Lambda}^{-1}_{s1}$ =  $\boldsymbol{\Lambda}^{-1}_{s3}$ = $\widetilde{\boldsymbol{E}}_v$, $\boldsymbol{\Lambda}^{-1}_{s2}$ =  $\boldsymbol{\Lambda}^{-1}_{s4}$ = $\widetilde{\boldsymbol{E}}_u$, and $\boldsymbol{\Lambda}^{-1}_{s5}$ =  $\widetilde{\boldsymbol{E}}_v \otimes \boldsymbol{I}_{L - 2} + \boldsymbol{I}_{P - 2}\otimes \widetilde{\boldsymbol{E}}_u$. For the PS-ANOVA model the precision matrix is then defined as

\begin{equation*}
\boldsymbol{G}_s^{-1} = \mbox{blockdiag}\left(\frac{1}{\sigma_{s1}^2}\boldsymbol{\Lambda}^{-1}_{s1}, \frac{1}{\sigma_{s2}^2}\boldsymbol{\Lambda}^{-1}_{s2}, \frac{1}{\sigma_{s3}^2}\boldsymbol{\Lambda}^{-1}_{s3}, \frac{1}{\sigma_{s4}^2}\boldsymbol{\Lambda}^{-1}_{s4}, \frac{1}{\sigma_{s5}^2}\boldsymbol{\Lambda}^{-1}_{s5}\right),
\label{g_inv_psp2d_anova}
\end{equation*}
and thus the variance-covariance matrix $\boldsymbol{G}_s$ is a linear function of variance parameters
\begin{equation}
\boldsymbol{G}_s =  \bigoplus_{k = 1}^{5}\sigma_{sk}^2\boldsymbol{\Lambda}_{sk} =  \bigoplus_{k = 1}^{5}\boldsymbol{G}_{sk} = \mbox{blockdiag}\left(\boldsymbol{G}_{s1}, \boldsymbol{G}_{s2}, \boldsymbol{G}_{s3}, \boldsymbol{G}_{s4}, \boldsymbol{G}_{s5}\right),
\label{g_psp2d_anova}
\end{equation}
where $\boldsymbol{G}_{sk} = \sigma_{sk}^2\boldsymbol{\Lambda}_{sk}$ ($k = 1, \ldots,5$). In other words, here the tensor product P-spline mixed model is represented as the sum of $5$ sets of mutually independent Gaussian random factors $\boldsymbol{c}_{sk}$ each depending on one variance $\sigma^2_{sk}$ ($k = 1, \ldots, 5$).
\section{Spatial models for field trials\label{s_trials}}
The tensor product P-spline presented in Section \ref{s_splines} constitutes the base for the analysis of agricultural field trials. As said, it allows the modeling of the random spatial variation typically presented in a field, providing an explicit estimate of the spatial random field. However, on top of this spatial field, we need to build up more complex models in order to account for the genetic variation, the presence of block and/or replication effects, or other sources of spatial variation as those due to the way the field was prepared. From now on, we therefore consider the following linear mixed model 
\begin{equation}
\boldsymbol{y} = \underbrace{\boldsymbol{X}_s\boldsymbol{\beta}_s + \boldsymbol{Z}_s\boldsymbol{c}_s}_{f\left(\boldsymbol{u}, \boldsymbol{v}\right)} + \boldsymbol{X}_d\boldsymbol{\beta}_d + \boldsymbol{Z}_d\boldsymbol{c}_d + \boldsymbol{\varepsilon},\;\mbox{with}\;\boldsymbol{c}_s\sim N\left(\boldsymbol{0},\boldsymbol{G}_{s}\right) \;\mbox{and}\;\boldsymbol{c}_d\sim N\left(\boldsymbol{0}, \boldsymbol{G}_d\right),
\label{mm_equation_orig}
\end{equation}
where $\boldsymbol{X}_s$, $\boldsymbol{Z}_s$ and $\boldsymbol{G}_{s}$ have been defined in (\ref{mm_XZ_psp2d}) and (\ref{g_psp2d_anova}). $\boldsymbol{X}_d$ and $\boldsymbol{Z}_d$ represent column-partitioned matrices, associated respectively with extra fixed and random components, as for instance, row, column, replicate and/or genotypic effects. We assume that $\boldsymbol{X}_d$ has full rank, $\boldsymbol{Z}_d = \left[\boldsymbol{Z}_{d1},\ldots,\boldsymbol{Z}_{db}\right]$, and $\boldsymbol{c}_{d} = \left(\boldsymbol{c}_{d1}^{t},\ldots,\boldsymbol{c}_{db}^{t}\right)^{t}$. Each $\boldsymbol{Z}_{dk}$ corresponds to the design matrix of the $k$-th random factor $\boldsymbol{c}_{dk}$, with $\boldsymbol{c}_{dk}$ being a $(m_{dk} \times 1)$ vector ($k = 1, \ldots, b$). We assume further that $\boldsymbol{c}_s$ and $\boldsymbol{c}_d$ are independent, and that the $b$ components of $\boldsymbol{c}_d$ are mutually independent with diagonal variance-covariance matrices $\sigma_{dk}^2\boldsymbol{\Lambda}_{dk}$, i.e., $\boldsymbol{G}_d =  \bigoplus_{k = 1}^{b}\boldsymbol{G}_{dk} = \bigoplus_{k = 1}^{b}\sigma_{dk}^2\boldsymbol{\Lambda}_{dk}$. In spite of this restriction, the estimation of rather complex mixed models can be accommodated, as it will be shown in Section \ref{s_app}. In order to keep the notation as simple as possible, we rewrite model (\ref{mm_equation_orig}) in a more compact way as follows
\begin{equation}
\boldsymbol{y} = \boldsymbol{X}\boldsymbol{\beta} + \boldsymbol{Z}\boldsymbol{c} + \boldsymbol{\varepsilon},\;\mbox{with}\;\boldsymbol{c}\sim N\left(\boldsymbol{0},\boldsymbol{G}\right)\;\mbox{and}\;\boldsymbol{\varepsilon}\sim N\left(\boldsymbol{0}, \sigma^2\boldsymbol{I}_n\right), 
\label{mm_equation}
\end{equation}
where $\boldsymbol{X} = \left[\boldsymbol{X}_s, \boldsymbol{X}_d\right]$, $\boldsymbol{Z} = \left[\boldsymbol{Z}_s, \boldsymbol{Z}_d\right] = \left[\boldsymbol{Z}_1,\ldots,\boldsymbol{Z}_q\right]$ $(q = 5 + b)$, and
\begin{equation}
\boldsymbol{G} = \mbox{blockdiag}\left(\boldsymbol{G}_s, \boldsymbol{G}_d\right) = \bigoplus_{k = 1}^{q}\boldsymbol{G}_{k} = \bigoplus_{k = 1}^{q}\sigma_{k}^2\boldsymbol{\Lambda}_{k}.
\label{G_mm_equation}
\end{equation}

As far as estimation of model (\ref{mm_equation}) is concerned, estimates of the fixed and random effect coefficients, for given values of the variance components, follow from standard mixed-model theory (see \ref{mixed_model_results}), and variance components can be obtained, as usual, by maximizing the REML log-likelihood function
\begin{equation}
l = - \frac{1}{2}\log|\boldsymbol{V}| - \frac{1}{2}\log|\boldsymbol{X^{t}V^{-1}X}| - \frac{1}{2}(\boldsymbol{y} - \boldsymbol{X}\widehat{\boldsymbol{\beta}})^{t}\boldsymbol{V}^{-1}(\boldsymbol{y} - \boldsymbol{X}\widehat{\boldsymbol{\beta}}). 
\label{REML_cri}
\end{equation}
Given that $\boldsymbol{G}$ is a linear function of variance components, estimation can be accommodated using standard mixed model procedures, as, e.g., those implemented in the \texttt{R}-packages \texttt{asreml-R}, \texttt{nlme} and \texttt{lme4}, or the \texttt{PROC MIXED} procedure in SAS$^\circledR$. In next Section we present the numerical procedure implemented in the \texttt{R}-package \texttt{SpATS} that accompany this paper. The procedure presents many appealing features, which make it a good candidate for the analysis of field trials: (a) it is fast and stable; (b) it is robust (i.e., it converges from almost any starting values); and (c) it always provides positive estimates of the variance components, although it is possible to obtain values very close to zero. For all these reasons it has been our choice.

\subsection{Variance component estimation\label{s_estproc}}
As said, REML estimates of the variance components are obtained by maximizing (\ref{REML_cri}). Taking derivatives with respect to the variance components $\sigma_k^2$ ($k = 1,\ldots,q$), we obtain \citep[see e.g.,][]{MXRA2014,Johnson1995}
\begin{equation*}
\frac{\partial{l}}{\partial{\sigma_k^2}} = - \frac{1}{2}trace\left(\boldsymbol{Z}^{t}\boldsymbol{Q}\boldsymbol{Z}\boldsymbol{G}\frac{\partial{\boldsymbol{G}^{-1}}}{\partial{\sigma_k^2}}\boldsymbol{G}\right) + \frac{1}{2}\widehat{\boldsymbol{c}}^{t}\frac{\partial{\boldsymbol{G}^{-1}}}{\partial{\sigma_k^2}}\widehat{\boldsymbol{c}},
\label{MX:rloglikder}
\end{equation*}
where $\boldsymbol{Q} = \boldsymbol{V}^{-1} - \boldsymbol{V^{-1}}\boldsymbol{X}\left(\boldsymbol{X^{t}V^{-1}X}\right)^{-1}\boldsymbol{X}^{t}\boldsymbol{V^{-1}}$ with $\boldsymbol{V} = \boldsymbol{R} + \boldsymbol{Z}\boldsymbol{G}\boldsymbol{Z}^{t}$ and $\boldsymbol{R} = \sigma^2\boldsymbol{I}_n$. By~(\ref{G_mm_equation}), it is easy to show that the former derivatives can be expressed as
\begin{equation*}
2\frac{\partial{l}}{\partial{\sigma_k^2}} = - \frac{1}{\sigma_k^2}trace\left(\boldsymbol{Z}_k^{t}\boldsymbol{Q}\boldsymbol{Z}_k\boldsymbol{G}_k\right) + \frac{1}{\sigma_k^4}\widehat{\boldsymbol{c}}_k^{t}\boldsymbol{\Lambda}_{k}^{-1}\widehat{\boldsymbol{c}}_k.
\end{equation*}
Then, REML estimates of the variance components are found by equating the former expression to zero, which gives
\begin{equation}
\widehat{\sigma}_k^{2} = \frac{\widehat{\boldsymbol{c}}_k^{t}\boldsymbol{\Lambda}_{k}^{-1}\widehat{\boldsymbol{c}}_k}{\mbox{ED}_k}, k = 1,\ldots,q,
\label{varcomp_esp}
\end{equation}
with 
\begin{equation}
\mbox{ED}_k = \mbox{trace}\left(\boldsymbol{Z}_k^{t}\boldsymbol{Q}\boldsymbol{Z}_k\boldsymbol{G}_k\right).
\label{ed_sap}
\end{equation}
An estimate of $\sigma^2$ can also be easily obtained following the same reasoning. Specifically, in this case we have
\begin{equation}
\widehat{\sigma}^{2} = \frac{\widehat{\boldsymbol{\varepsilon}}^{t}\widehat{\boldsymbol{\varepsilon}}}{\mbox{ED}_{\boldsymbol{\varepsilon}}},
\label{varcomp_sigma}
\end{equation}
where $\widehat{\boldsymbol{\varepsilon}} = \boldsymbol{y} - \boldsymbol{X}\widehat{\boldsymbol\beta} + \boldsymbol{Z}\widehat{\boldsymbol{c}}$ and 
\begin{equation}
 \mbox{ED}_{\boldsymbol{\varepsilon}} = \mbox{trace}\left(\boldsymbol{R}\boldsymbol{Q}\right) = n - \mbox{rank}(\boldsymbol{X})- \sum_{k = 1}^{q}\mbox{ED}_k.
\label{ed_sigma}
\end{equation}
As can be seen, the right-hand side of eqn. (\ref{ed_sap}) depends on the unknown variance components. Hence, eqns. (\ref{varcomp_esp}) and (\ref{varcomp_sigma}) need to be solved with an iterative procedure. Given some starting values for the variance components, estimation of model (\ref{mm_equation}) is thus obtained by iterating, until convergence, among (a) estimating the fixed and random effect coefficients (linear system (\ref{MX:linearsystem})); (b) evaluating the right-hand side of eqn. (\ref{ed_sap}); and (c) updating the variances by means of eqns. (\ref{varcomp_esp}) and (\ref{varcomp_sigma}). In this work, the REML-deviance was used as the convergence criterion. 

To the best of our knowledge, this iterative algorithm was originally proposed by Henderson in an unpublished manuscript, and discussed in detail by \cite{Harville77} and \cite{Engel1990}, among others. \cite{Schall1991} further extended the algorithm for the estimation of generalized linear mixed models. In the P-spline literature, the algorithm has been also successfully used \citep[e.g.,][]{Schnabel2009, Lee2013}, and usually referred to as Schall's algorithm.

From a computational point of view, the traces in (\ref{ed_sap}) may involve the computation and manipulations of several large matrices. However, there are several ways this computation can be relaxed. For instance, using some results on mixed models \cite[see, e.g.,][]{Johnson1995} we have that
\begin{equation}
\boldsymbol{Z}_k^{t}\boldsymbol{Q}\boldsymbol{Z}_k\boldsymbol{G}_k  = \boldsymbol{I}_{m_k} - \boldsymbol{G}_k^{-1}\boldsymbol{C}^{-1}_{kk} = \boldsymbol{I}_{m_k} - \frac{1}{\sigma_k^2}\boldsymbol{\Lambda}_k^{-1}\boldsymbol{C}^{-1}_{kk},
\label{equivalence_hat_matrices} 
\end{equation}
where $m_k$ is the number of random coefficients in the vector $\boldsymbol{c}_k$ and $\boldsymbol{C}^{-1}_{kk}$ is that partition of the inverse of $\boldsymbol{C}$ in (\ref{MX:linearsystem}) corresponding to $\boldsymbol{c}_k$. An alternative approach would be to use the result given in eqn. (5.3) in \cite{Harville77}
\[
\boldsymbol{Z}^{t}\boldsymbol{Q}\boldsymbol{Z} = \boldsymbol{G}^{-1}\boldsymbol{C}^{-1}_{m}\left[\boldsymbol{X},\boldsymbol{Z}\right]^{t}\boldsymbol{R}^{-1}\boldsymbol{Z},
\]
where $\boldsymbol{C}^{-1}_{m}$ denotes the matrix formed by the last $m$ rows of the inverse of $\boldsymbol{C}$ (with $m = \sum_{k= 1}^{q}m_k$). Here, the block-diagonal elements of $\boldsymbol{Z}^{t}\boldsymbol{Q}\boldsymbol{Z}$ correspond to $\boldsymbol{Z}_k^{t}\boldsymbol{Q}\boldsymbol{Z}_k$. Despite the apparent complexity of this expression, in order to compute (\ref{ed_sap}) only the diagonal of $\boldsymbol{Z}^{t}\boldsymbol{Q}\boldsymbol{Z}$ needs to be explicitly obtained, since $\boldsymbol{G}$ is a diagonal matrix \citep[see][ for further details]{MXRA2014}.

\subsection{Effective dimensions}\label{eff_dim}
Denoting the denominator of (\ref{varcomp_esp}) and (\ref{varcomp_sigma}) as $\mbox{ED}_{\{\cdot\}}$ (from ``effective dimension'') has not been done without purpose. In the smoothing context, the notion of effective dimension or effective degrees of freedom is well known \citep[see, e.g.,][]{Hastie1990}. The effective dimension (denoted by $\mbox{ED}$) of a ``smooth'' model is defined as the trace of the so-called ``hat'' matrix $\boldsymbol{H}$, defined as $\widehat{\boldsymbol{y}} = \boldsymbol{H}\boldsymbol{y}$. In this setting, $\mbox{ED}$ can be interpreted as a measure of the complexity of the model: the larger the $\mbox{ED}$, the more complex (or the less smooth) the model \cite[see also][]{Ye1998, Eilers2015}. 

In recent years, several new definitions and generalizations of the concept of effective dimension that are applicable to (generalized) linear mixed models have been proposed in the statistical literature \cite[see, e.g.,][and references therein]{You2016}. In almost all cases, the aim has been to provide a complexity measure that allows models' comparison and selection (via, for instance, the AIC). For our application, however, we are more interested in obtaining a separate complexity measure for each component in model (\ref{mm_equation}), that can give us insights about the contribution of that effect when explaining the response (phenotypic) variation. This issue has been already discussed by \cite{Cui2010}. In that paper the authors define the effective dimension of a model's component as the trace of the ratio of the component modeled variance matrix to the total variance matrix. This new definition can be interpreted as the fraction of response variation attributed to individual components. Besides, it allows explaining how components compete with one another to explain that variation.   

In line with previous work in the smoothing context \citep[e.g.,][]{Hastie1990, Ruppert2003}, this paper considers defining the effective dimension of a model's component as the trace of the corresponding hat matrix. \ref{ed_equivalence} shows the equivalence between this definition and that by \cite{Cui2010}. Note first that for the linear mixed model (\ref{mm_equation}) we might define two different hat matrices: one for the fixed part of the model and one for the random part. From standard mixed-model theory (see \ref{mixed_model_results} for further details) it follows that
\[
\boldsymbol{H}\boldsymbol{y} = \widehat{\boldsymbol{y}} = \boldsymbol{X}\widehat{\boldsymbol{\beta}} + \boldsymbol{Z}\widehat{\boldsymbol{c}} = \boldsymbol{H}_ {F}\boldsymbol{y} + \boldsymbol{H}_ {R}\boldsymbol{y},
\]
where $\boldsymbol{H}_ {F} = \boldsymbol{X}\left(\boldsymbol{X}^{t}\boldsymbol{V}^{-1}\boldsymbol{X}\right)^{-1}\boldsymbol{X}^{t}\boldsymbol{V}^{-1}$ and $
\boldsymbol{H}_ {R} = \boldsymbol{Z}\boldsymbol{G}\boldsymbol{Z}^{t}\boldsymbol{Q}$. Thus,
\[
\mbox{ED} = \mbox{trace}\left(\boldsymbol{H}\right) = \mbox{trace}\left(\boldsymbol{H}_{F}\right) + \mbox{trace}\left(\boldsymbol{H}_{R}\right).
\]
However, we can even go one step further. The block structure of both the random design matrix $\boldsymbol{Z}$ and the variance-covariance matrix $\boldsymbol{G}$ in (\ref{mm_equation}) implies that (see \ref{mixed_model_results})
\[
\boldsymbol{H}_{R} = \sum_{k=1}^{q}\boldsymbol{H}_{k}
\]
where $\boldsymbol{H}_{k} = \boldsymbol{Z}_k\boldsymbol{G}_k\boldsymbol{Z}_k^{t}\boldsymbol{Q}$, with $\boldsymbol{Z}_k$ and $\boldsymbol{G}_k$ denoting the $k$-th block of $\boldsymbol{Z}$ and $\boldsymbol{G}$ respectively. As shown in \ref{mixed_model_results}, each of these hat matrices corresponds to an individual random component in model (\ref{mm_equation}), either coming from the PS-ANOVA spatial field or a ``pure'' random factor, i.e., 
\begin{equation*}
\widehat{f}_v(\boldsymbol{v}) = \boldsymbol{Z}_1\widehat{\boldsymbol{c}}_1 = \boldsymbol{H}_1\boldsymbol{y}
\qquad
\widehat{f}_u(\boldsymbol{u}) = \boldsymbol{Z}_2\widehat{\boldsymbol{c}}_2 = \boldsymbol{H}_2\boldsymbol{y}
\qquad
\boldsymbol{u}\odot\widehat{h}_v(\boldsymbol{v}) = \boldsymbol{Z}_3\widehat{\boldsymbol{c}}_3 = \boldsymbol{H}_3\boldsymbol{y},
\end{equation*}
\begin{equation*}
\boldsymbol{v}\odot\widehat{h}_u(\boldsymbol{u}) = \boldsymbol{Z}_4\widehat{\boldsymbol{c}}_4 = \boldsymbol{H}_4\boldsymbol{y}
\qquad
\widehat{f}_{u,v}(\boldsymbol{u}, \boldsymbol{v}) = \boldsymbol{Z}_5\widehat{\boldsymbol{c}}_5 = \boldsymbol{H}_5\boldsymbol{y}.
\end{equation*}
\[
\boldsymbol{Z}_k\widehat{\boldsymbol{c}}_k = \boldsymbol{H}_k\boldsymbol{y}, k = 6,\ldots,q.
\]
This results suggests defining the effective dimension for $\boldsymbol{c}_k$ as the trace of $\boldsymbol{H}_k$. Using trace properties, we have
\[
\mbox{trace}\left(\boldsymbol{H}_{k}\right) = \mbox{trace}\left(\boldsymbol{Z}_k\boldsymbol{G}_k\boldsymbol{Z}_k^{t}\boldsymbol{Q}\right) = \mbox{trace}\left(\boldsymbol{Z}_k^{t}\boldsymbol{Q}\boldsymbol{Z}_k\boldsymbol{G}_k\right) = \mbox{ED}_k,
\]
and the total effective dimension of model (\ref{mm_equation}) is thus decomposed as the sum of independent contributions
\begin{align*}
\mbox{ED} & = \mbox{trace}\left(\boldsymbol{H}\right) \\
					& = \mbox{trace}\left(\boldsymbol{H}_F\right) + \mbox{trace}\left(\boldsymbol{H}_R\right) \\
					& = \mbox{trace}\left(\boldsymbol{H}_F\right) + \sum_{k = 1}^{q}\mbox{trace}\left(\boldsymbol{H}_k\right)\\
					& = \mbox{rank}(\boldsymbol{X}) + \sum_{k = 1}^{q}\mbox{ED}_k.
\end{align*}
Besides, eqn. (\ref{hat_residuals}) in \ref{mixed_model_results} also suggests defining the effective dimension for the residuals as the trace of $\boldsymbol{H}_{\boldsymbol{\varepsilon}}$ and thus (see eqn. (\ref{ed_sigma}))
\[
\mbox{trace}\left(\boldsymbol{H}_{\varepsilon}\right) = \mbox{trace}\left(\boldsymbol{R}\boldsymbol{Q}\right) = \mbox{ED}_{\varepsilon}.
\]
This is in concordance with the definition given by \cite{Cui2010} (see \ref{ed_equivalence}). As extensively discussed by the authors, $\mbox{ED}_{\boldsymbol{\varepsilon}}$ avoids the problem that is posed by most of the traditional definitions for the residual effective dimension in the smoothing context, that is to say, that $\mbox{ED}$ + $\mbox{ED}_{\boldsymbol{\varepsilon}}$ is not equal to the number of observations $n$.

For all components $\mbox{ED}_k$ ($k = 1,\ldots,q$) will vary between $0$ and $\left(m_k - \zeta_k\right)$, where $\zeta_k$ is the number of zero eigenvalues of $\boldsymbol{Z}_k^{t}\boldsymbol{Q}\boldsymbol{Z}_k\boldsymbol{G}_k$. In \cite{Cui2010} the authors show that this upper bound, $\left(m_k - \zeta_k\right)$, can also be expressed as: $\mbox{rank}\left(\left[\boldsymbol{X}, \boldsymbol{Z}_k\right]\right) - \mbox{rank}\left(\boldsymbol{X}\right)$. The signal-to-noise ratio $\sigma_k^2/\sigma^2$ modulates the value of $\mbox{ED}_k$: when $\sigma_k^2/\sigma^2 \rightarrow 0$ then $\mbox{ED}_k \rightarrow 0$; and when $\sigma_k^2/\sigma^2\rightarrow \infty$, then $\mbox{ED}_k \rightarrow \left(m_k - \zeta_k\right)$. Arguably, $\mbox{ED}_k$ can be therefore be interpreted as a measure of the complexity of the corresponding component. A value of zero would indicate that this component does not contribute to the response variability. 

A nice property of the PS-ANOVA spatial field is that we have a separate effective dimension for each of the five smooth components, which in turns gives a separate measure of the contribution of that component. In this case, $\mbox{ED}_k$ ($k = 1,\ldots,5$) can be interpreted as a measure of the smoothness of the corresponding term: the larger the effective dimension, the less smooth the effect. For the other random factors, $\mbox{ED}_k$ ($k = 6,\ldots,q$) will also give a measure of the complexity, but here, it is worth interpreting it as a measure of the shrinkage. What is more, when the genetic effect is included in model (\ref{mm_equation}) as random, the associated effective dimension corresponds to the generalized heritability proposed by \cite{Oakey2006}. A formal derivation and deeper discussion is provided in Section~\ref{eff_dim_h2}.

To finish this part we would like to emphasize that, at convergence, the algorithm presented in Section \ref{s_estproc} explicitly provides an estimate of the effective dimension associated to each random component in model (\ref{mm_equation}). Moreover, as can be seen in eqn. (\ref{varcomp_esp}), a beautiful result is that the algorithm furnishes variance components estimates that are the ratio of the sum of squares of the BLUPs of the components of $\boldsymbol{c}_k$ (weighted according to their precision, $\boldsymbol{\Lambda}_k^{-1}$) to the individual effective dimension $\mbox{ED}_k$ (and the same applies for the residual variance).

\subsubsection{Motivating example revisited}
Let us now come back to the uniformity barley data discussed in Section \ref{s_motexample}. For fitting model (\ref{tpspline_model_uni}), we used cubic B-spline bases of dimension $L = 18$ and $P = 51$, jointly with nested basis for the columns with $P_N = 27$. Table \ref{results_uniformity_effdim} shows the model (i.e., the number of coefficients), and effective dimensions associated to the row and column random factors, the smooth spatial field $f(u,v)$, and each of the PS-ANOVA components. If we focus on the random effects for the rows and columns, we have that the estimated effective dimensions are $5.5$ (out of $14$) and $38.0$ (out of $47$) respectively. This result suggests that the column effect is stronger than the row effect, and that these two components are probably needed. For the PS-ANOVA spatial field (excluding the intercept), the total effective dimension is $78.7$, with the smooth-by-smooth interaction trend being responsible for the strongest contribution, with an effective dimension of $53.1$. As can be observed on the graphical results depicted in Section \ref{s_motexample}, the smooth trend along the rows, $f_u(u)$, is more complex (or rougher) than the one along the columns, $f_v(v)$, and this fact is made evident on the effective dimension related to each of these components, $6.2$ and $3.7$ respectively. Besides, as could have also been expected, the effective dimension associated to $vh_u(u)$ is also larger than that associated $uh_v(v)$. These results suggest and emphasize the need of modeling spatial trends by means of bivariate surfaces. Here the additive assumption -- only based on main smooth effects -- would have not been flexible enough to recover the spatial trend variation present in the data.  
\begin{table}
\caption{Model and effective dimension of the smooth spatial component, and the ANOVA-type decomposition components for the barley uniformity trial.}\label{results_uniformity_effdim}
	\center
	\begin{tabular}{ccccccccc}\hline
	\multirow{3}{*}{Dimensions} & \multicolumn{8}{c}{Spatial components}\\
	& \multicolumn{2}{c}{Random} & \multicolumn{6}{c}{Smooth}\\
	& $\boldsymbol{c}_r$ & $\boldsymbol{c}_c$ & Global - $f(u,v)$ & $f_u(u)$ &$f_v(v)$ & $vh_u(u)$ & $uh_v(v)$ & $f_{u,v}(u,v)$ \\\cline{2-9}
	Model ($m_k$) & 48 & 15 & 533 & 16 & 49 & 16 & 49 & 400\\
	Effective (ED$_k$) & 38.0 & 5.5 & 78.7 & 6.2 & 3.7 & 8.2 & 4.5 & 53.1\\\hline
	\end{tabular}
\end{table}	
\subsection{Heritability and effective dimension}\label{eff_dim_h2}
To introduce the standard definition of heritability, let us start with the classical quantitative genetic model, in which $m_g$ genotypes, each replicated $r$ times, are evaluated and no other model components (either spatial, fixed or random) are considered
\[
\boldsymbol{y} = \boldsymbol{1}_n\beta_0 + \boldsymbol{Z}_g\boldsymbol{c}_g + \boldsymbol{\varepsilon}.
\]
We assume that the observations are ordered according to the genotypes, i.e., $\boldsymbol{Z}_g = \boldsymbol{I}_{m_g}\otimes\boldsymbol{1}_{r}$, and that $\boldsymbol{c}_g \sim N(\boldsymbol{0},\sigma_g^2\boldsymbol{I}_{m_g})$ and $\boldsymbol{\varepsilon} \sim N(\boldsymbol{0},\sigma^2\boldsymbol{I}_n)$. Under this model, the 
standard heritability measure is defined as 
\begin{equation}
H_s^2 = \frac{\sigma_g^2}{(\sigma_g^2 + \sigma^2/r)}.
\label{H2_class}
\end{equation}
That is to say, the standard heritability is the proportion of the total (phenotypic) variation explained by the genetic component.

We turn now to the notion of effective dimension defined above. For the genetic effects $\boldsymbol{c}_g$, the associated effective dimension is $\mbox{ED}_g = \mbox{trace}\left(\boldsymbol{Z}_g^{t}\boldsymbol{Q}\boldsymbol{Z}_g\boldsymbol{G}_g\right)$, where $\boldsymbol{G}_g = \sigma_g^2\boldsymbol{I}_{m_g}$. By eqn. (3.7) in \cite{Harville77}, we have
\[
\boldsymbol{Z}_g^{t}\boldsymbol{Q}\boldsymbol{Z}_g\boldsymbol{G}_g = \left(\frac{1}{\lambda}\boldsymbol{I}_{m_g} + \boldsymbol{Z}_g^{T}\boldsymbol{S}\boldsymbol{Z}_g\right)^{-1}\boldsymbol{Z}_g^{T}\boldsymbol{S}\boldsymbol{Z}_g,
\]
where $\lambda = \frac{\sigma_g^2}{\sigma^2}$ and $\boldsymbol{S} = \boldsymbol{I}_{n} - \boldsymbol{1}_n(\boldsymbol{1}_n^{t}\boldsymbol{1}_n)^{-1}\boldsymbol{1}_n^{t}$. It can be shown that $\boldsymbol{Z}_g^{T}\boldsymbol{S}\boldsymbol{Z}_g$ has $(m_g - 1)$ eigenvalues equal to $r$ (the number of replicates), and $1$ eigenvalue equal to zero. Moreover, $\frac{1}{\lambda}\boldsymbol{I}_{m_g}$ has $m_g$ eigenvalues all equal to $\frac{1}{\lambda}$. Using the property that the trace of a matrix is the sum of its eigenvalues, we have 
\[
\mbox{ED}_g = \sum_{i = 1}^{m_g}\lambda_i = \sum_{i = 1}^{(m_g - 1)}\left(\frac{1}{\lambda} + r\right)^{-1}r = (m_g - 1)\frac{\sigma_g^2}{(\sigma_g^2 + \sigma^2/r)},
\]
where $\lambda_i$ are the eigenvalues of $\boldsymbol{Z}_g^{t}\boldsymbol{Q}\boldsymbol{Z}_g\boldsymbol{G}_g$. Hence, there is direct link between the standard heritability measure $H_s^2$ defined in (\ref{H2_class}) and the genetic effective dimension, with 
\[
H_s^2 = \frac{\mbox{ED}_g}{m_g - 1}.
\]
When the statistical analysis of a field trial experiment involves modeling more sources of variation (as, e.g., spatial and/or extraneous variation), the standard definition of heritability given in (\ref{H2_class}) does not longer apply, and several generalizations have been proposed in the literature \citep[e.g.,][]{Cullis2006, Oakey2006}. For instance, \cite{Cullis2006} present a generalized definition of heritability -- applicable whenever $\boldsymbol{G}_g = \sigma_g^2\boldsymbol{I}_{m_g}$ -- based on the pairwise prediction error variance of genetic effects
\[
H^2_c = 1 - \frac{1}{\sigma_g^2}\sum_{i = 1}^{m_g}\frac{pev(c_{gi})}{m_g}.
\]
where $pev(c_{gi}) = var(\widehat{c}_{gi} - c_{gi})$. Using the equivalence given in (\ref{equivalence_hat_matrices}) and noting that $\boldsymbol{C}^{-1}_{kk}$ corresponds to $\mathbb{C}\mbox{ov}\left(\widehat{\boldsymbol{c}}_g -\boldsymbol{c}_g\right)$ (i.e., the prediction error variance-covariance matrix for the genetic effects), the generalized heritability measure proposed by \cite{Cullis2006} can also be expressed in terms of the genetic effective dimension
\[
H^2_c  = 1 - \frac{1}{\sigma_g^2}\sum_{i = 1}^{m_g}\frac{pev(c_{gi})}{m_g}  = \frac{\mbox{trace}(\boldsymbol{I}_{m_g} - \frac{1}{\sigma_g^2}\boldsymbol{C}^{-1}_{kk})}{m_g} = \frac{\mbox{trace}(\boldsymbol{Z}_g^{t}\boldsymbol{Q}\boldsymbol{Z}_g\boldsymbol{G}_g)}{m_g} = \frac{\mbox{ED}_g}{m_g}.
\]
In \cite{Oakey2006} a more general definition of heritability is presented that can be used regardless of the structure of genetic variance-covariance matrix $\boldsymbol{G}_g$. As can be seen in eqn. (7) of that paper, the definition the authors propose turns out to be the ratio between the genetic effective dimension and the number of genetic effects $m_g$ minus the number of zero eigenvalues of $\boldsymbol{Z}_g^{t}\boldsymbol{Q}\boldsymbol{Z}_g\boldsymbol{G}_g$, i.e., 
\begin{equation}
H^2_g = \frac{\mbox{ED}_g}{m_g - \zeta_g}.
\label{H2_oakey}
\end{equation}
As said before, the denominator of ($\ref{H2_oakey}$) represents the upper bound of the genetic effective dimension, and $1 - H^2_g$ can therefore be interpreted as a shrinkage factor. The generalized heritability proposed by \cite{Cullis2006} would be a special case, but ignoring the number of zero eigenvalues. 

On the basis of the estimation procedure presented above, an estimate of the generalized heritability can be thus obtained as 
\[
\widehat{H}^2_g = \frac{\widehat{\mbox{ED}}_g}{m_g - \zeta_g}.
\]
where $\widehat{\mbox{ED}}_g$ denotes the estimated effective.
\section{Simulation studies}\label{sim_studies}
This section is devoted to present the results of several studies performed to evaluate the behavior of our SpATS model under controlled scenarios, and its comparison with the separable autoregressive (AR$\times$AR) model proposed by \cite{Gilmour1997}. In the context of single-trial experiments, this proposal has become the standard modeling strategy, specially among applied breeders, and therefore it has been chosen as the benchmark model. In Section \ref{mot_example_supp_mat}, the uniformity trial presented in Section \ref{s_motexample} is used to introduce the proposal by \cite{Gilmour1997}, and comparisons between both approaches when including genotypic effects are reported. Section \ref{simulation_study_supp_mat} presents the results of a simulation study when the underlying simulated model follows an AR$\times$AR process. In both cases, simulations were done using the $\texttt{R}$-packages \texttt{SpATS} and \texttt{ASreml-R}.
\subsection{Barley uniformity trial}\label{mot_example_supp_mat}
As said, the uniformity trial presented in Section \ref{s_motexample} was analyzed using the AR$\times$AR model proposed by \cite{Gilmour1997}. Model selection was performed by means of the sample variogram and plots of the residuals as suggested by \cite{Stefanova2009}. Starting with the simplest model, including only the separable Gaussian AR process of order $1$, we further evaluated the need of extra model components. In total, $9$ different models were considered, with the best being
\begin{equation}
\boldsymbol{y} = \boldsymbol{1}_n\beta_0 + \boldsymbol{u}\beta_1 + \boldsymbol{v}\beta_2 + f_u(\boldsymbol{u}) + f_v(\boldsymbol{v}) + \boldsymbol{Z}_c\boldsymbol{c}_c + \boldsymbol{\boldsymbol{\xi}} + \boldsymbol{\varepsilon}.
\label{gilmour_model_uni}
\end{equation}
Here $\boldsymbol{\xi}$ is a {($720 \times 1$)} spatially dependent random vector, for which a separable Gaussian AR process of order $1$ in the row and column directions is assumed. Accordingly, $cov\left(\xi_l, \xi_p\right) = \sigma_s^2\rho_r^{\mid u_l - u_p\mid}\rho_c^{\mid v_l - v_p\mid}$, where $\rho_r$ and $\rho_c$ are the autocorrelation parameters for row and column, respectively. As in our approach, $f_u\left(\cdot\right)$ and $f_v\left(\cdot\right)$ represent smooth-trend functions over the row and column direction respectively. We note that in the geostatistics literature, $\boldsymbol{\varepsilon}$ is usually referred to as measurement error or nugget effect.  

As postulated by \cite{Gilmour1997}, model (\ref{gilmour_model_uni}) accounts for three sources of spatial variation: the global trend variation, the local trend variation; and the so-called extraneous variation. Here, (a) the global trend variation is modeled by the linear effect along the rows ($\beta_1$) and columns ($\beta_2$) as well as by the smooth-effect functions $f_u\left(\cdot\right)$ and $f_v\left(\cdot\right)$; and (b) the local trend variation by means of the spatially dependent random error $\boldsymbol{\xi}$. The extraneous variation, related to the experimental procedure, is accounted for by the column random factor $\boldsymbol{c}_c$. Under this framework, we can see our SpATS model (see eqn. (\ref{tpspline_model_uni})) as that based on aggregating both the local and global trend variation in one component, and modeling it by means of a smooth bivariate surface. Table \ref{results_uniformity} shows the REML estimates of the variance components based on model (\ref{tpspline_model_uni}) and model (\ref{gilmour_model_uni}). Based on this table, we find it difficult to compare both approaches. Thus, to gain more insights in the performance of these two models, we designed a simulation study in which, on top of the uniformity data, genotypic effects were included, i.e, 
\[
\boldsymbol{y}^{*} = \boldsymbol{y} + \boldsymbol{Z}_g\boldsymbol{c}_ g, 
\]    
where $\boldsymbol{c}_{g}$ denotes the genotypic effects, with $\boldsymbol{c}_{g} \sim N(\boldsymbol{0}, \sigma_g^2\boldsymbol{I}_{m_g})$. For the results reported here, we considered $\sigma_g^2 = 144$, and a total of $m_g = 360$ genotypes, each replicated twice. The $360$ genotypes were allocated to the plots following an alpha design, in blocks of size $15$ (the number of rows in the field).

For each data set simulated as described above, we fitted our SpATS model (see eqn. (\ref{tpspline_model_uni})), including the genetic random factor. For \cite{Gilmour1997}'s approach we considered model (\ref{gilmour_model_uni}) (with and without the nugget) plus the genetic random factor. For comparison purposes, we also fitted a model only with the correction for rows and columns (eqn. (\ref{model_uni})). The procedure was repeated a total of $R = 500$ times. For SpATS, we used cubic B-spline bases of dimension $L = 18$ and $P = 51$, jointly with nested basis for the columns with $P_N = 27$.

For measuring models' performance, the discrepancy between the BLUPs for the genotypic effects and the corresponding true (simulated) quantities was measured in terms of the empirical version of the global root mean squared error (RMSE):
\[
\mbox{RMSE} = \sqrt{\frac{1}{360}\sum_{i=1}^{360}\left(\widehat{c}_{gi} - c_{gi}\right)^2}.
\]
As far as the REML estimates of $\sigma_g^2$ is concerned, the behavior was evaluated in terms of the bias.

\begin{figure}
\begin{center}
\includegraphics[width=6.5cm]{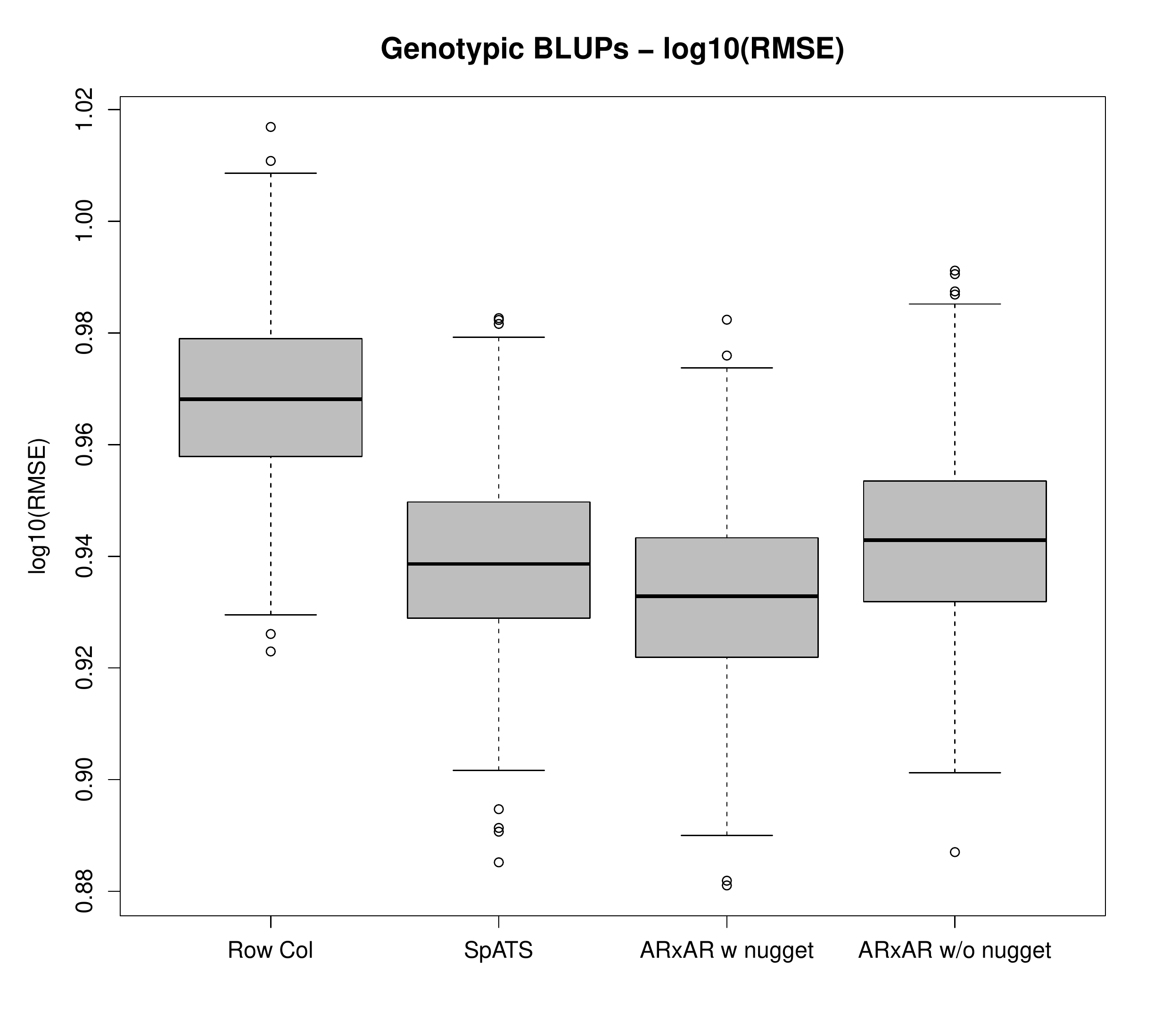}
\includegraphics[width=6.5cm]{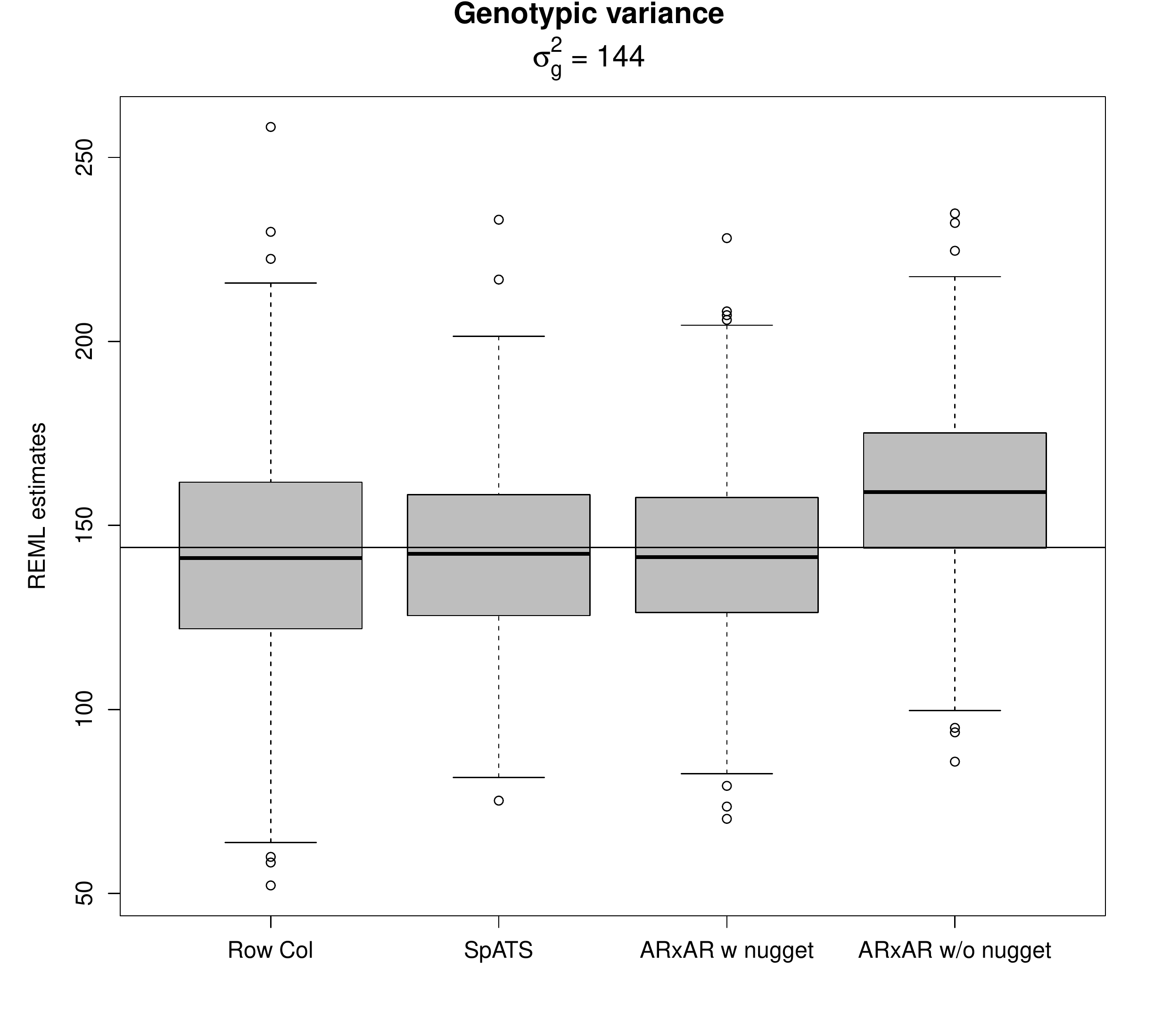}
\end{center}
\caption{For the barley uniformity data simulation study: Boxplots, based on $500$ simulated data sets, of the log10(RMSE) associated to the genotypic random factor and the REML estimates of $\sigma_g^2$. ``Row Col'' corresponds to the model including only the correction for rows and columns, ``SpATS'' to our proposal, ``AR$\times$AR w nugget'' and ``AR$\times$AR w/o nugget'' to the  AR$\times$AR model with and without the nugget respectively.}
\label{RMSE_Uniformity}
\end{figure}

Figure \ref{RMSE_Uniformity} shows the boxplots of log10(RMSE) associated to the genotypic BLUPs and the REML estimates of $\sigma_g^2$, for each of the four models considered. In terms of the log10(RMSE), and as could have been expected, the worst performance corresponds to the model including only the correction for rows and columns. The remaining three models present a similar behavior, with the best approach being the AR$\times$AR model including the nugget, followed by our SpATS model. However, if we focus on the REML estimates of $\sigma_g^2$, we observe that the AR$\times$AR model excluding the nugget tends to overestimate the genetic signal. This behavior can explain the larger heritability (on average) provided by this model in comparison with SpATS or the AR$\times$AR model including the nugget (see Table \ref{sim_ar_ar}). Surprisingly, the model including only the correction for rows and columns provides good estimates of $\sigma_g^2$. However, it also produces the lowest heritabilities. Note that this is the expected behavior, as the variance associated to the measurement error will be inflated in this case.  

\begin{table}
\caption{REML estimates of the variance parameters for the barley uniformity trial based on both AR$\times$AR and SpATS approaches.}\label{results_uniformity}
\center
\begin{tabular}{ccccccc}\hline
\multirow{2}{*}{Model} & \multicolumn{6}{c}{Parameter}\\
& $\sigma_s^2$ & $\sigma_r^2$ & $\sigma_c^2$ & $\rho_r$ & $\rho_c$ & $\sigma^2$\\\hline
AR$\times$AR & 265.67 & - & 144.20 & 0.383 & 0.834 & 173.12 \\ 
SpATS & - & 20.38 & 145.14 & - & - & 238.94 \\\hline
\end{tabular}
\end{table}

\begin{table}
\caption{Numerical results associated to the simulation study based on the barley uniformity data. For the genotypic BLUPs, the $\log_{10}$(RMSE) is shown. For the REML estimates of $\sigma_g^2$ the results show the bias. In all cases, averages and standard deviations over $500$ simulated data sets are presented. ``Row Col'' corresponds to the model including only the correction for rows and columns, ``SpATS'' to our proposal, ``AR$\times$AR w nugget'' and ``AR$\times$AR w/o nugget'' to the  AR$\times$AR model with and without the nugget respectively.}\label{results_uniformity_sim}
\center
\begin{tabular}{ccccc}\hline
	& \multicolumn{4}{c}{Model}\\
	& Row Col & SpATS & AR$\times$AR w nugget & AR$\times$AR w/o nugget\\\cline{2-5}
	\begin{tabular}{@{}c@{}}Genotypic\\RMSE \end{tabular} & 0.968 (0.016) & 0.939 (0.016) & 0.933 (0.016) & 0.943 (0.016)\\
	$\sigma_g^2 = 144$ & -2.425 (30.061) & -1.516 (24.538) & -1.563 (24.142) & 15.281 (24.391)\\ 
	H$_g^2$ & 0.399 (0.062) & 0.481 (0.054) & 0.494 (0.053) & 0.534 (0.051)\\\hline 
\end{tabular}
\end{table}

\subsection{Separable Gaussian Autoregressive Process}\label{simulation_study_supp_mat}
In this study, data was generated assuming a separable Gaussian AR process of order $1$ in the row and column directions. Specifically, the following model was considered:
\begin{equation}
\boldsymbol{y} = \boldsymbol{Z}_g\boldsymbol{c}_ g + \underbrace{\boldsymbol{\xi} + \boldsymbol{\varepsilon}}_{\boldsymbol{\epsilon}},
\label{AR_AR_Process}
\end{equation}
where $\boldsymbol{c}_{g}$ denotes the genotypic effects, with $\boldsymbol{c}_{g} \sim N(\boldsymbol{0}, \sigma_g^2\boldsymbol{I}_{m_g})$, $\boldsymbol{\xi} \sim N\left(\boldsymbol{0},\sigma_s^2\boldsymbol{\Sigma}\right)$ where $\boldsymbol{\Sigma}_{lp} = \rho_r^{\mid u_l - u_p\mid}\rho_c^{\mid v_l - v_p\mid}$, $\boldsymbol{\varepsilon} \sim N\left(\boldsymbol{0},\sigma^2\boldsymbol{I}_{n}\right)$, and $\boldsymbol{\epsilon} = \boldsymbol{\xi} + \boldsymbol{\varepsilon}$. 

Different levels of genetic variation $\left(\sigma_g^2 \in \left\{0.25; 1; 4\right\}\right)$ and autocorrelations $\left(\rho_r = \rho_c \in \left\{0.1; 0.5; 0.9\right\}\right)$ were studied. In all cases, $\sigma^2 = \sigma^2_s = 1$ and a total of $m_g = 100$ genotypes, each replicated twice, were considered. The $100$ genotypes were allocated to the plots following an alpha design, in blocks of size $10$. The field layout thus consisted of $10$ blocks (rows) and $20$ columns ($n = 200$). 

For each data set simulated as described above, we fitted our SpATS model, including the PS-ANOVA spatial field and the genetic random factor. For \cite{Gilmour1997}'s approach we considered a model including the genetic random factor, the AR$\times$AR process and the nugget ($\boldsymbol{\varepsilon}$). The procedure was repeated a total of $R = 1000$ times. For SpATS, cubic B-spline bases of dimension $13$ and $23$ were assumed for the row and column positions, respectively, and nested B-spline bases, with half the dimension, were used. 

As for the simulation reported in Section \ref{mot_example_supp_mat}, models' performance was measured in terms of the RMSE (for the genotypic BLUPs), and the bias for the REML estimates of the variances ($\sigma_g^2$, $\sigma^2$ and $\sigma_s^2$). 
 
Table \ref{sim_ar_ar} shows, for those runs for which both SpATS and AR$\times$AR models converged, the results for all levels of genetic variation and autocorrelations considered in the study. The table lists the averages and standard deviations of the RMSEs for the genotypic BLUPs and the bias and associated standard deviation for $\sigma_g^2$, $\sigma^2$ and $\sigma_s^2$. The percentage of fitting models that converged (over $1000$ runs) is also shown in the table, jointly with the effective dimension associated to the PS-ANOVA spatial field (excluding the intercept). If we focus on the genotypic random factor, both approaches behave similarly for all scenarios considered. As could have been expected, the RMSE decreases as the genotypic signal increases. A similar performance between both approaches is also observed for $\sigma_g^2$. The interpretation of the results for $\sigma^2$, $\sigma_s^2$ and $\mbox{ED}_s$ requires however a more detailed analysis. First note that when both $\rho_r$ and $\rho_c$ $\rightarrow 1$, then $\mathbb{C}\mbox{ov}\left(\boldsymbol{\xi}\right) \rightarrow \sigma_s^2\mathbb{I}_{200}$, where $\mathbb{I}_{n}$ is a $n \times n$ matrix of ones. In this case, for each realization of the stochastic process (\ref{AR_AR_Process}), the spatially dependent random error $\boldsymbol{\xi}$ would be a constant vector, and no spatial variation would be therefore present. On the other hand, when both $\rho_r$ and $\rho_c$ $\rightarrow 0$ then  $\mathbb{C}\mbox{ov}\left(\boldsymbol{\xi}\right) \rightarrow \sigma_s^2\boldsymbol{I}_{200}$. Thus, $\boldsymbol{\xi}$ is confounded with the nugget or independent error $\boldsymbol{\varepsilon}$, and $\mathbb{C}\mbox{ov}\left(\boldsymbol{\epsilon}\right) \rightarrow \left(\sigma_s^2 + \sigma^2\right)\boldsymbol{I}_{200}$. To some extent, it would also imply that no spatial variation is present and that all is measurement error. As pointed out by \cite{Piepho2015}, these two extreme cases may cause convergence problems when fitting the AR$\times$AR model. Our results are in concordance with those previous findings, and we refer to that paper for a more comprehensive discussion. Besides, it could also explain the systematic bias (of opposite sign but similar magnitude) in the estimates of $\sigma_s^2$ and $\sigma^2$ provided by the AR$\times$AR model. As can be observed, the AR$\times$AR model tends to underestimate $\sigma^2$ (negative bias), and this is counteracted by overestimating $\sigma_s^2$. This effect is especially remarkable for low autocorrelations. As far as the SpATS model is concerned, for large autocorrelation values the approach performs as expected providing unbiased estimates of $\sigma^2$. When the autocorrelations decrease, the bias also increases. However, it is worth remembering that in this case $\mathbb{C}\mbox{ov}\left(\boldsymbol{\epsilon}\right) \rightarrow \left(\sigma_s^2 + \sigma^2\right)\boldsymbol{I}_{200}$. Thus, our SpATS model is correct when estimating the variance associated to the measurement error. Note that for all simulations $\sigma_s^2 = 1$ was considered, and that the bias associated with $\sigma^2$ approaches this value as $\rho_r$ and $\rho_c$ decrease. Finally, the estimated effective dimension for the PS-ANOVA spatial field also reflects the expected performance of the model: for low and large autocorrelations (small spatial variation), the model provides smaller $\mbox{ED}_s$ than for medium values.      
\begin{table}
\scriptsize
\centering
\caption{Numerical results associated to the study when the underlying simulated model follows an AR$\times$AR process. For the REML estimates of $\sigma_g^2$, $\sigma^2$ and $\sigma_s^2$ the results show the bias. For the genotypic BLUPs, the $\log_{10}$(RMSE) is shown. $\mbox{ED}_s$ denotes the effective dimension associated to the PS-ANOVA spatial field. In all cases, averages and standard deviations are presented. The results correspond to those runs for which both SpATS and AR$\times$AR models converged.}\label{sim_ar_ar}
	\begin{tabular}{ccccccccc}
    \hline
	\multirow{2}{*}{$\sigma_g^2$} & \multirow{2}{*}{$\rho_c = \rho_r$} & \multirow{2}{*}{Model} & Conv. & Genotypic & \multirow{2}{*}{$\sigma_g^2$} & \multirow{2}{*}{$\sigma^2$} & \multirow{2}{*}{$\sigma_s^2$} & \multirow{2}{*}{$\mbox{ED}_s$}\\
	& & & (\%) & RMSE & & & \\\hline
	\multirow{6}{*}{0.25} & \multirow{2}{*}{0.9} & SpATS & 100.0 & -0.375 (0.032) & -0.014 (0.132) &  0.021 (0.159) & - & 9.10 (2.82)\\
																			& & AR$\times$AR & 68.8  & -0.374 (0.032) & -0.017 (0.134) & -0.136 (0.285) & 0.199 (1.485) & - \\
	
	& \multirow{2}{*}{0.5} & SpATS & 100.0 & -0.355 (0.030) &  -0.008 (0.159) & 0.306 (0.211)  & - & 15.27 (5.03)\\
								& & AR$\times$AR & 96.2  & -0.356 (0.030) &  -0.018 (0.159) & -0.112 (0.316) & 0.168 (0.486) & -\\
	
	& \multirow{2}{*}{0.1} & SpATS & 100.0 & -0.341 (0.032) & 0.002 (0.190) &  0.769 (0.265) & - & 8.96 (4.16)\\
								& & AR$\times$AR & 93.8  & -0.341 (0.032) &-0.013 (0.190) & -0.368 (0.632) & 0.376 (0.628)  & - \\\hline
	
	\multirow{6}{*}{$1$} & \multirow{2}{*}{0.9} & SpATS & 100.0  & -0.220 (0.031) & -0.076 (0.192)  &  0.022 (0.162) & - & 8.35 (2.88)\\
																		 & & AR$\times$AR & 63.7   & -0.220 (0.031) & -0.076 (0.189)  & -0.201 (0.329) & 0.117 (1.299) & - \\
	
	& \multirow{2}{*}{0.5} &  SpATS & 100.0 & -0.180 (0.030) & -0.080 (0.233)  & 0.327 (0.230)  & - & 13.54 (4.67)\\
								 & & AR$\times$AR & 94.1  & -0.186 (0.030) & -0.085 (0.225)  & -0.153 (0.367) &  0.231 (0.634)  & - \\
	
	& \multirow{2}{*}{0.1} & SpATS & 100.0 & -0.152 (0.030) & -0.083 (0.276) & 0.797 (0.293)  & - & 8.10 (3.73)\\
								& & AR$\times$AR & 90.8  & -0.155 (0.030) & -0.085 (0.273) & -0.408 (0.663) & 0.442 (0.640) & - \\\hline
	
	\multirow{6}{*}{$4$} &\multirow{2}{*}{0.9} & SpATS & 100.0 & -0.143 (0.032) & -0.299 (0.332) & 0.021 (0.164) & - & 7.89 (2.76)\\
																		& & AR$\times$AR & 61.8  & -0.143 (0.033) & -0.305 (0.324) &-0.272 (0.368) &  0.187 (1.513) & -\\
	
	& \multirow{2}{*}{0.5} &  SpATS & 100.0 & -0.088 (0.031) & -0.308 (0.388) & 0.334 (0.235)  & - & 12.01 (4.22)\\
								 & & AR$\times$AR & 90.7  & -0.095 (0.031) & -0.312 (0.375) & -0.217 (0.406) &  0.295 (0.548) & - \\
	
	& \multirow{2}{*}{0.1} & SpATS & 100.0 & -0.048 (0.031) & -0.315 (0.439) & 0.792 (0.298)  & - & 7.35 (3.36) \\
								& & AR$\times$AR & 88.5  & -0.053 (0.031) & -0.324 (0.446) & -0.492 (0.637) &  0.515 (0.635) & - \\\hline
	\end{tabular}
\end{table}
\section{Applications\label{s_app}}
To further illustrate our proposal in large-scale experiments, we considered in this Section two data sets, one of a field trial on wheat conducted in Chile and discussed in the paper by \cite{Lado2013}, and the other on sugar beet from a big field in France. This material was kindly provided by the breeding company SESVanderHave (Tienen, Belgium).
\subsection{Sugar beet data, France}
In this section a big field of sugar beet is analyzed. The field experiment was located in France, in the year $2011$. Data were recorded for eight traits, and the results will be presented for the trait $\alpha$-amino nitrogen in millimol per liter. All material was provided by the breeding company SESVanderHave (Tienen, Belgium).

The field experiment consisted of $31$ trials, with a total of $2411$ plots and $1095$ genotypes. All the trials contained four common checks, the other $1091$ genotypes were observed in one of the trials. The trials were different in size, with the most common layout an alpha design with $36$ genotypes and two replicates. Four of the trials were unreplicated, $26$ trials had an alpha design with two replicates, and one trial had an alpha design with three replicates.

\begin{figure}
    \begin{center}
    \subfigure[Field layout]{
		\includegraphics[width=4cm]{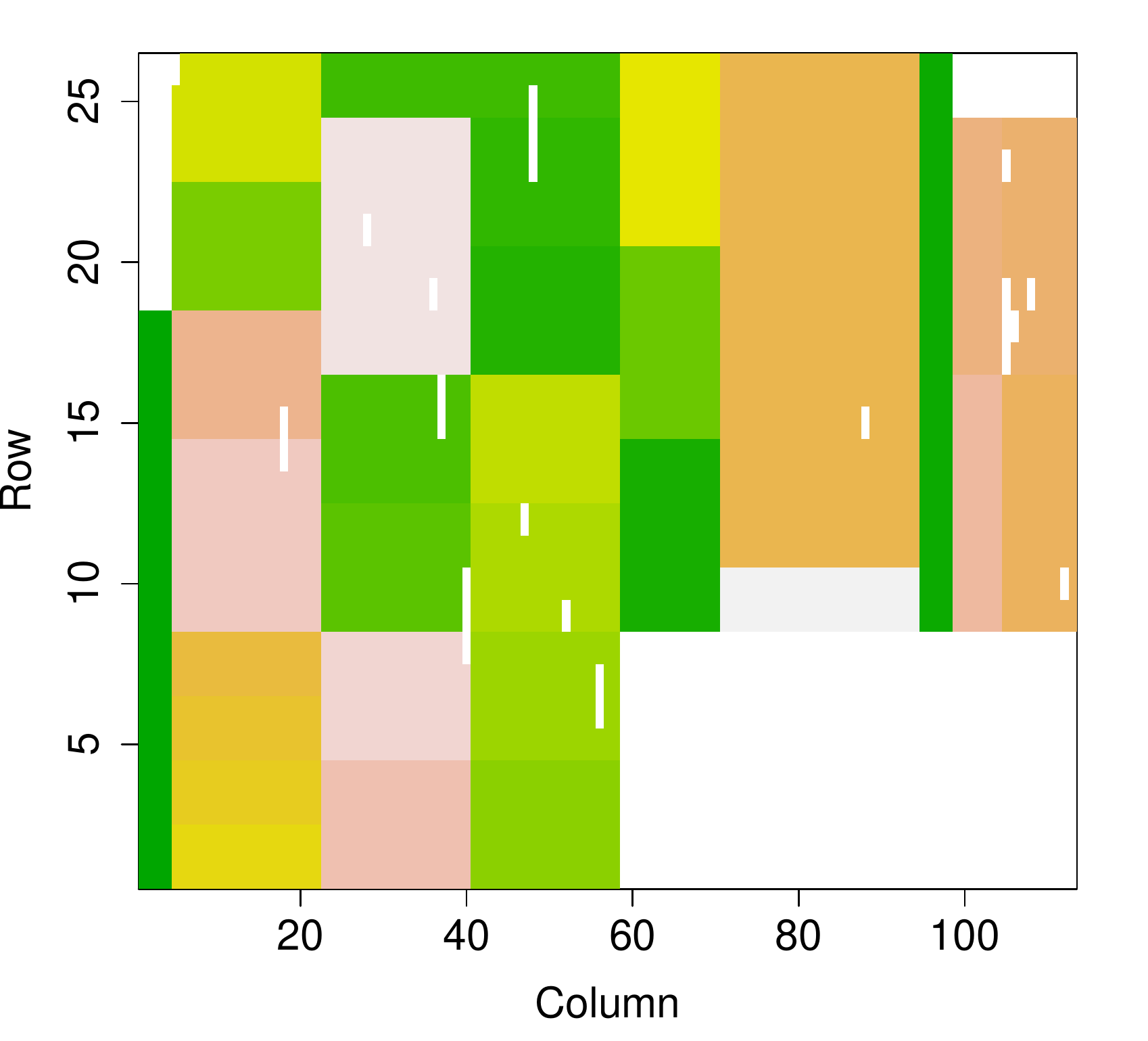}
		\label{SES_layout}}
    \subfigure[Raw yield data]{
		\includegraphics[width=4cm]{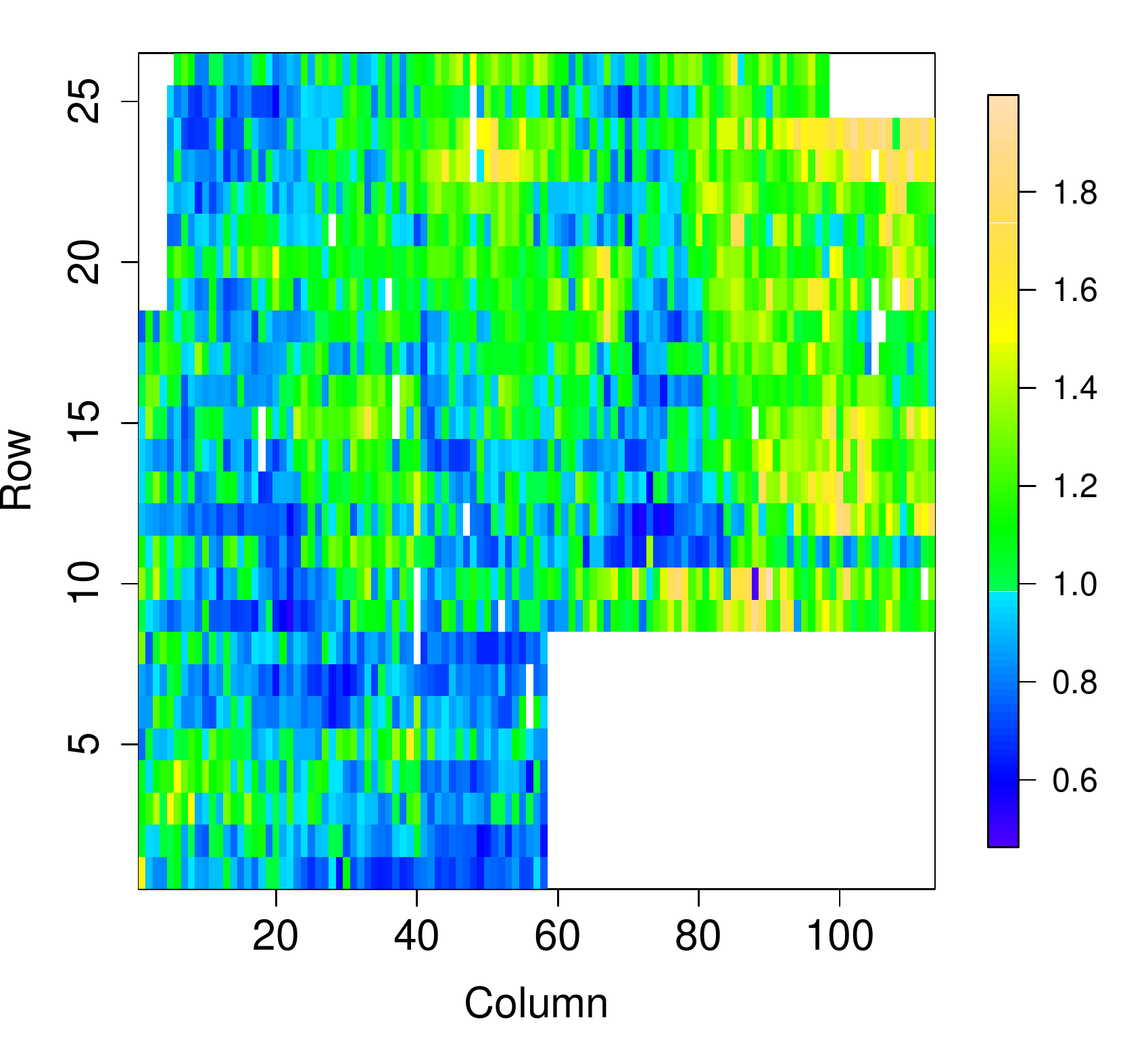}
		\label{SES_raw}}
		\subfigure[Fitted values]{
		\includegraphics[width=4cm]{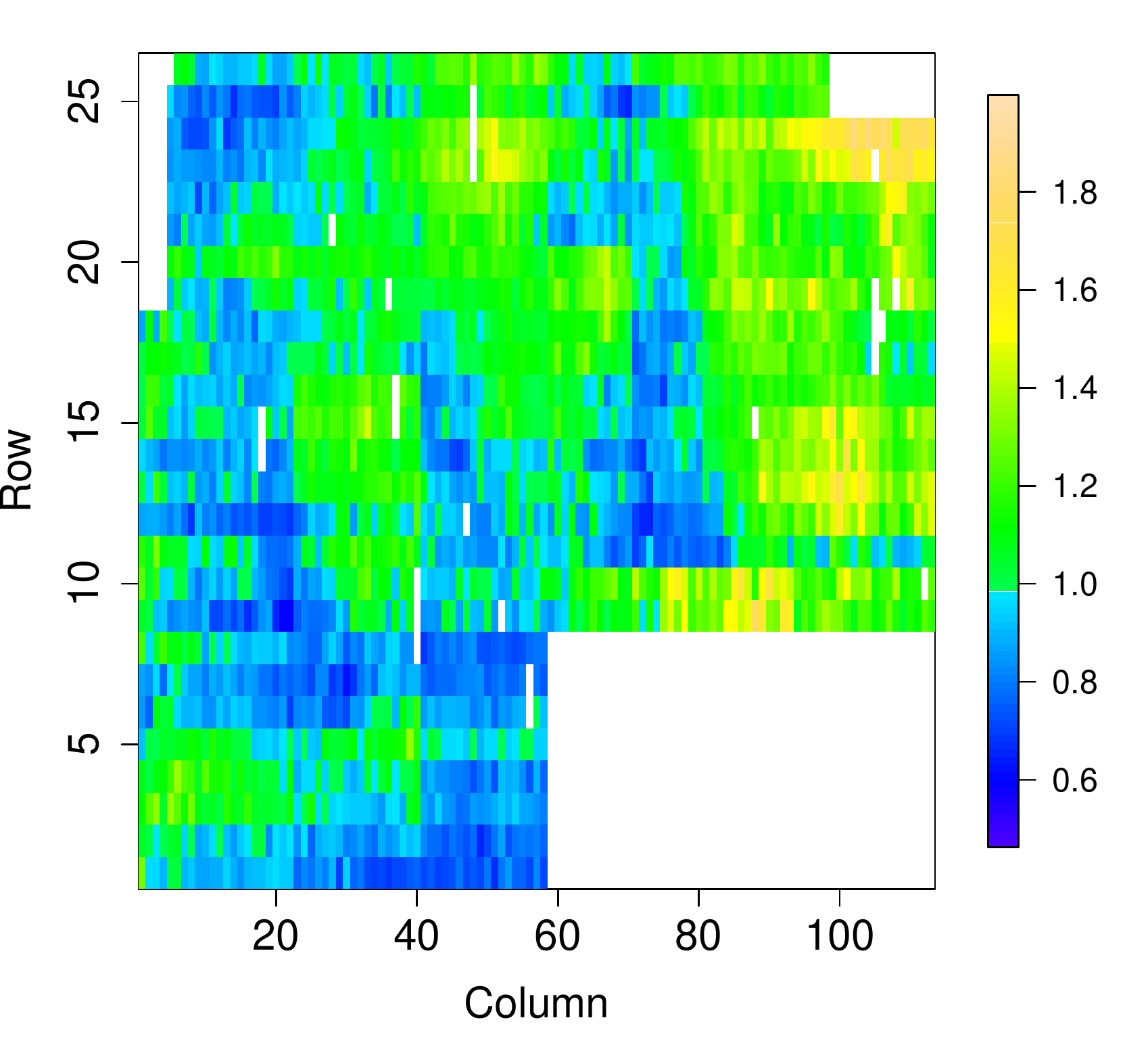}
		\label{SES_Fitted}}
		\subfigure[Residuals' spatial plot]{
		\includegraphics[width=4cm]{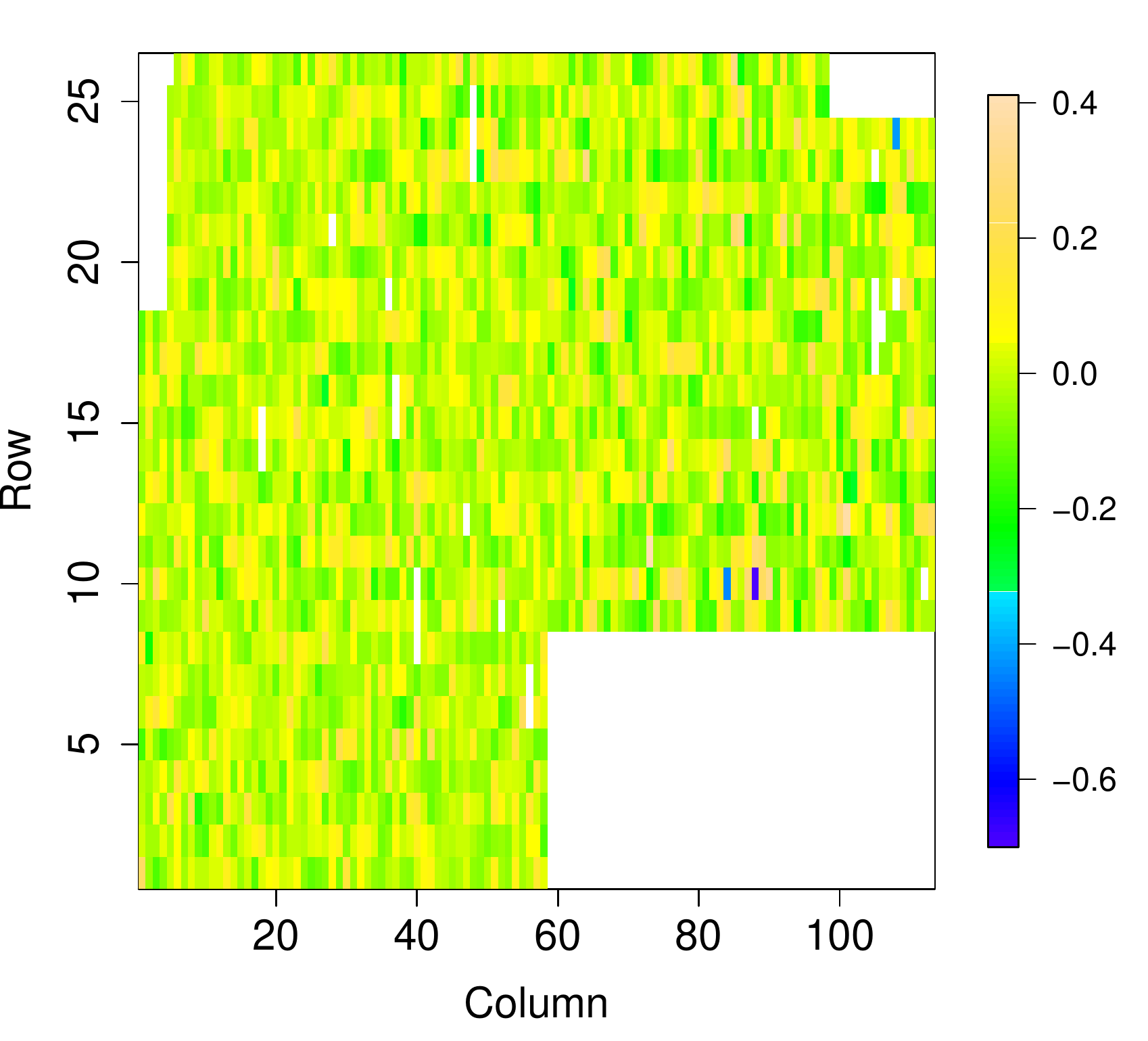}
		\label{SES_residuals}}
		\subfigure[Fitted spatial trend]{
		\includegraphics[width=4cm]{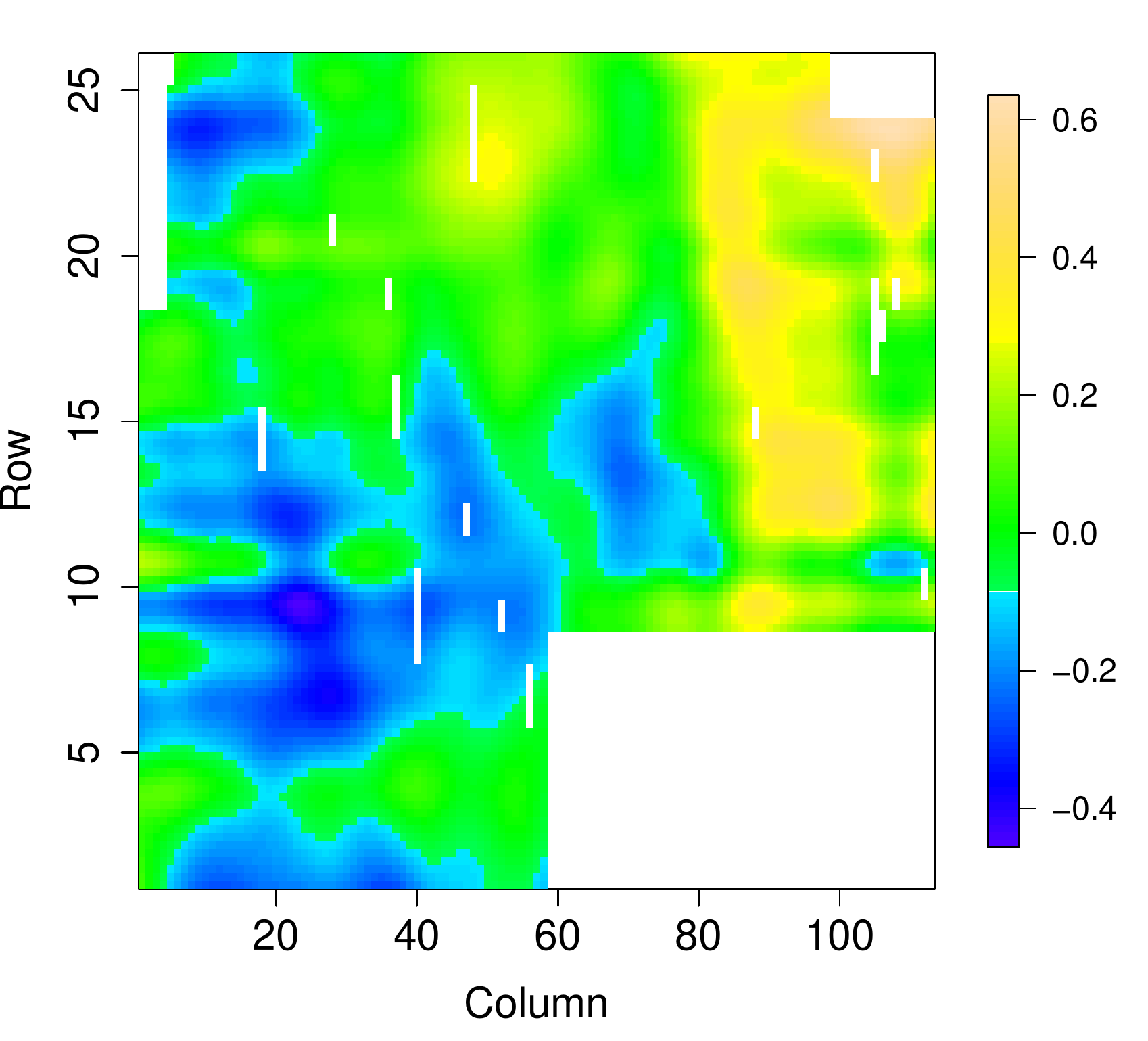}
		\label{SES_trend}}
		\subfigure[Genotypic BLUPs]{
		\includegraphics[width=4cm]{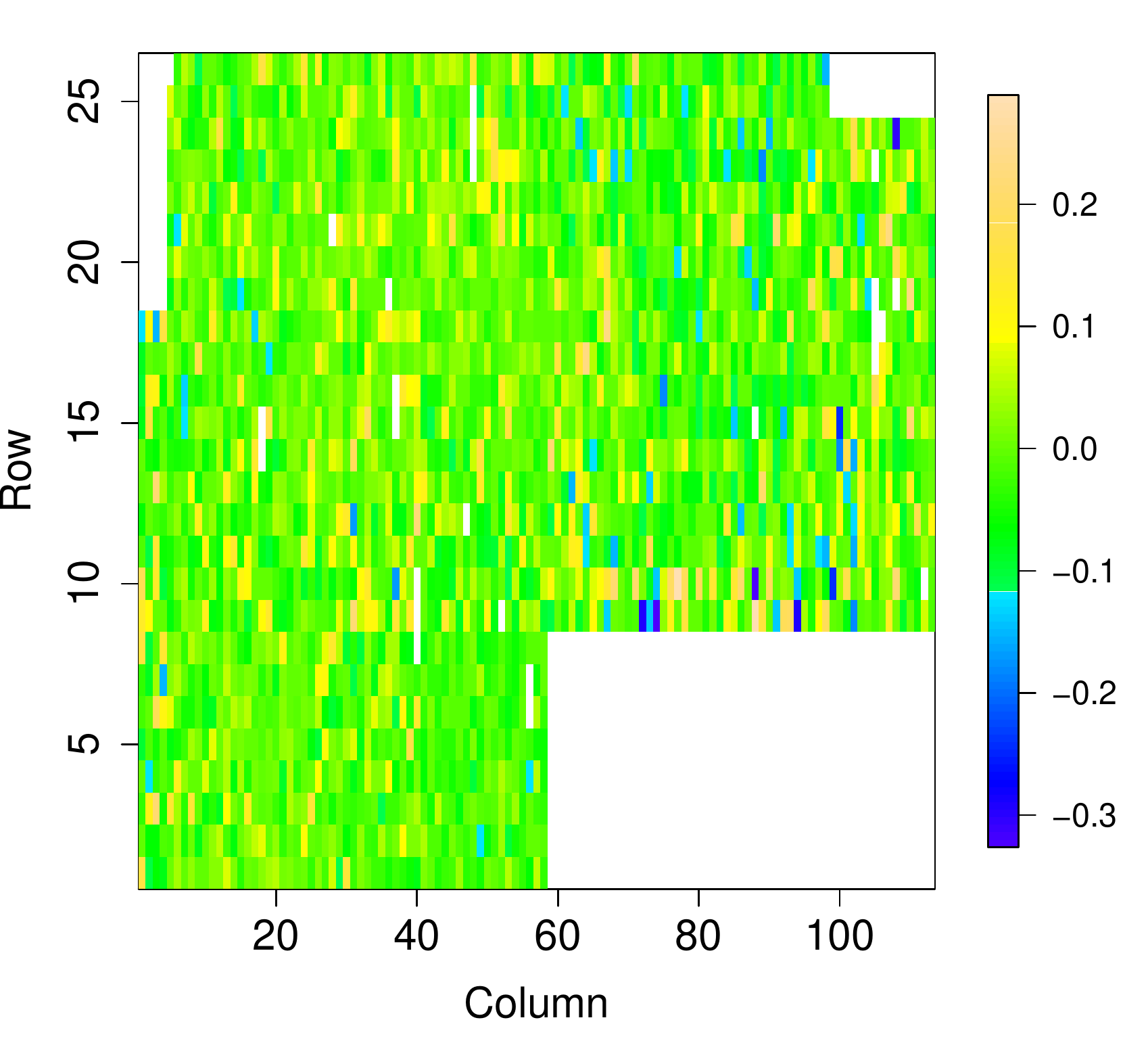}
		\label{SES_BLUPs}}
    \end{center}
    \caption{For the sugar beet experiment in France: field layout, raw yield data, fitted values, fitted spatial trend, residuals' spatial plot and genotypic BLUPs.}
		\label{SES_results}
  \end{figure}

The field layout is shown in Figure \ref{SES_layout}. The $31$ trials are represented by a different color. Note that the layout of the field is irregular, that there are no trials in the lower right corner (white areas). Nonetheless, no data manipulation, such as e.g. filling in the missing values to make the layout regular, is required to use our approach.   

For this experiment, we assumed a model including random factors for rows ($\boldsymbol{c}_r$) columns ($\boldsymbol{c}_c$) and trials ($\boldsymbol{c}_t$). The genetic lines were also included in the model as random ($\boldsymbol{c}_g$) but the genetic checks were included as fixed ($\boldsymbol{\beta}_c$). To be more precise, and using the same notation as in Section \ref{s_motexample}, the following mixed model was fitted
\[
\boldsymbol{y} = f(\boldsymbol{u}, \boldsymbol{v}) + \boldsymbol{X}_c\boldsymbol{\beta}_c + \boldsymbol{Z}_g\boldsymbol{c}_g + \boldsymbol{Z}_r\boldsymbol{c}_r + \boldsymbol{Z}_c\boldsymbol{c}_c + \boldsymbol{Z}_t\boldsymbol{c}_t  + \boldsymbol{\varepsilon}.
\]
where $\boldsymbol{X}_c$ is the $(n \times 4)$ design matrix assigning observations to genetic checks and $\boldsymbol{Z}_g$ is the $(n \times 1091)$ design matrix associated to genetic lines, with $\boldsymbol{c}_g = \left(c_{g1},\ldots,c_{g1090}\right)^{t} \sim N\left(\boldsymbol{0}, \sigma_g^2\boldsymbol{I}_{1090}\right)$. The rows of these two matrices corresponding to genetic lines and genetic checks, respectively, have all their elements equal to zero. For the rest of the random factors, we assumed $\boldsymbol{c}_r \sim N\left(\boldsymbol{0}, \sigma_r^2\boldsymbol{I}_{26}\right)$, $\boldsymbol{c}_c \sim N\left(\boldsymbol{0}, \sigma_c^2\boldsymbol{1}_{113}\right)$ and $\boldsymbol{c}_t \sim N\left(\boldsymbol{0}, \sigma_t^2\boldsymbol{I}_{31}\right)$. For the tensor-product P-spline, a basis dimension of $29$ and $53$ was chosen for the row and column positions, respectively, and nested bases, with half the dimension, were used. Accordingly, the model has $1789$ coefficients to be estimated and $2411$ observations. Despite the large dimension, the fitting process needed around $18$ seconds.  

Figure \ref{SES_results} depicts the raw yield data, the fitted values, the residuals, the fitted spatial trend (i.e., the PS-ANOVA component but excluding the intercept), and the genotypic BLUPs. As can be observed, the fitted spatial trend is successful in recovering the complex spatial pattern across the field, and the residual plot suggest that the spatial independence assumption for the error vector $\boldsymbol{\varepsilon}$ might be appropriate. Table \ref{results_SES_effdim} shows the model and (estimated) effective dimensions associated to each of the five smooth components of the PS-ANOVA spatial trend and each random factor. Note that the estimation procedure presented in Section~\ref{s_trials} provides, in some cases, effective dimensions that are exactly zero, meaning that this model component does not contribute or have an impact on the trait of interest. If we focus on the genetic signal, we have an effective dimension of about $590.5$. In this case, it is easy to show that $\mbox{rank}\left(\left[\boldsymbol{X}, \boldsymbol{Z}_g\right]\right) - \mbox{rank}\left(\boldsymbol{X}\right) = 1090$. Equivalently, there is only one eigenvalue equal to zero, which implicitly induces a zero-mean constraint on the BLUPs for $\boldsymbol{c}_g$. As a consequence, an estimate of the generalized heritability is $\widehat{H}_g = 590.5/ (1091 - 1) = 0.54$.

	\begin{table}
\caption{For the sugar beet experiment in France: model and effective dimensions associated to the PS-ANOVA spatial trend and each random factor. The letter $u$ denotes the row position, $v$ the column position, and $\boldsymbol{c}_r$, $\boldsymbol{c}_r$, $\boldsymbol{c}_t$ and $\boldsymbol{c}_g$ the row, column, trial and genetic line random factors, respectively.}\label{results_SES_effdim}
	\center
	\begin{tabular}{cccccccccc}\hline
	\multirow{2}{*}{Dimensions} & \multicolumn{9}{c}{Model components}\\
	& $f_u(u)$ & $f_v(v)$ & $vh_u(u)$ & $uh_v(v)$ & $f_{u,v}(u,v)$ & $\boldsymbol{c}_r$ & $\boldsymbol{c}_c$ & $\boldsymbol{c}_t$ & $\boldsymbol{c}_g$ \\\cline{2-10}
	Model ($m_k$) & 27 & 51 & 27 & 51 & 364 & 26 & 113 & 31 & 1091\\
	Effective (ED$_k$) & 0.0 & 12.4 & 21.1 & 0.0 & 120.1 & 18.3 & 13.1 & 22.8 & 590.5\\\hline
	\end{tabular}
\end{table}	   
\subsection{Wheat data, Chile}
The study consisted on $384$ advanced lines from two breeding programs in Chile and Uruguay (South America). The lines were evaluated in two different environments, Santa Rosa (SR) and Cauquenes (CQ), under three different levels of water supply and in two consecutive years, $2011$ and $2012$. More precisely, for both $2011$ and $2012$, the lines were evaluated in Santa Rosa under mild water stress (MWS) and fully irrigated (FI) conditions, and for $2012$ in Cauquenes, a dry region, under severe water stress (WS). Different traits of interest were considered, as grain yield (GY), thousand kernel weight (TKW), number of kernels per spike (NKS) and days to heading (DH). It should be noted that, in $2012$, only GY was considered. For each trial, an alpha-lattice design with $20$ incomplete blocks, each containing $20$ genotypes, was used, and two replicates were sown. Each block comprised $1$ column and $40$ rows, yielding a total of $800$ $(20 \times 40 \times 2)$ plots on the field.

Based on these data, the final aim pursued in the paper by \cite{Lado2013} was to perform genomic selection on the basis of genotyping-by-sequencing methods. As noted by the authors, the modeling of the spatial variation of the phenotypic data in each trial has an impact on the prediction accuracy of the subsequent genomic selection. As could have been expected, the better the modeling, the larger the prediction accuracy. In this paper we thus re-analyzed the phenotypic data using our approach. For the sake of simplicity we only present here the results for $2011$. For each trait (GY, TKW, NKS and DH) and water condition (MWS and FI) the following model was assumed
\begin{equation*}
\boldsymbol{y} = f\left(\boldsymbol{v}, \boldsymbol{u}\right) + \boldsymbol{Z}_g\boldsymbol{c}_g + \boldsymbol{Z}_r\boldsymbol{c}_r + \boldsymbol{Z}_c\boldsymbol{c}_c + \boldsymbol{\varepsilon},
\end{equation*}
with $\boldsymbol{c}_g \sim N\left(\boldsymbol{0}, \sigma_g^2\boldsymbol{I}_{384}\right)$, $\boldsymbol{c}_r \sim N\left(\boldsymbol{0}, \sigma_r^2\boldsymbol{I}_{40}\right)$ and $\boldsymbol{c}_c \sim N\left(\boldsymbol{0}, \sigma_c^2\boldsymbol{I}_{20}\right)$. For the tensor-product P-spline, a basis dimension of $43$ and $23$ was assumed for the row and column positions, respectively, and, as usual, we used nested bases, with half the dimension. Under this representation, each model has about $802$ coefficients to be estimated and $800$ observations (there are missing values), but the fitting processes needed between $2$ and $10$ seconds.    

\begin{figure}
    \begin{center}
    \subfigure[Raw data]{
		\includegraphics[width=3.5cm]{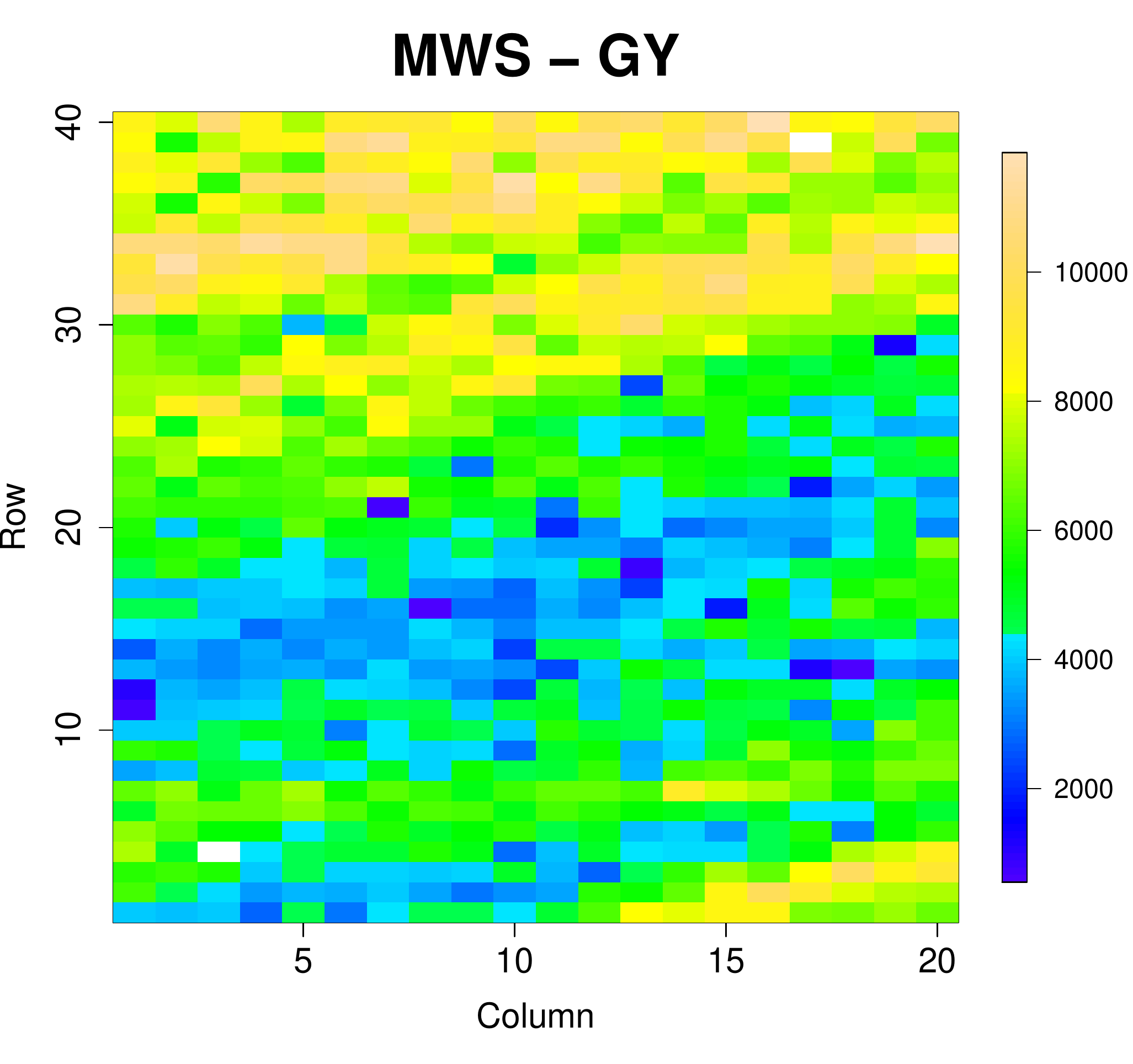}
		\includegraphics[width=3.5cm]{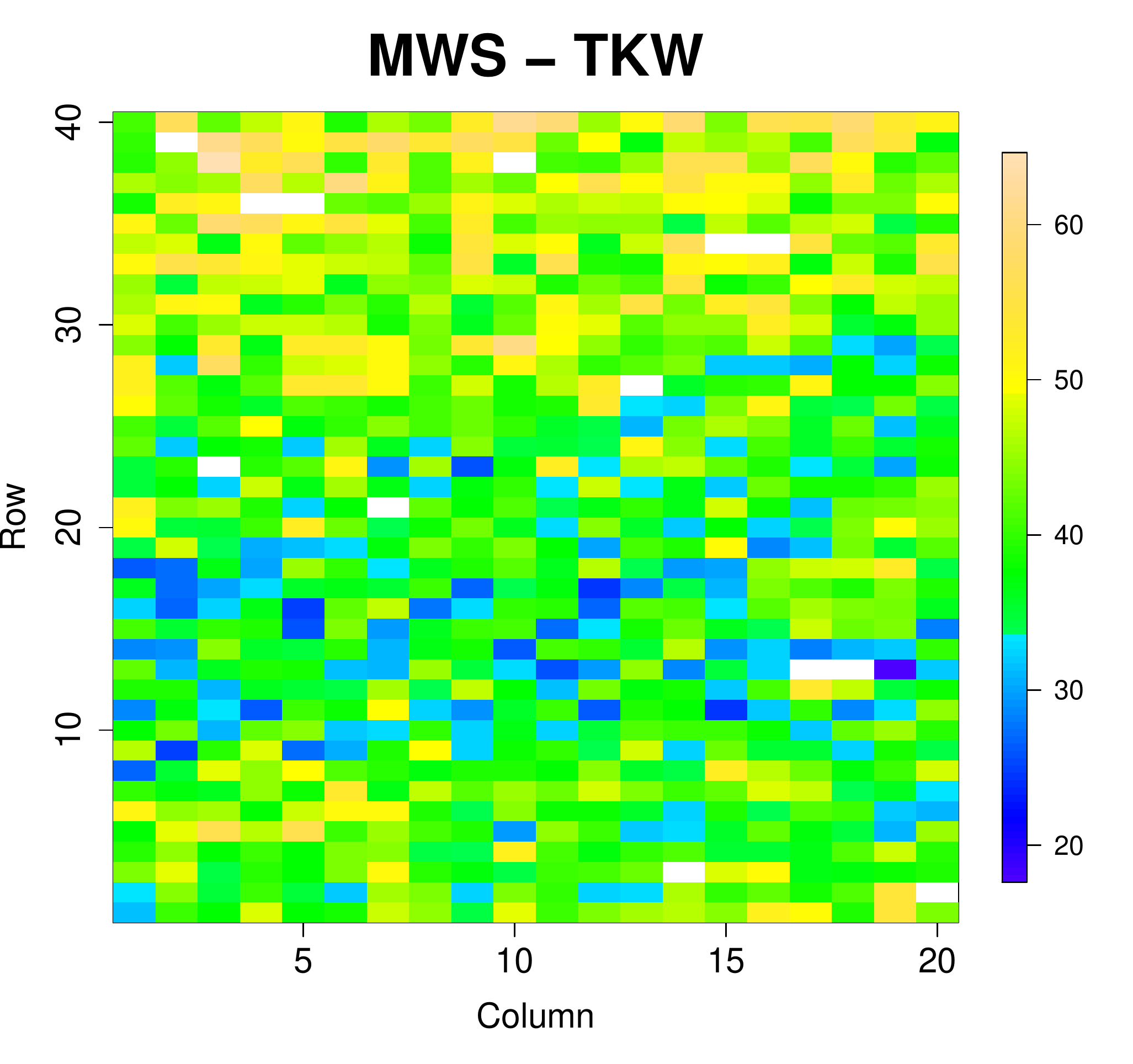}
		\includegraphics[width=3.5cm]{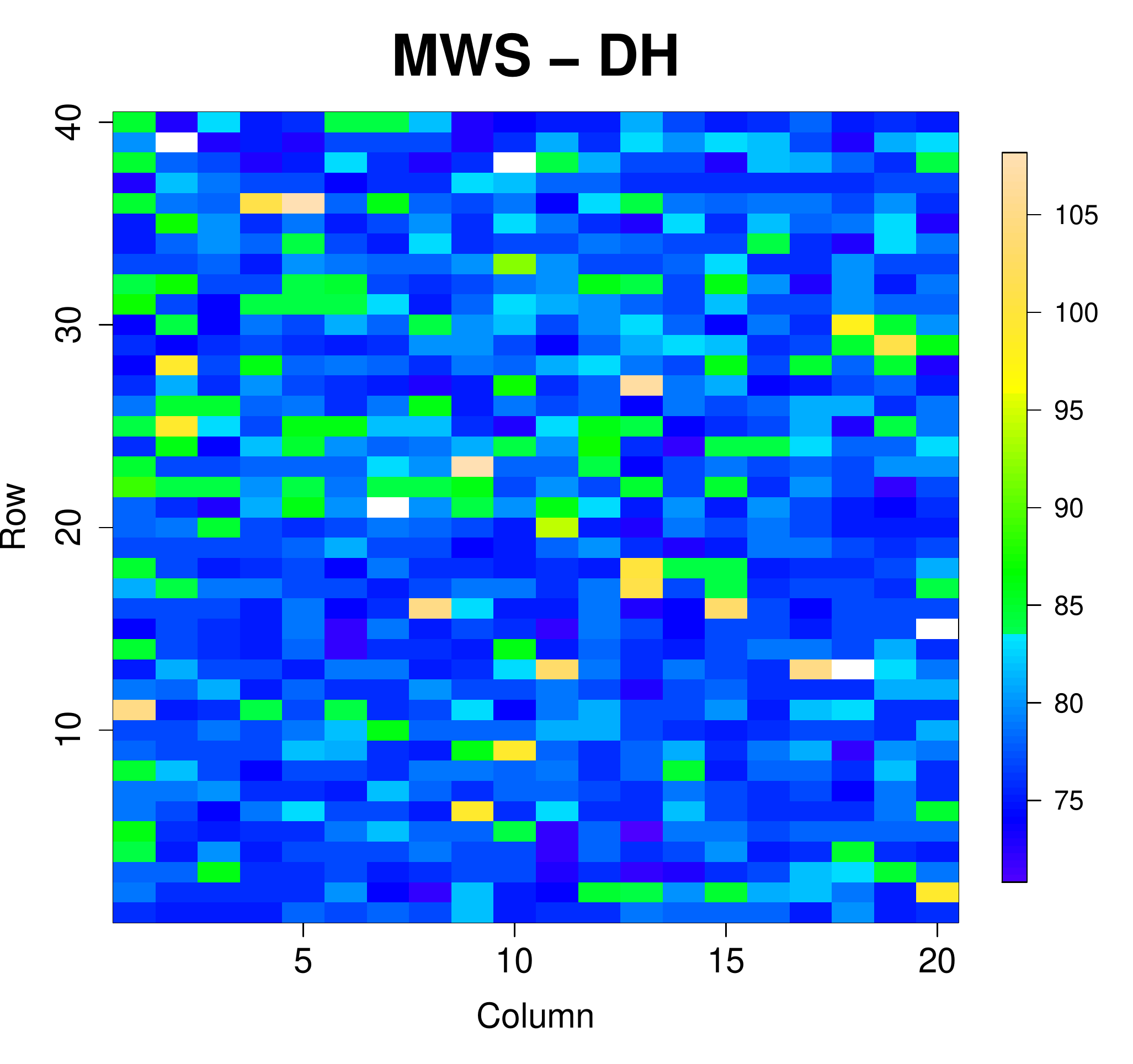}
		\includegraphics[width=3.5cm]{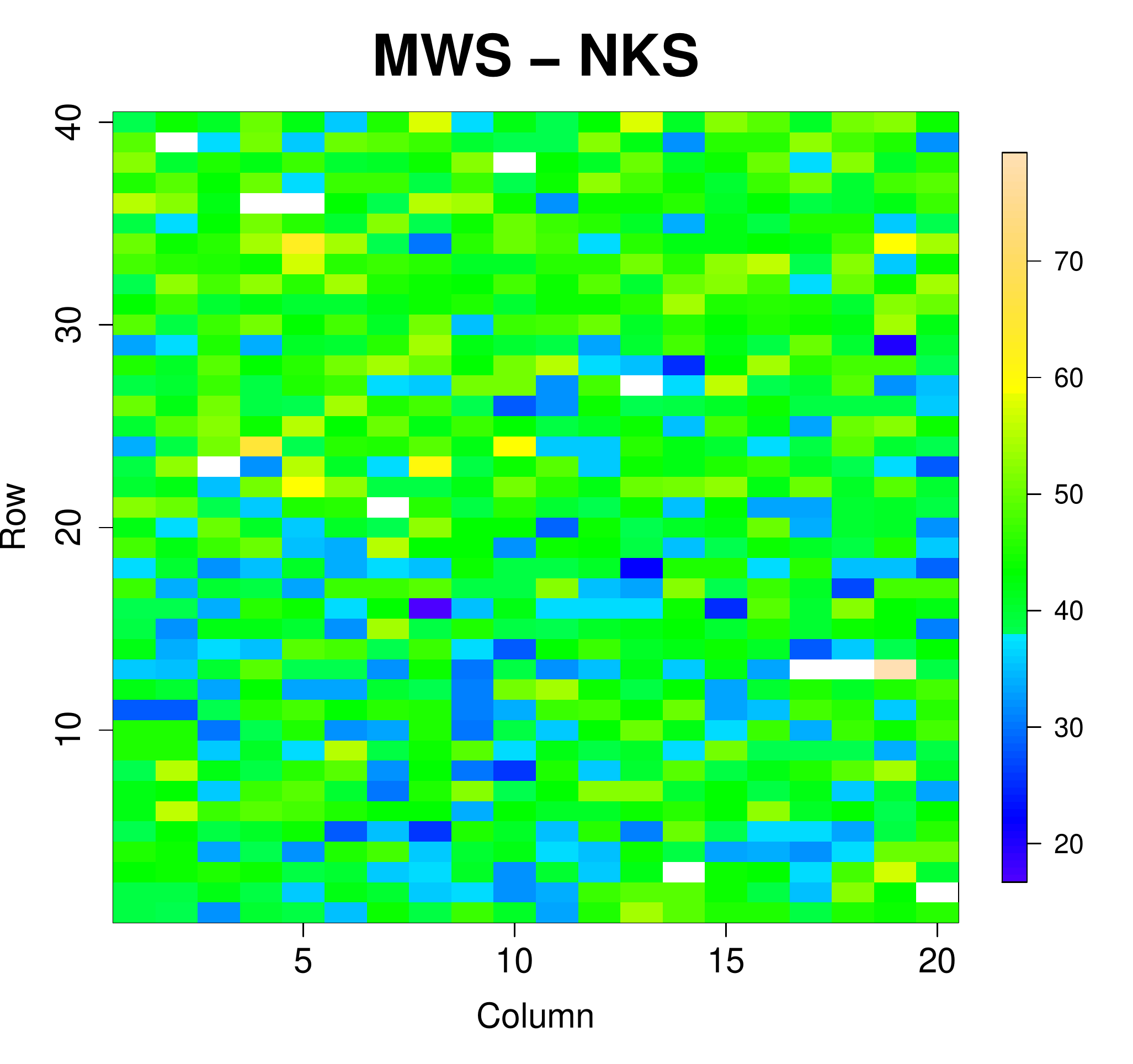}
		\label{Lado_results_raw_mws}}
    \subfigure[Fitted spatial trend]{
		\includegraphics[width=3.5cm]{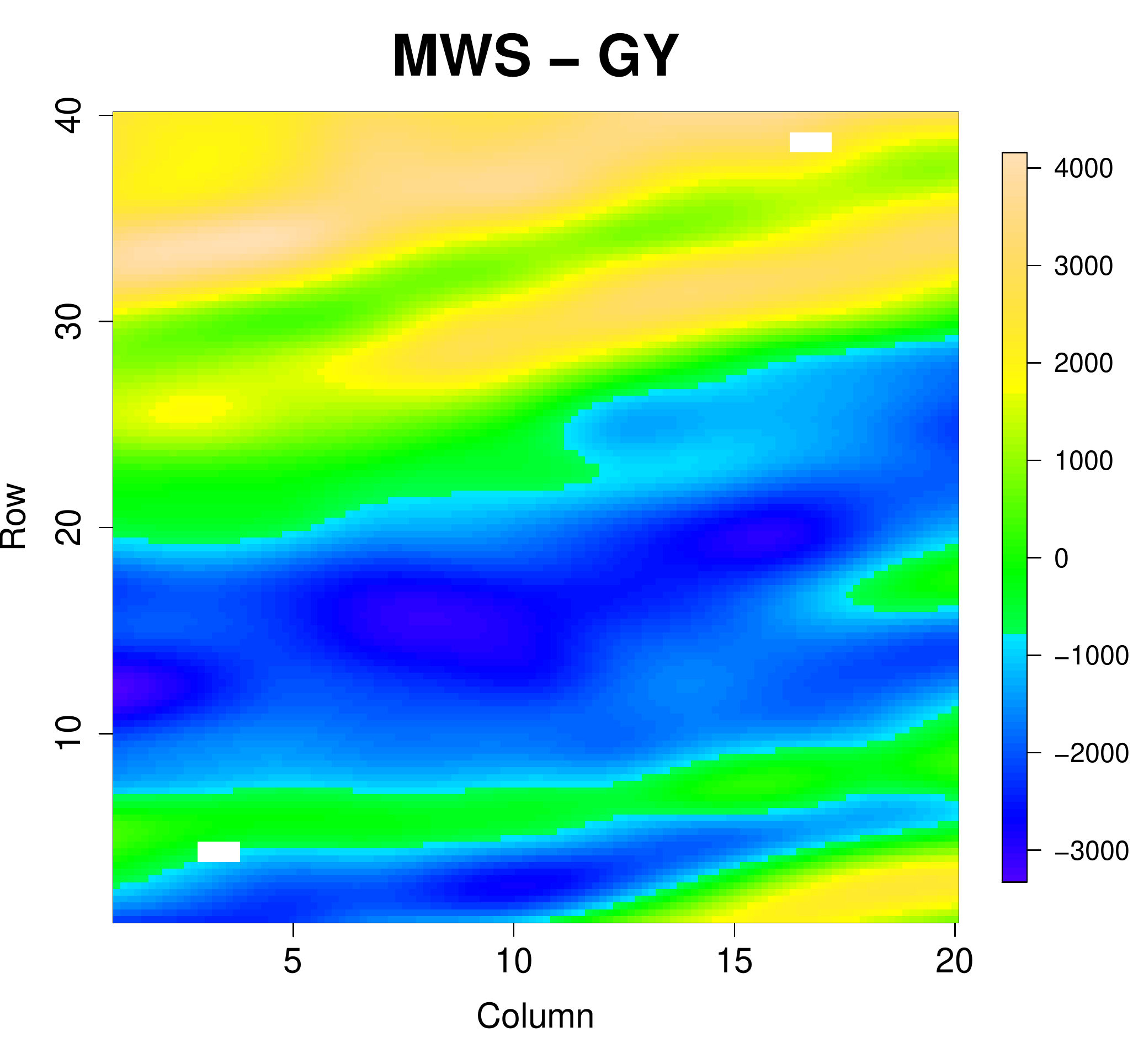}
		\includegraphics[width=3.5cm]{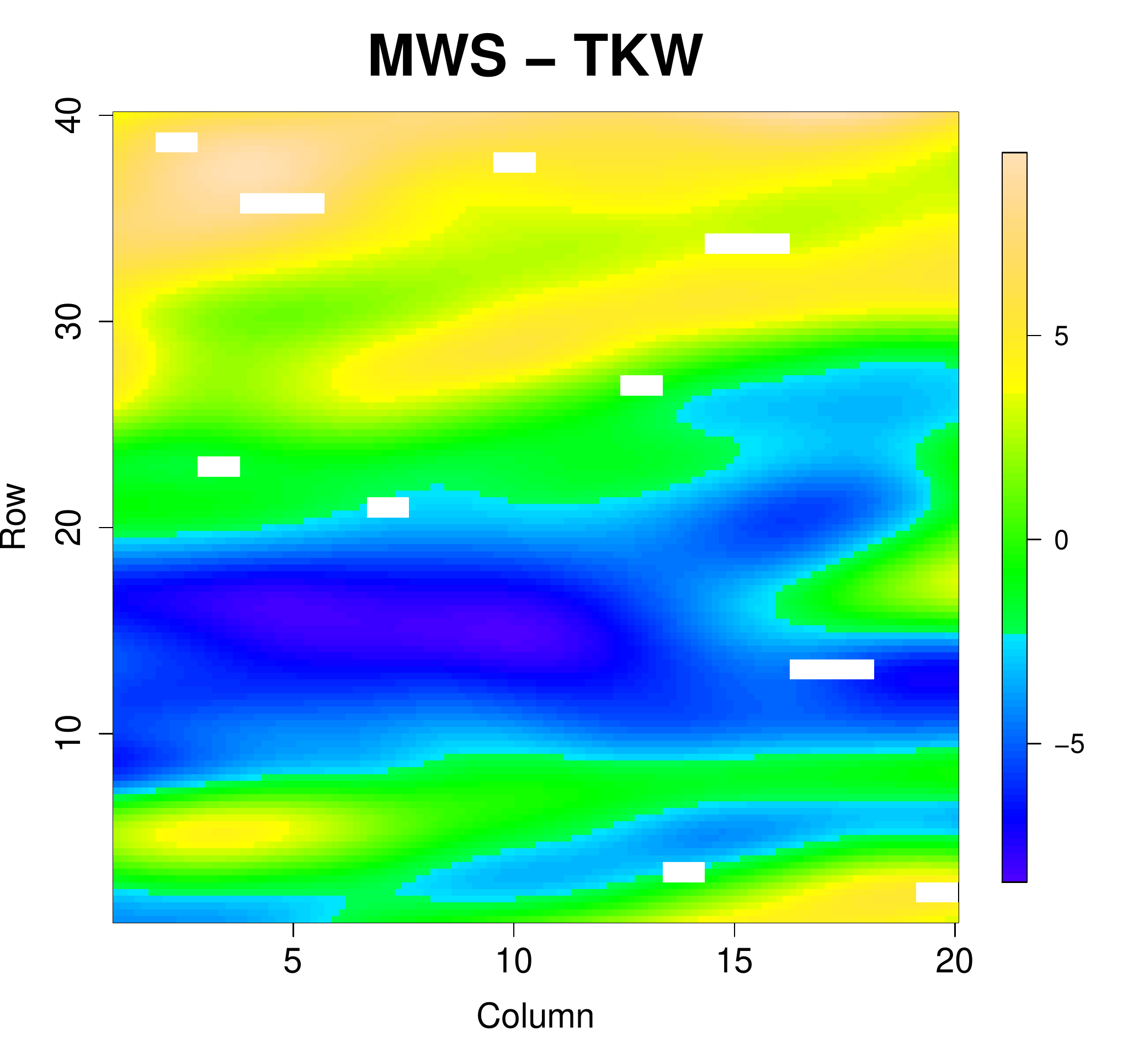}
		\includegraphics[width=3.5cm]{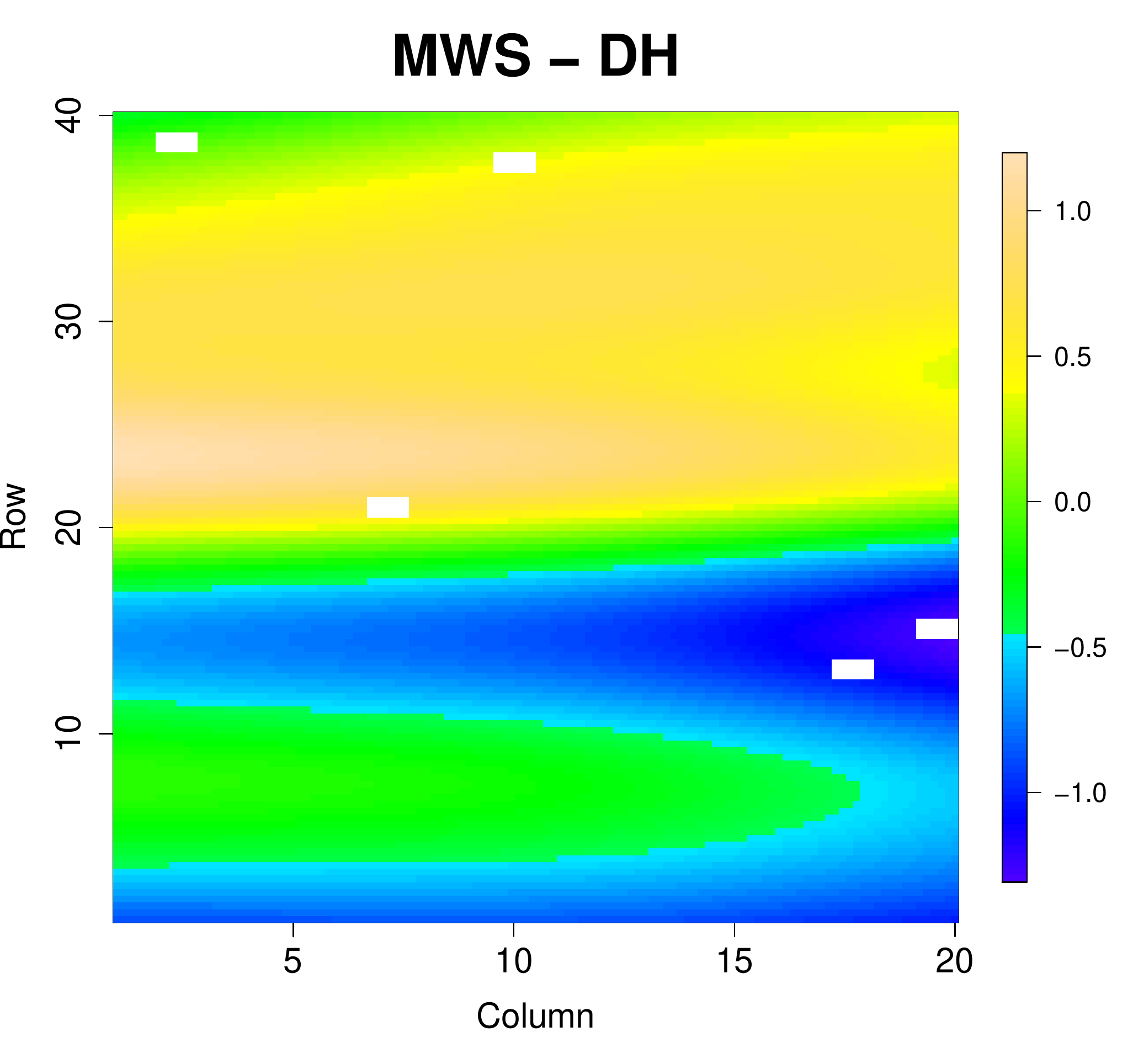}
		\includegraphics[width=3.5cm]{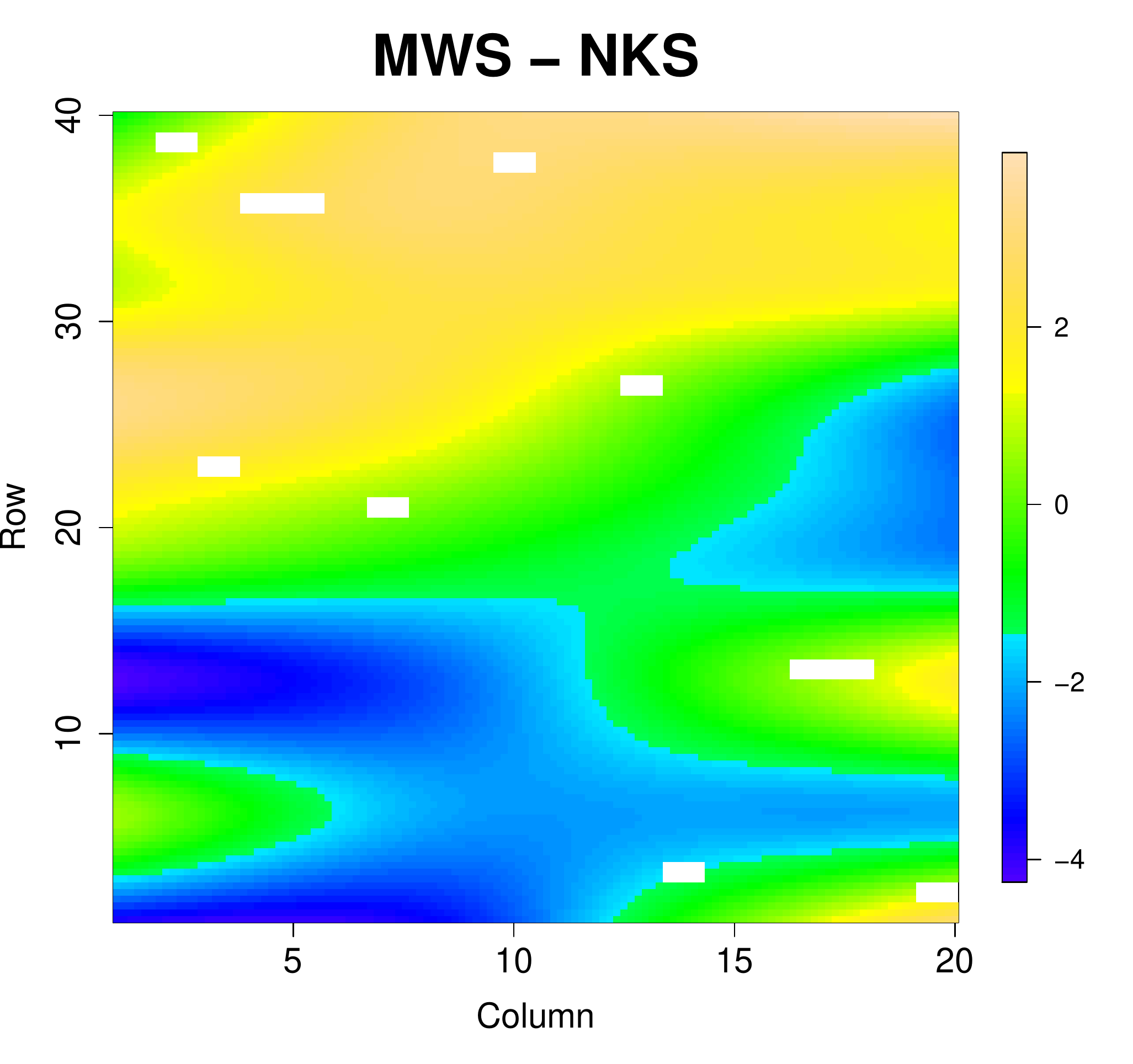}
		\label{Lado_results_trend_mws}}
		\subfigure[Residuals' spatial plot]{
		\includegraphics[width=3.5cm]{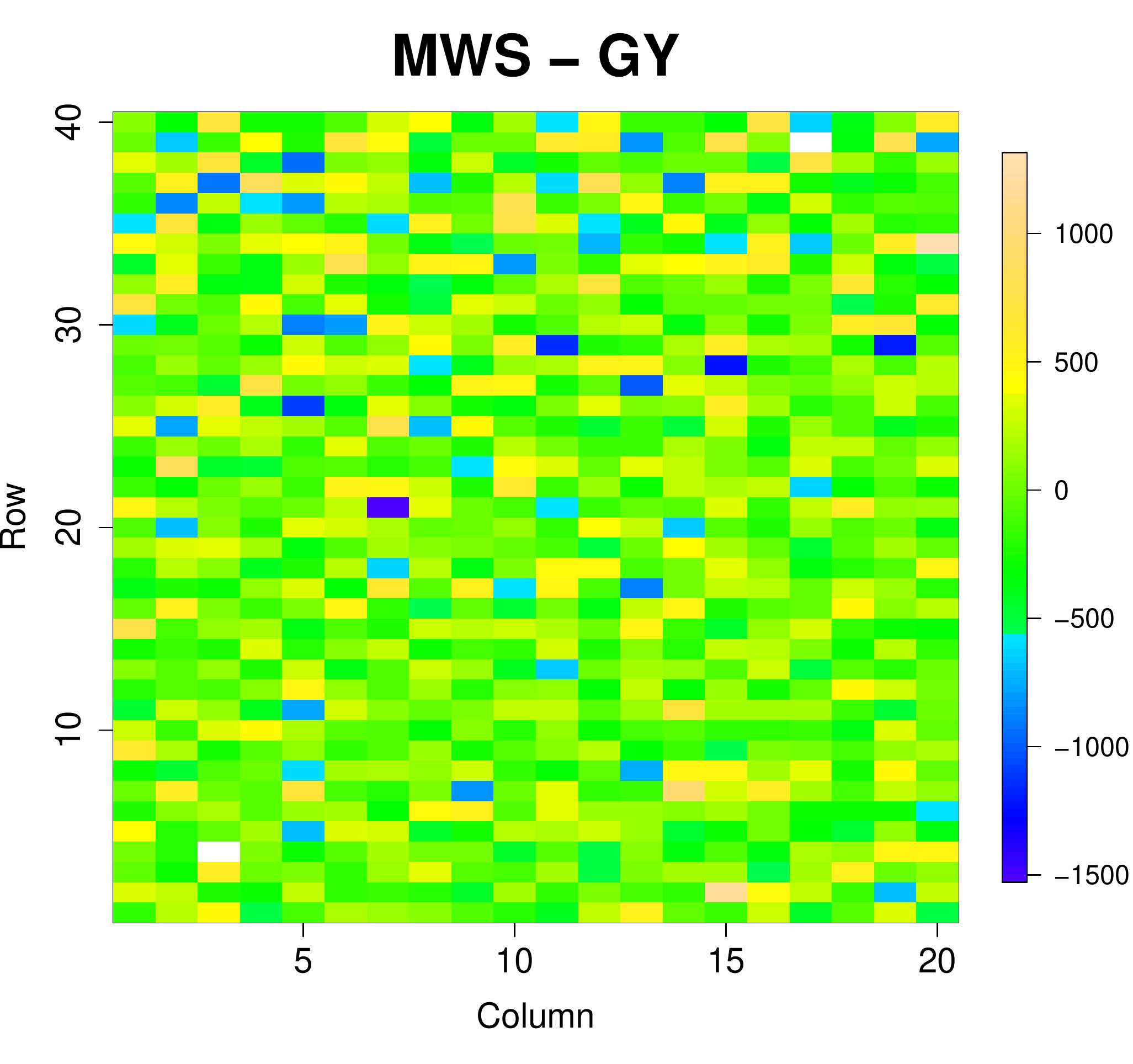}
		\includegraphics[width=3.5cm]{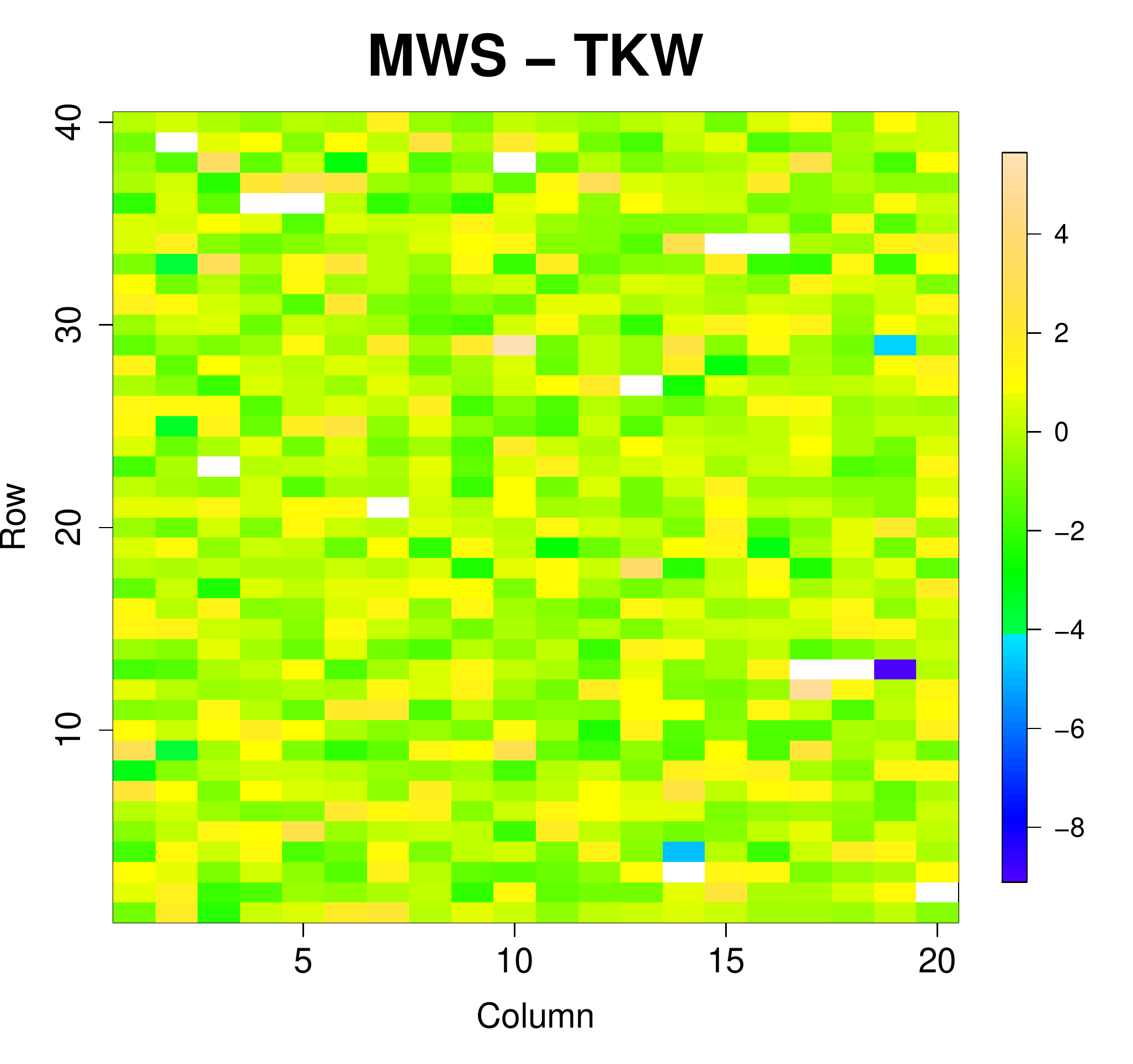}
		\includegraphics[width=3.5cm]{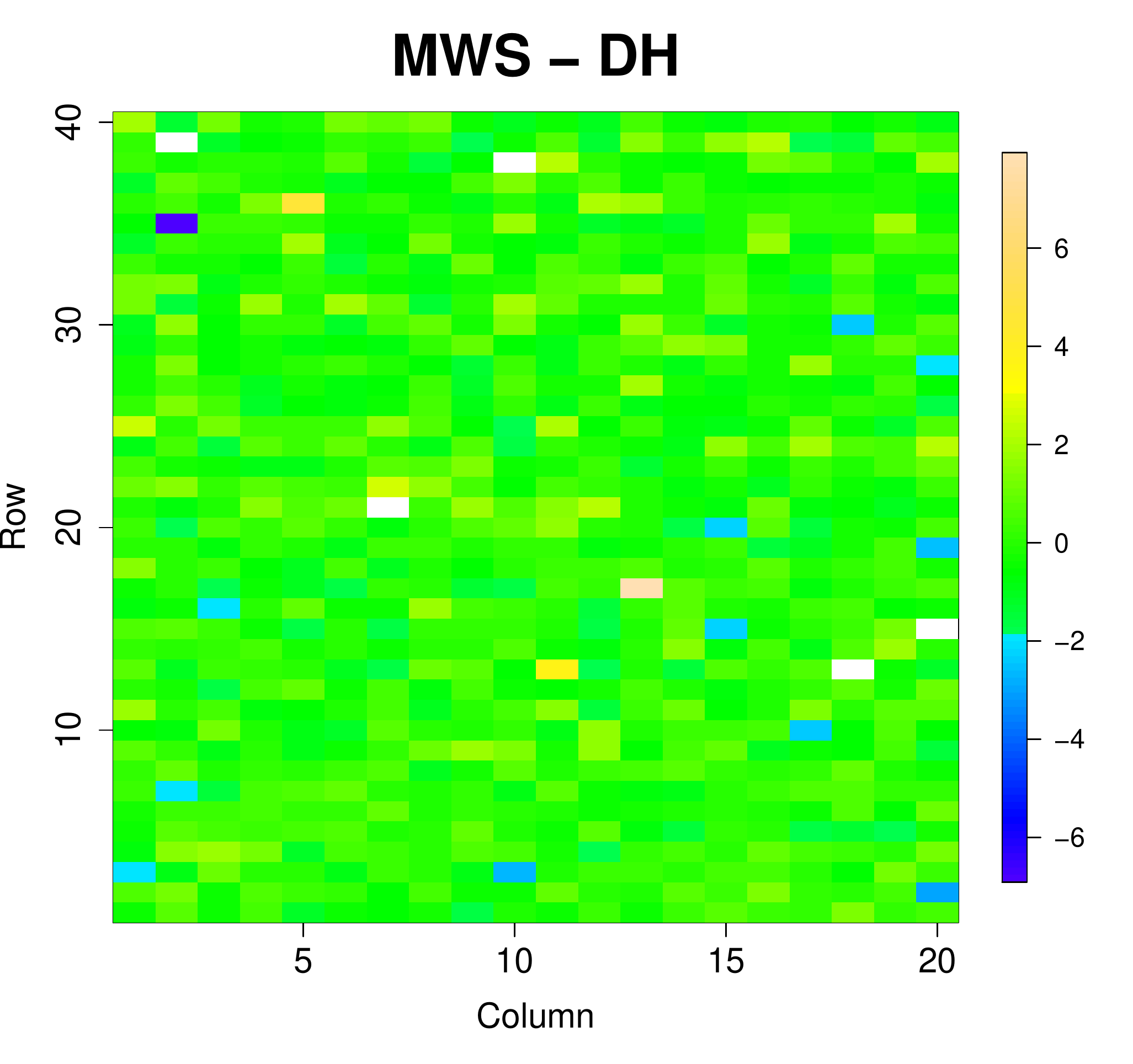}
		\includegraphics[width=3.5cm]{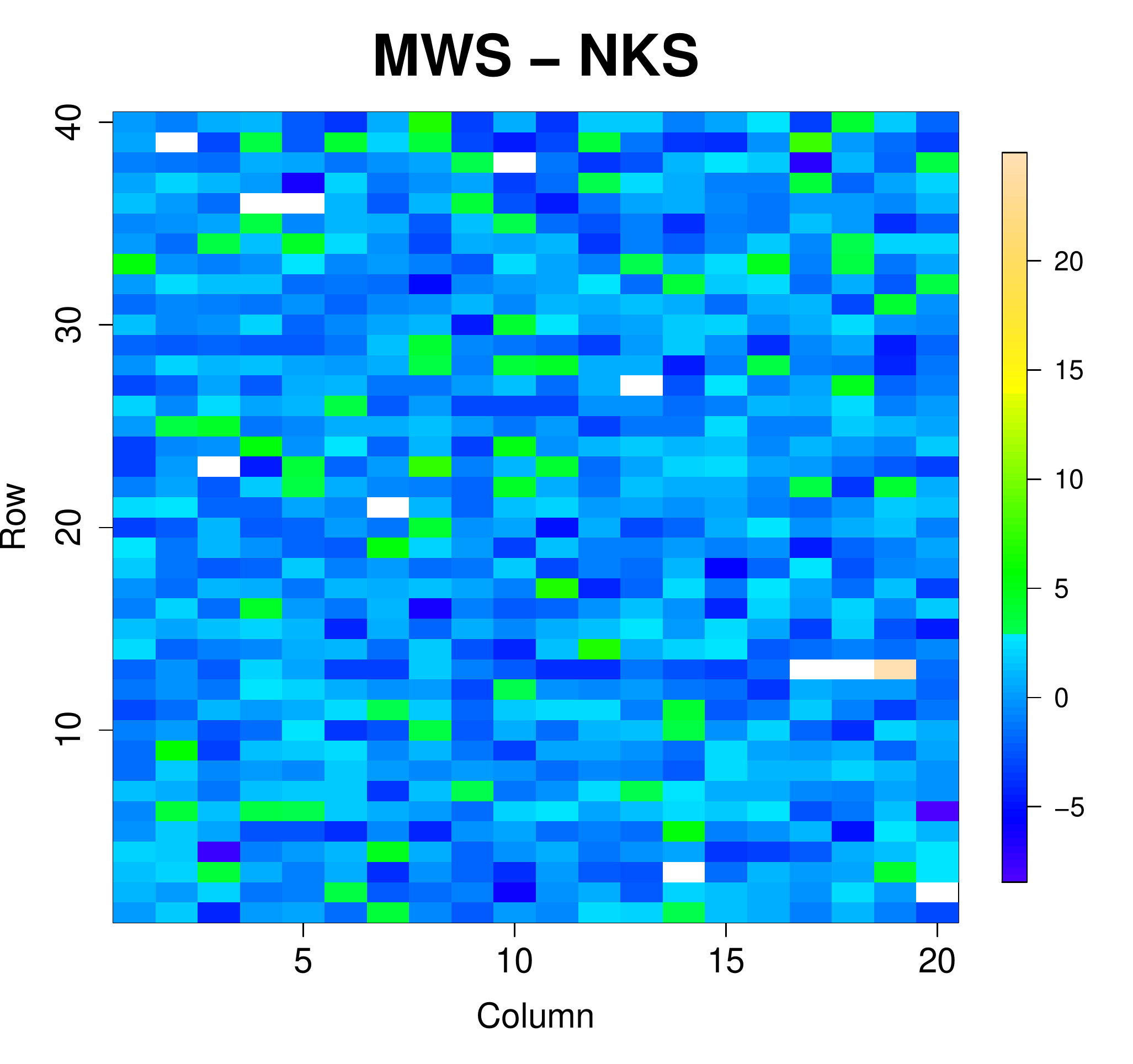}
		\label{Lado_results_residuals_mws}}
    \end{center}
    \caption{Raw data, fitted spatial trend and residuals' spatial plot for the Chilean wheat data in Santa Rosa, 2011, for each trait (GY: Grain Yield; TKW: thousand kernel weight; DH: days to heading; NKS: number of kernels per spike) and mild water stress (MWS) condition.}
		\label{Lado_results_raw_trend_mws}
  \end{figure}

\begin{figure}
    \begin{center}
    \subfigure[Raw data]{
		\includegraphics[width=3.5cm]{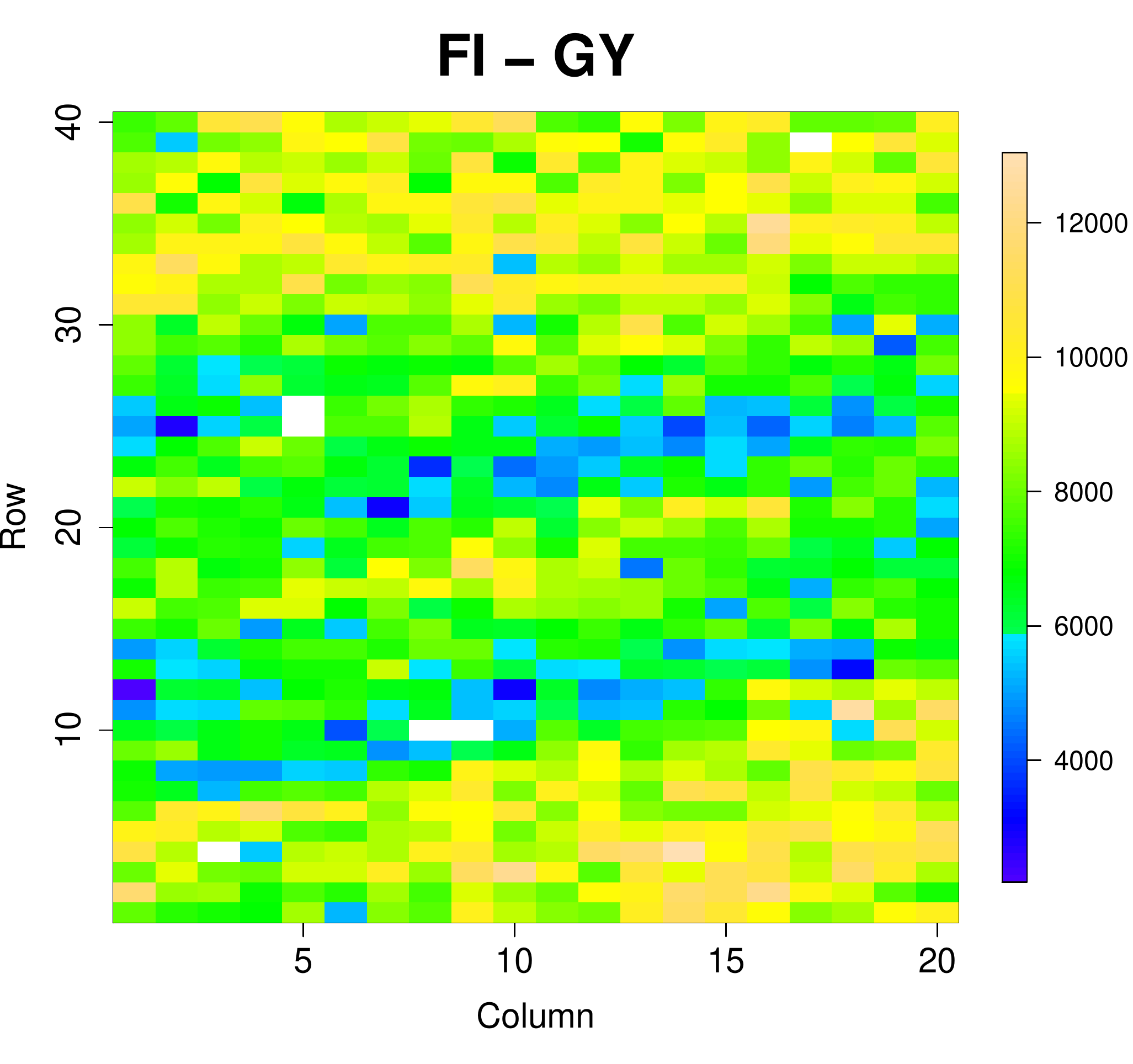}
		\includegraphics[width=3.5cm]{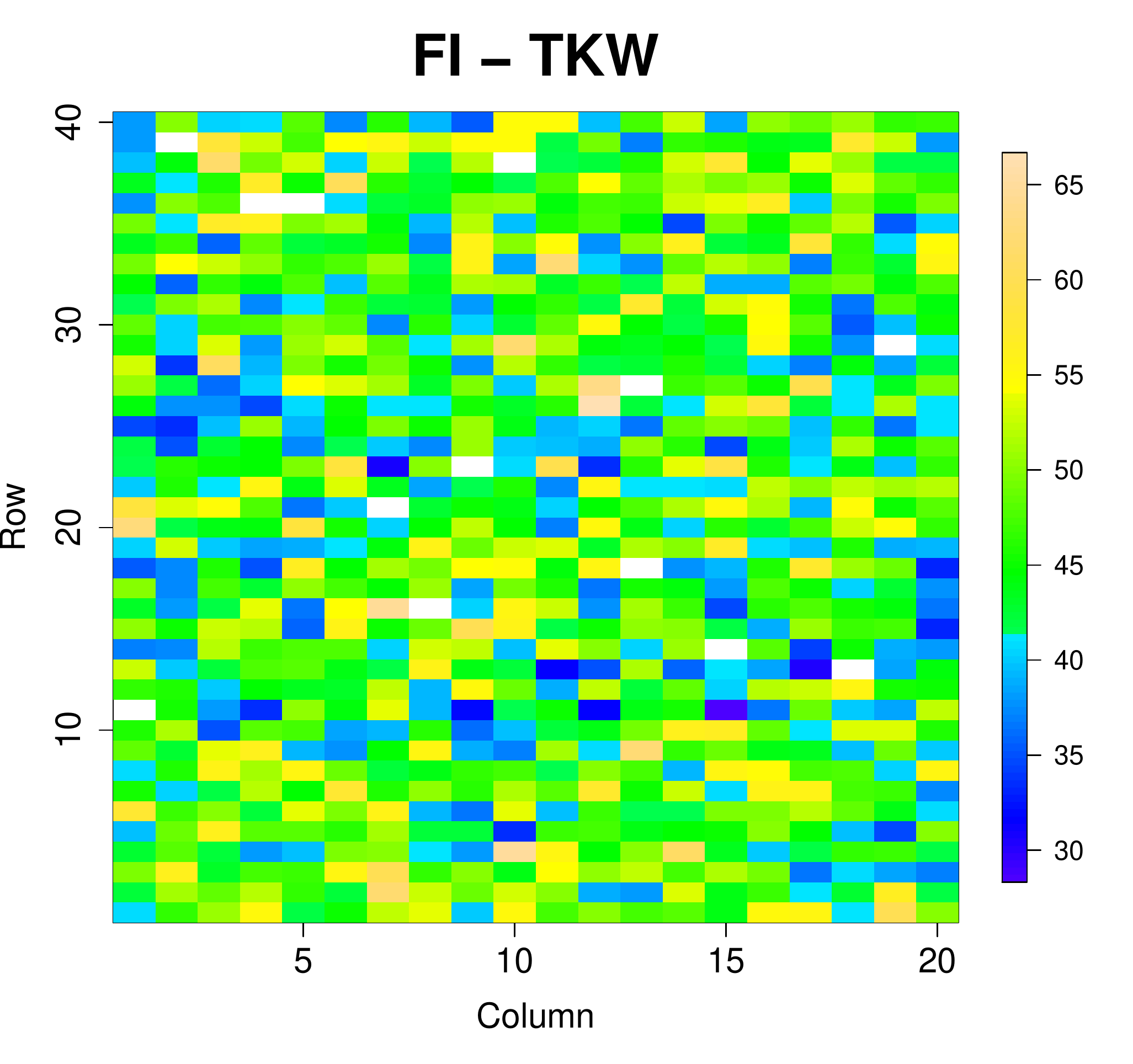}
		\includegraphics[width=3.5cm]{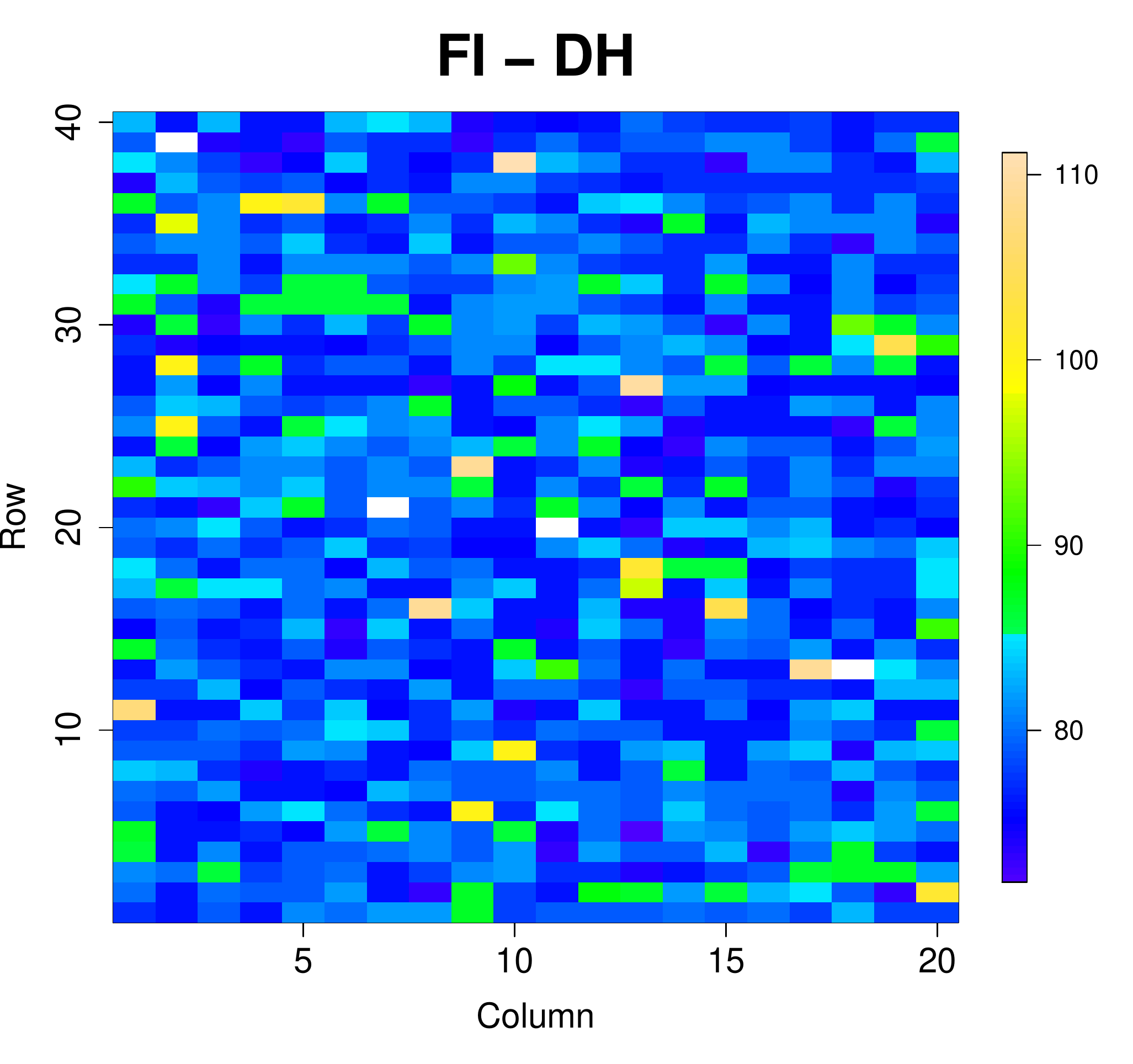}
		\includegraphics[width=3.5cm]{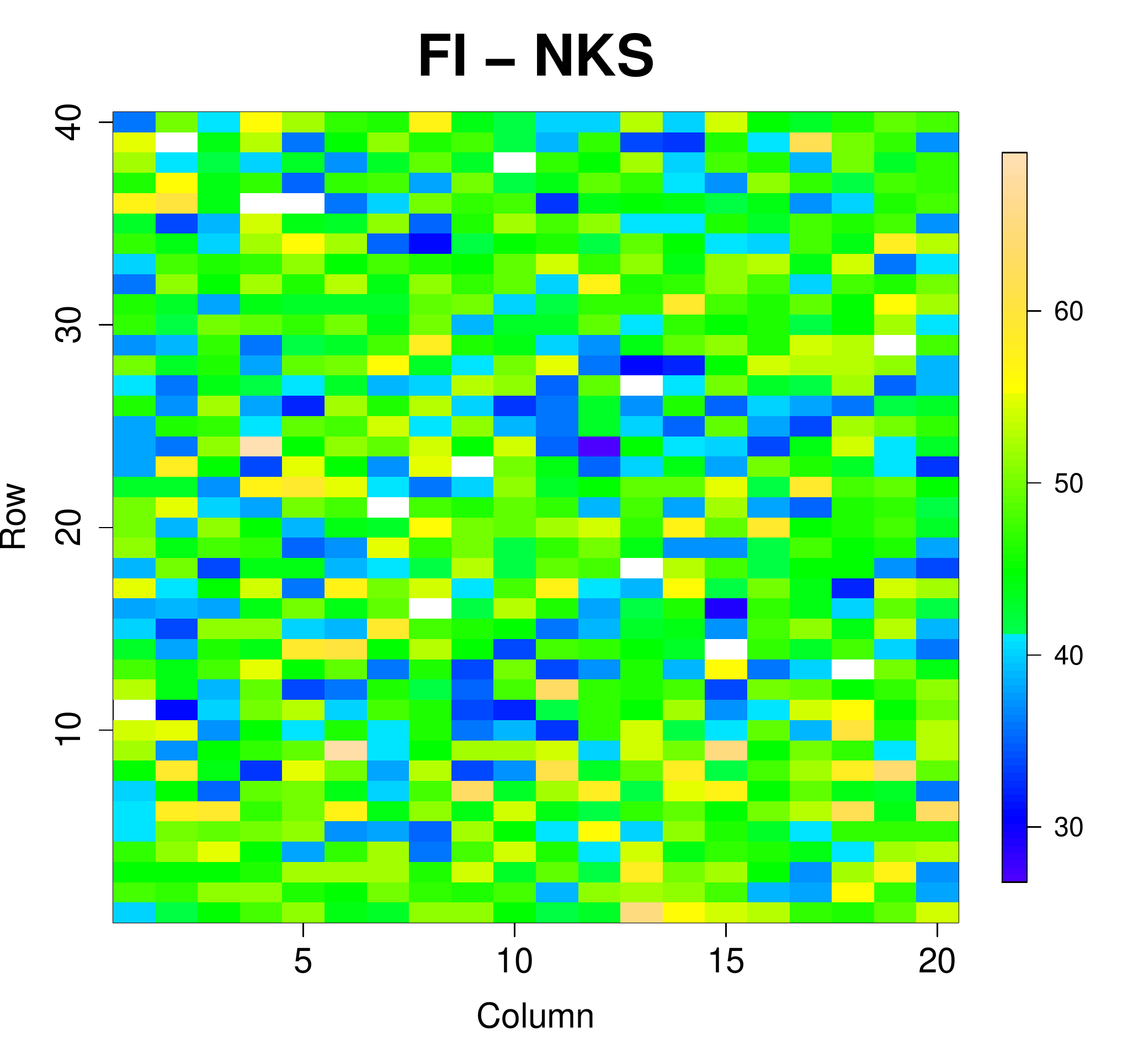}
		\label{Lado_results_raw_fi}}
    \subfigure[Fitted spatial trend]{
		\includegraphics[width=3.5cm]{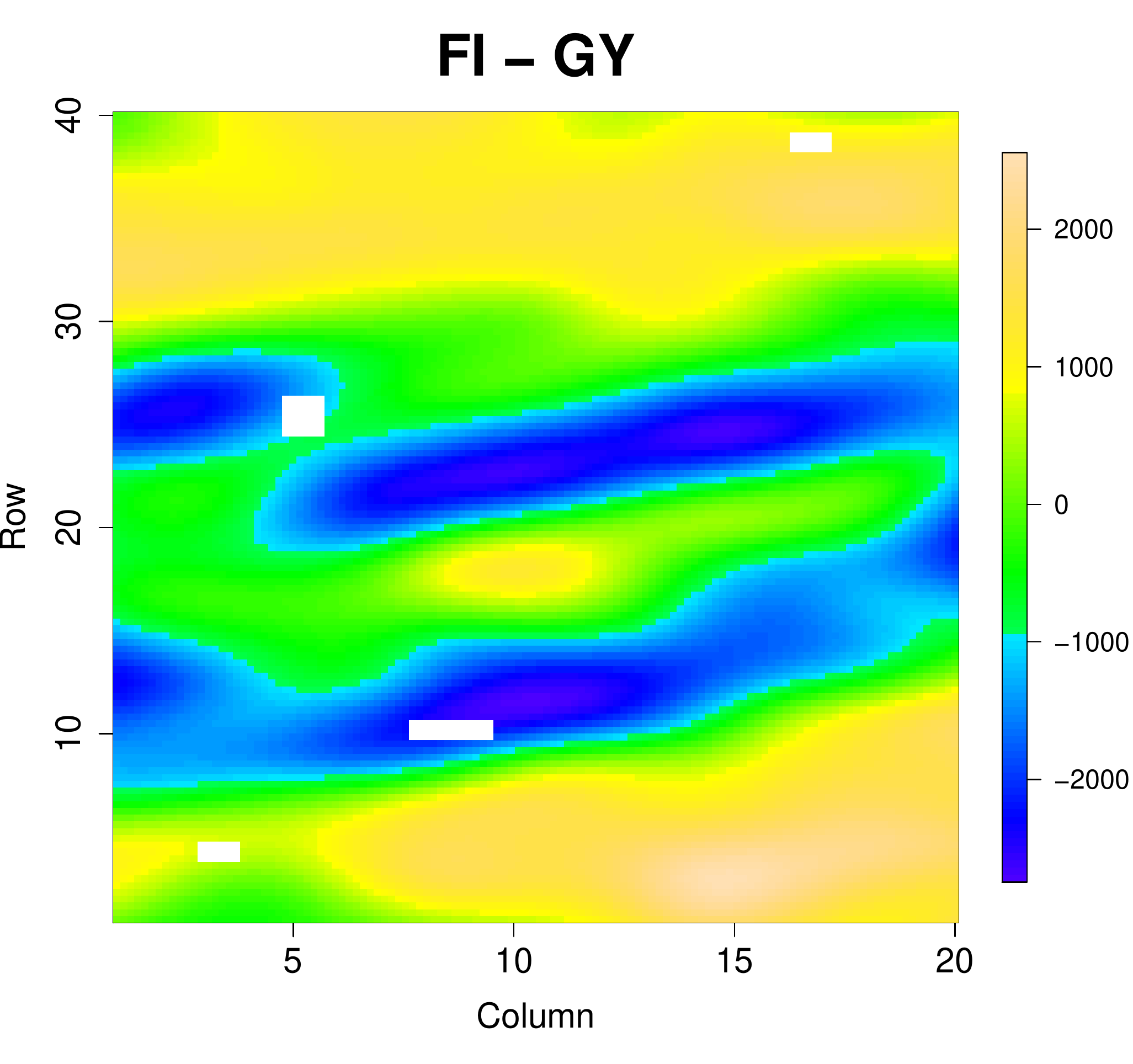}
		\includegraphics[width=3.5cm]{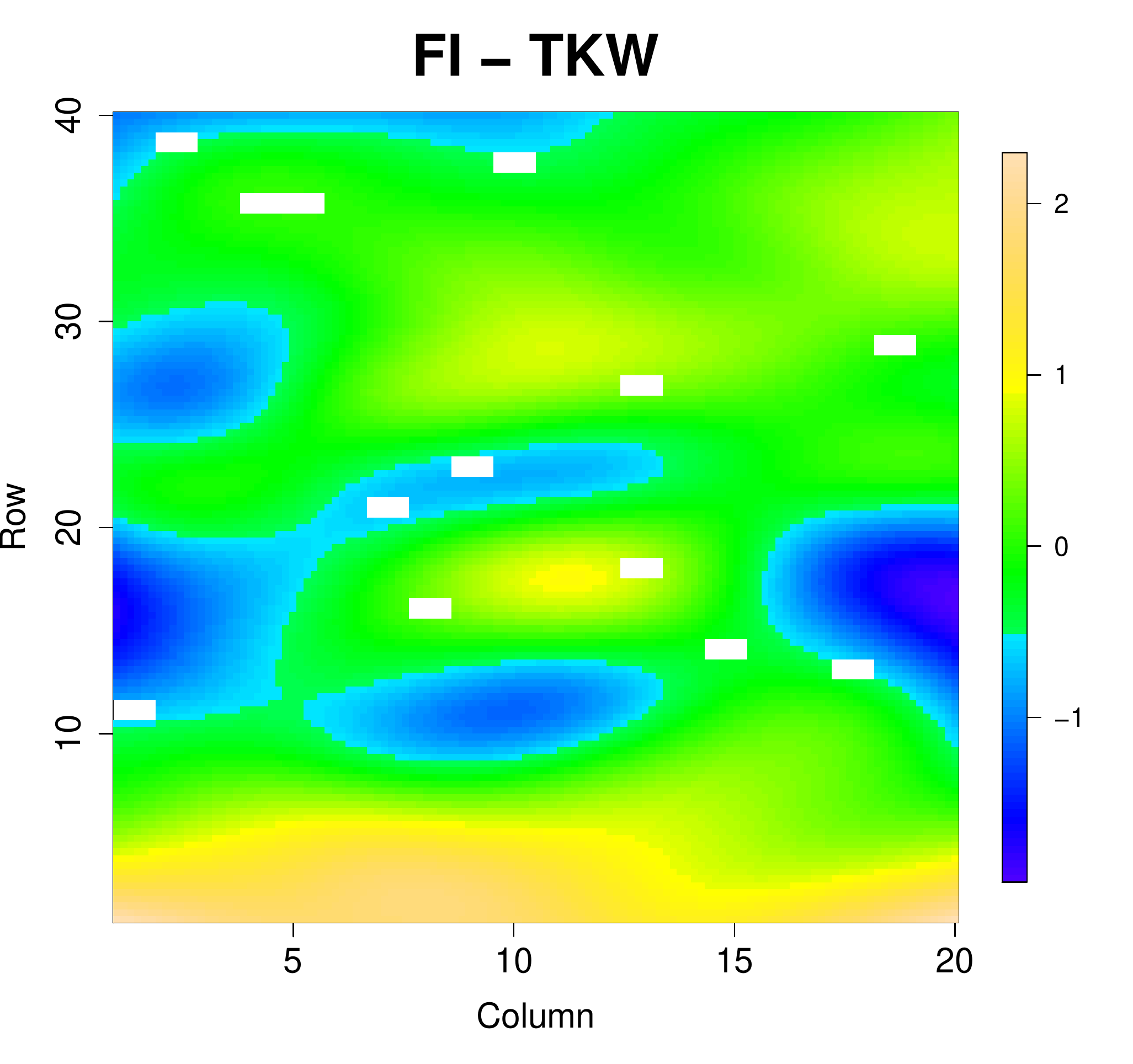}
		\includegraphics[width=3.5cm]{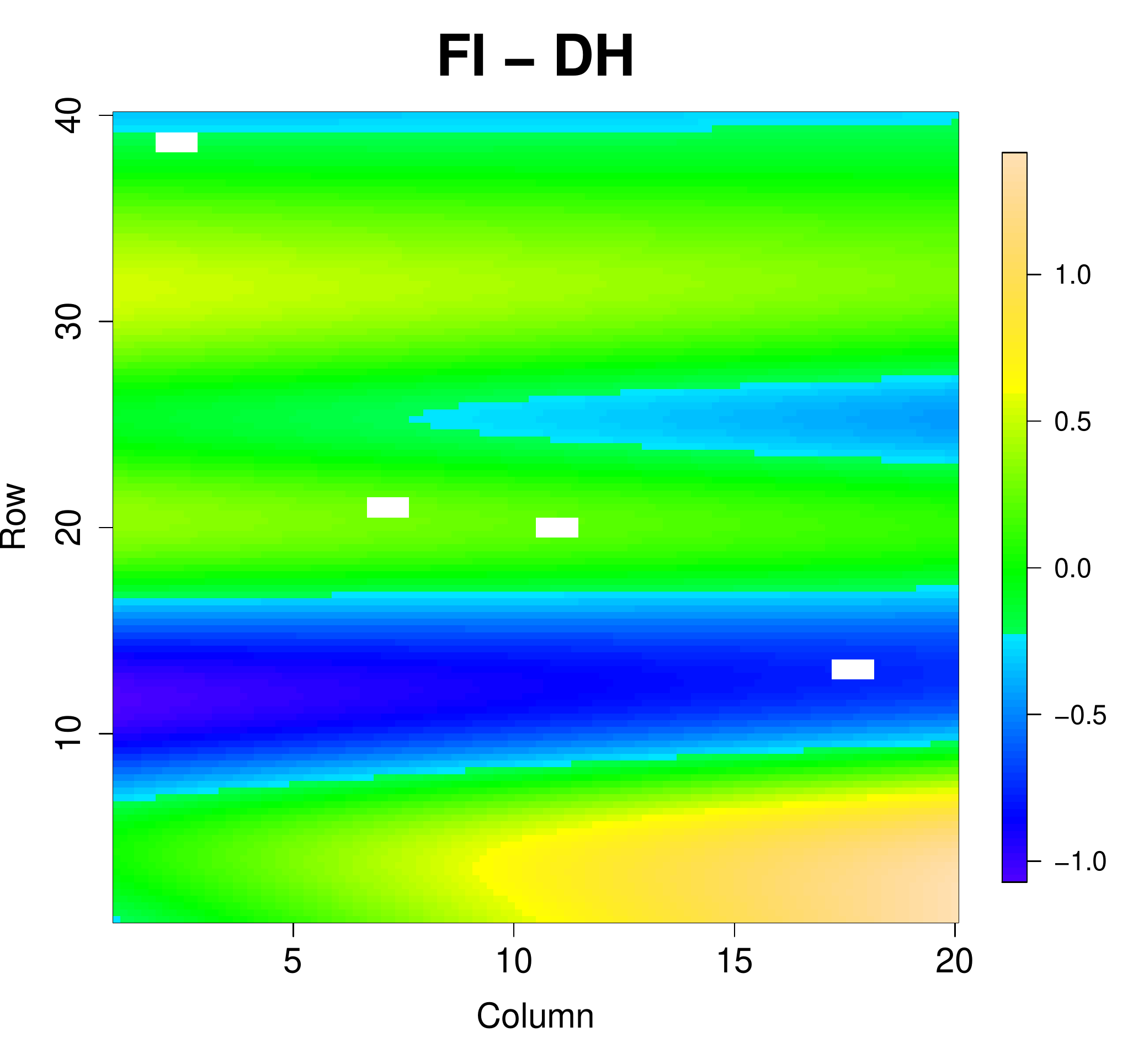}
		\includegraphics[width=3.5cm]{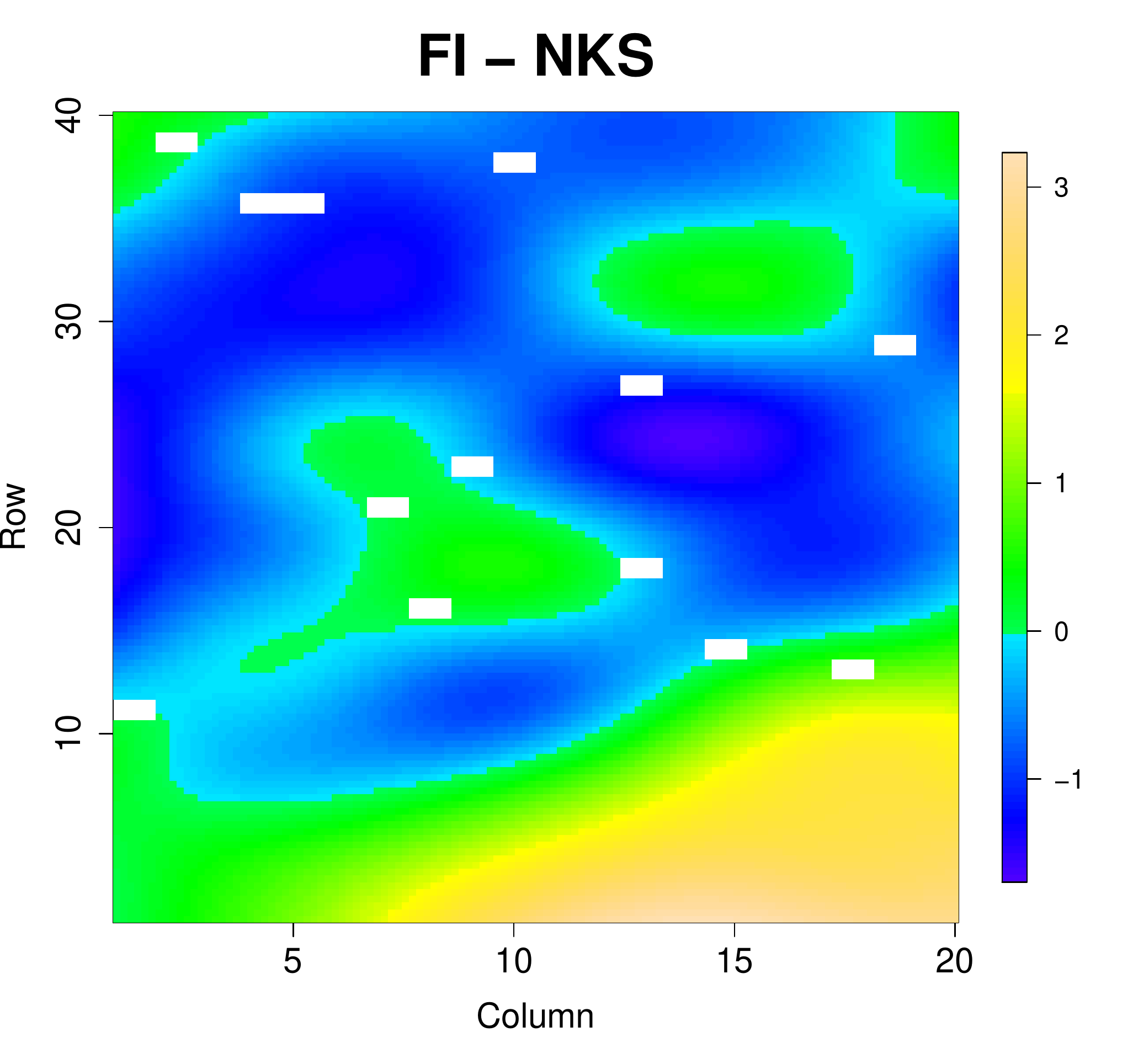}
		\label{Lado_results_trend_fi}}
		\subfigure[Residuals' spatial plot]{
		\includegraphics[width=3.5cm]{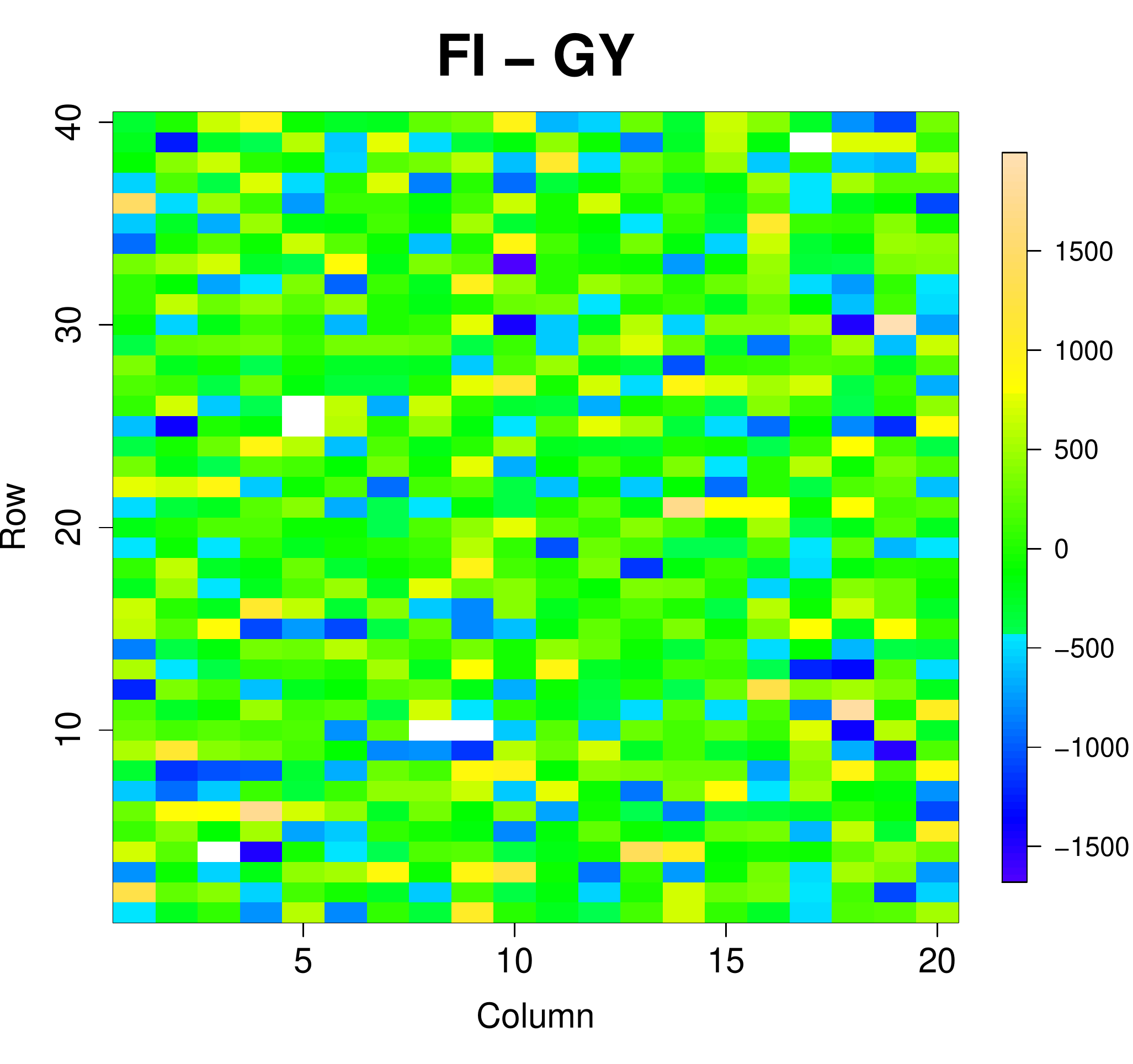}
		\includegraphics[width=3.5cm]{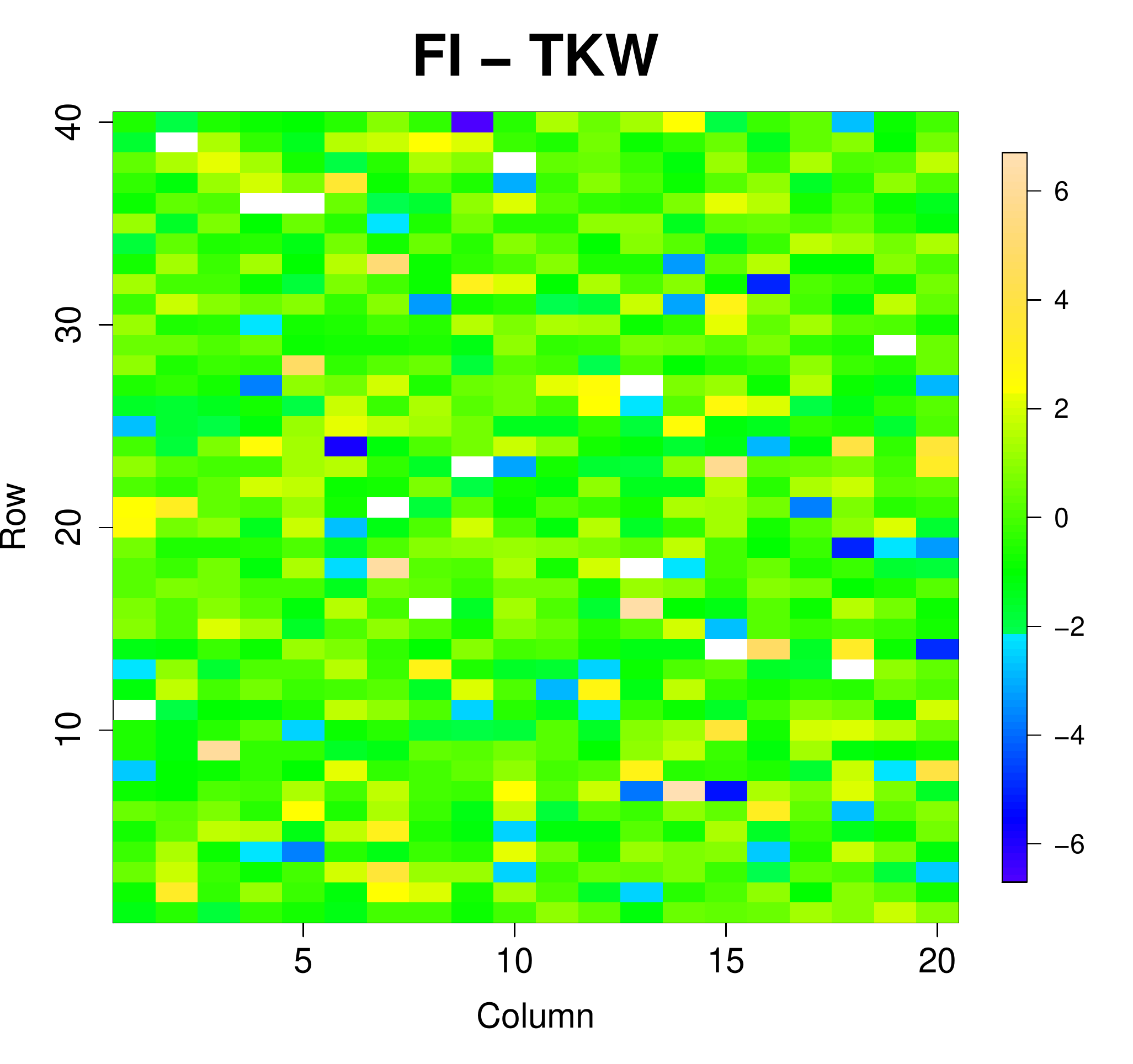}
		\includegraphics[width=3.5cm]{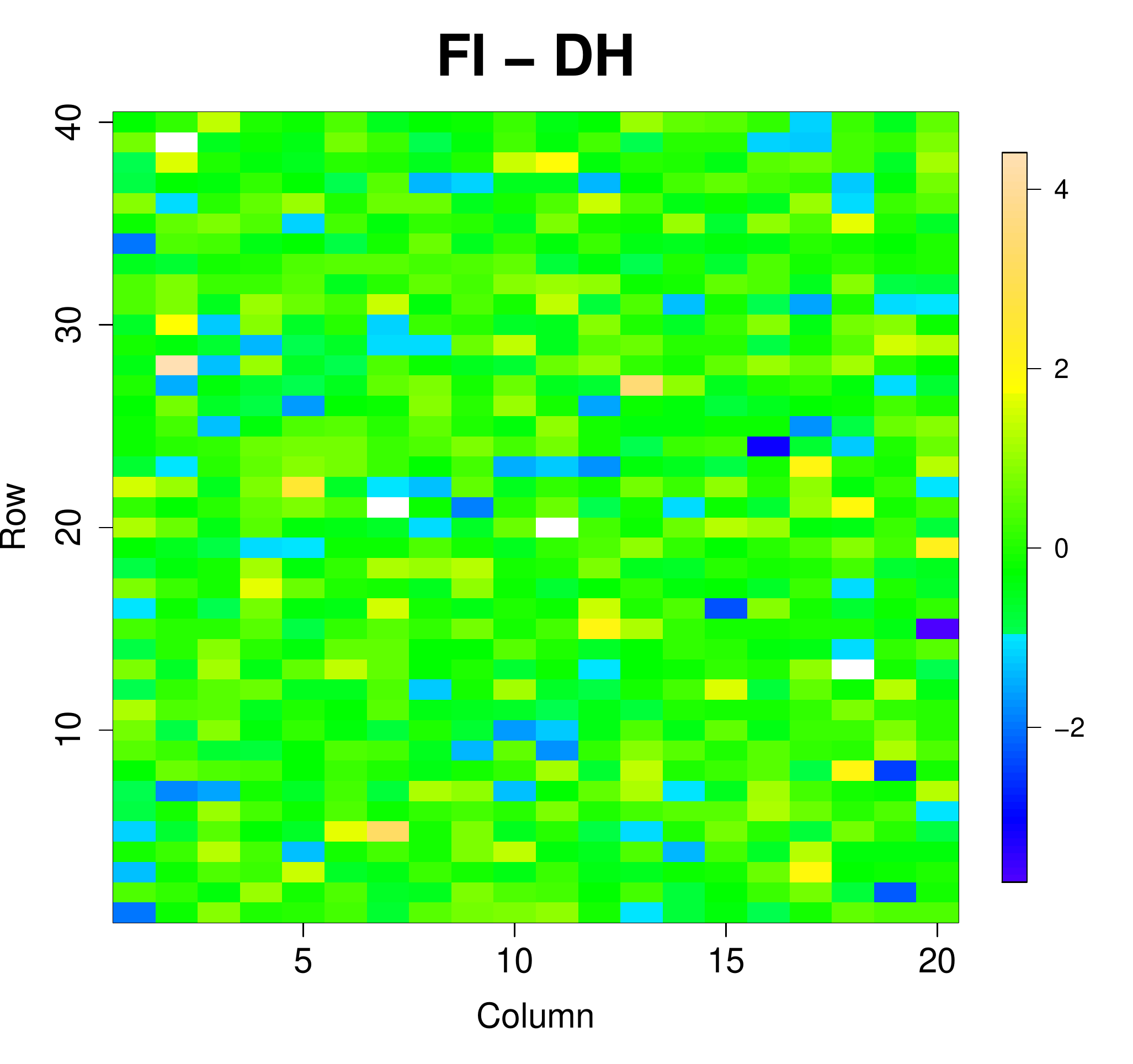}
		\includegraphics[width=3.5cm]{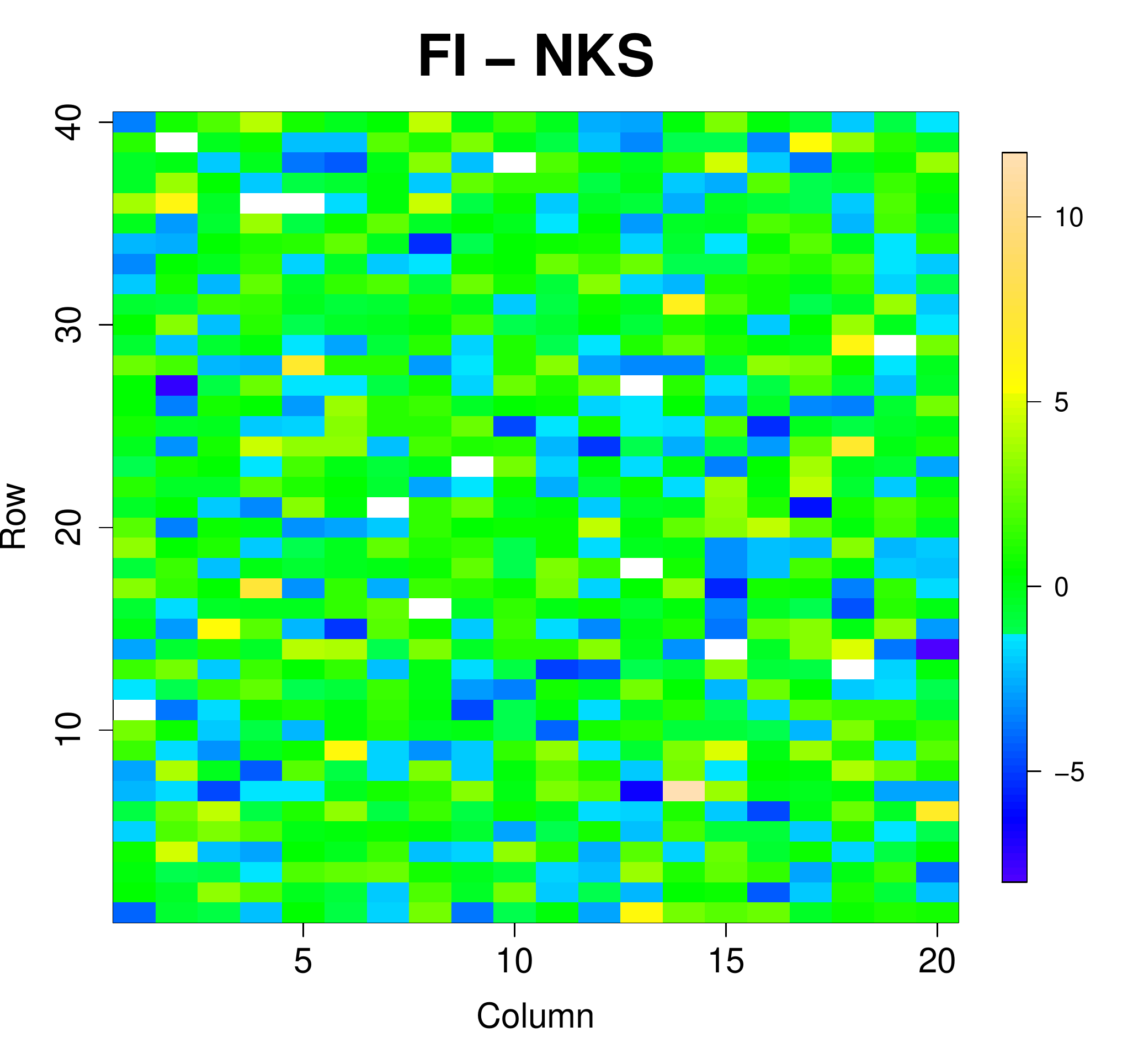}
		\label{Lado_results_residuals_fi}}
    \end{center}
    \caption{Raw data, fitted spatial trend and residuals' spatial plot for the Chilean wheat data in Santa Rosa, 2011, for each trait (GY: Grain Yield; TKW: thousand kernel weight; DH: days to heading; NKS: number of kernels per spike) and fully irrigated (FI) condition.}
		\label{Lado_results_raw_trend_fi}
  \end{figure}

Figures \ref{Lado_results_raw_mws} and \ref{Lado_results_raw_fi} depict the raw data for each trait and water condition. The fitted spatial trends are shown in Figures \ref{Lado_results_trend_mws} and \ref{Lado_results_trend_fi}, and Figures \ref{Lado_results_residuals_mws} and \ref{Lado_results_residuals_fi} shows the spatial plot of the residuals.                                                                                       Table \ref{eff_dim_her_lado} shows the effective dimensions related to each of the five smooth components of the PS-ANOVA spatial trend as well as those associated to the row and column random factors. As for the previous example, some $\mbox{ED}_k$ are zero or close to zero. For instance, for the DH and FI condition, the spatial variation is mainly modeled by the smooth trend over the rows and the row random factor, with the rest of components having a rather low or even null impact. All results suggest that for both, the MWS and FI conditions, the GY is the trait presenting the largest spatial variation, and DH the one with the lowest. This is in concordance with the estimated heritability also presented in Table \ref{eff_dim_her_lado}, with the largest and the lowest heritability having been obtained for DH and GY, respectively. It should be noted that, in all cases, the heritability estimate obtained using SpATS is larger than the broad sense heritability reported in the paper by \cite{Lado2013}.
	
\begin{table}
\caption{For the Chilean wheat data: Effective dimensions associated to the spatial trend and the row and column random factors, and generalized heritability. The letter $u$ denotes the row position, $v$ the column position, and $\boldsymbol{c}_r$ and $\boldsymbol{c}_r$ the row and column random factors, respectively. The results shown are for Santa Rosa in 2011 and for each trait (GY: Grain Yield; TKW: thousand kernel weight; DH: days to heading; NKS: number of kernels per spike) and condition (MWS: Mild water stress; FI: Fully irrigated).}\label{eff_dim_her_lado}
\center
\begin{tabular}{clcccccccc}\hline
& & \multicolumn{8}{c}{Condition and Trait}\\\cline{3-10}
& & \multicolumn{4}{c}{MWS} & \multicolumn{4}{c}{FI}\\
& & GY & TKW & DH & NKS & GY & TKW & DH & NKS\\\cline{3-10}
\multirow{8}{*}{\begin{tabular}{@{}c@{}}Effective dimension\\(ED$_k$)\end{tabular}} & $f_u(u)$ & 8.8 & 8.5 & 6.6 & 2.4 & 10.2 & 2.9 & 7.2 & 1.3\\
																		 & $f_v(v)$ & 0.0 & 0.7 & 0.5 & 0.0 & 0.6 & 0.7 & 0.0 & 0.0\\
																		 & $vh_(u)$ & 16.8 & 16.7 & 1.1 & 7.6 & 8.4 & 0.4 & 1.4 & 0.5\\
																		 & $uh_v(v)$ & 0.0 & 0.0 & 0.0 & 1.8 & 1.6 & 0.0 & 0.1 & 0.7\\
																		 & $f_{u,v}(u,v)$ & 90.6 & 49.5 & 0.2 & 0.6 & 64.9 & 17.9 & 0.0 & 12.2\\
																		 & $\boldsymbol{c}_r$ & 20.2 & 5.2 & 3.7 & 12.5 & 13.7 & 17.8 & 7.7 & 23.7\\
																		 & $\boldsymbol{c}_c$ & 6.9 & 0.4 & 5.6 & 3.6 & 6.3 & 0.0 & 0.0 & 0.3 \\
																		 & Total & 143.3 & 81.0 & 17.7 & 28.5 & 105.7 & 39.7 & 16.4 & 38.8\\\hline
\multicolumn{2}{c}{$H_g^2$} & 0.74 & 0.94 & 0.96 & 0.84 & 0.69 & 0.94 & 0.98 & 0.86\\\hline															
\end{tabular}
\end{table} 
\section{Discussion}
We have presented a powerful and efficient new approach to the modeling of field trials. Effects of genotypes are described by a mixed model in a standard way, on top of tensor product P-splines to fit the spatial field. 

Our approach breaks with the current tradition to model the spatial component as correlated noise. We believe that this offers large advantages. First, as shown in the simulations, modeling spatial correlation by e.g. autoregressive processes may need a lot of manual ad hoc tuning. In contrast, our SpATS model is very robust and runs without user intervention, as our testing on very many real trials has shown. A second advantage is that model selection is substantially simplified. Furthermore, the applications showed that our method can easily deal with irregular field layouts, and local patches with special behavior become easily visible. We believe that having an explicit estimate of the spatial field variation is very helpful. In many cases it is more complex than an autoregressive process. Besides, the ANOVA-type decomposition of tensor product P-splines provides interesting insights on the field trial being analyzed. 

An important result of this paper is the expression of the standard as well as generalized heritability on the basis of the genetic effective dimension. This result provides, on the one hand, a common definition of heritability that can be used for a broad range of statistical models used for the analysis of field trials. On the other hand, we believe that the link between the genetic effective dimension and the (generalized) heritability measure brings new insights to the interpretation of both quantities. To the best of our knowledge, this is the first time that this relation is presented in the literature. The study of the appropriateness of this result to non-Gaussian responses represents an interesting area of research. 

What has been said here about genetic variation (and the associated heritability measure) also applies to other random effects incorporated into a mixed model. This suggests moving the interpretation of the results obtained by fitting a mixed model from estimated variances components to estimated effective dimensions. Moreover, evaluating the results in terms of effective dimensions also furnishes a common scale that will allow determining those random model components that contribute the most to explaining the observed phenotypic variation. 

A current line of research is focused on the extension of our SpATS model to the analysis of multi-environment trials (METs, trials carried out in multiple environments or contexts, \citealp[see, e.g., ][]{Smith2001}). In the analysis of METs we are faced with several challenges. First, the modeling of a different spatial trend for each environment, and second the need to account for the (possible) interaction between the genotype and the environment. The first challenge may be approached by including an interaction between a factor (the environment) and a smooth surface (the tensor product P-spline). The second challenge is related to the inclusion of correlated random effects (whenever the genotype effect is treated as random). This possibility will allow assuming a different genetic variance for each environment, and, at the same time, modeling the (within environments) correlation between genotypes sharing the same environment and the (across environments) correlation of genotype effects along the different environments. Further work is warranted to develop computationally-efficient estimation procedures in this setting. 

Our calculations were done with the \texttt{R}-package \texttt{SpATS} that can be freely downloaded from \url{https://cran.r-project.org/package=SpATS}. However, it is worth remembering that estimation can also be accommodated using any standard mixed model software.
\section*{Acknowledgements}
This research was supported by the Spanish Ministry of Economy and Competitiveness MINECO grants MTM2014-55966-P and BCAM Severo Ochoa excellence accreditation SEV-2013-0323, and by the Basque Government through the BERC 360 2014-2017. The first author thanks the Agrupamento INBIOMED from DXPCTSUG-FEDER unha maneira de facer Europa (2012/273). We thank SESVanderHave for providing the sugar beet data. We are grateful to Cajo ter Braak, Mar\'ia Durb\'an, Dae-Jin Lee and Julio Velazco for useful discussion. 
\appendix
\section{Nested bases\label{s_p2mix}}
With large data sets the computation of $\boldsymbol{Z}_s$ in (\ref{mm_XZ_psp2d}), as well as its inner product, can demand a lot of time, especially for large values of $L$ and $P$. \cite{Lee2013} propose to speed up computation by using nested bases. The idea is to reduce the dimension of the marginal B-spline bases $\invbreve{\boldsymbol{B}}$ and $\breve{\boldsymbol{B}}$ (and therefore the associated number of coefficients to be estimated), but only for the smooth-by-smooth interaction term
, i.e., $f_{u,v}$. As pointed out by the authors, this simplification can be justified by the fact that the main effects, $f_u$ and $f_v$, and the smooth varying coefficient terms, $h_v$ and $h_u$, would in fact explain most of the structure (or spatial trend) presented in the data, and so a less rich representation of the interaction term could be needed. 

Let $\invbreve{\boldsymbol{B}}_{N}$ and $\breve{\boldsymbol{B}}_{N}$ be two reduced marginal B-spline basis of dimension $n \times L_{N}$ ($L_{N} < L$) and $n \times P_{N}$ ($P_{N} < P$) with associated penalty matrices $\invbreve{\boldsymbol{D}}^{t}_{N}\invbreve{\boldsymbol{D}}_{N}$ and $\breve{\boldsymbol{D}}_{N}^{t}\breve{\boldsymbol{D}}_{N}$, respectively. Then, the reduced mixed model matrix $\boldsymbol{Z}_s$ for the PS-ANOVA model is constructed as follows
\begin{equation*}
\boldsymbol{Z}_s = \left[\boldsymbol{Z}_v,\boldsymbol{Z}_u,\boldsymbol{Z}_v\Box\boldsymbol{u},\boldsymbol{v}\Box\boldsymbol{Z}_u,\widetilde{\boldsymbol{Z}}_v\Box\widetilde{\boldsymbol{Z}}_u\right],
\label{mm_Z_psp2d_reor_nested}
\end{equation*}
where $\widetilde{\boldsymbol{Z}}_v = \invbreve{\boldsymbol{B}}_{N}\widetilde{\boldsymbol{U}}^{N}_v$ and $\widetilde{\boldsymbol{Z}}_u = \breve{\boldsymbol{B}}_{N}\widetilde{\boldsymbol{U}}^{N}_v$, with $\widetilde{\boldsymbol{U}}^{N}_v$ and $\widetilde{\boldsymbol{U}}^{N}_v$ being the matrices containing the eigenvectors associated to the non-zero eigenvalues of $\invbreve{\boldsymbol{D}}^{t}_{N}\invbreve{\boldsymbol{D}}_{N}$ and $\breve{\boldsymbol{D}}_{N}^{t}\breve{\boldsymbol{D}}_{N}$, respectively. 

\begin{figure}
    \begin{center}
    \includegraphics[width=9cm]{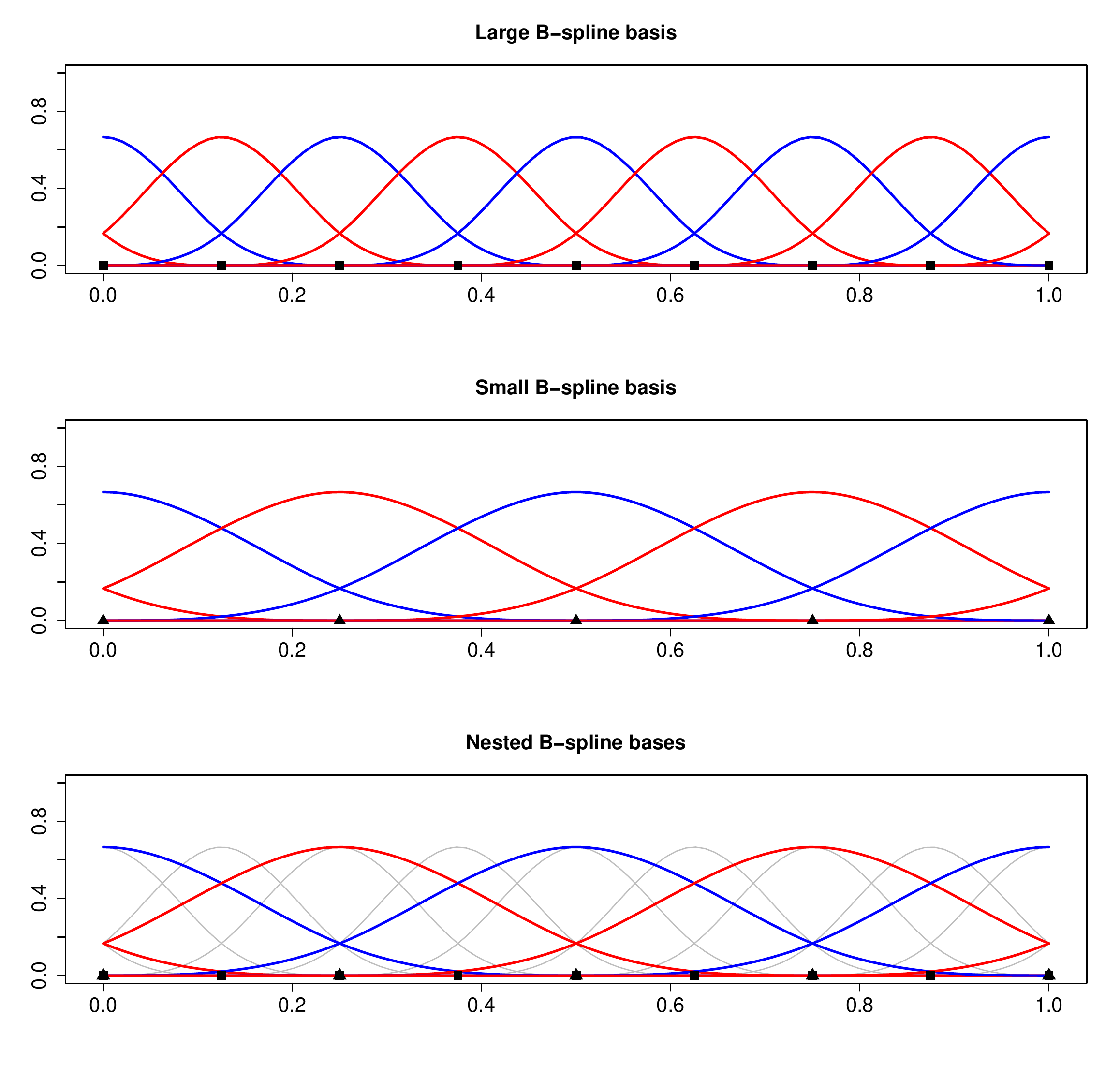}
    \end{center}
    \caption{Example of nested B-spline bases.}
    \label{f_nested_bases}
  \end{figure}

In order to ensure that the reduced model is in fact nested in the model including only the main effects, \cite{Lee2013} showed that the number of segments that define $\invbreve{\boldsymbol{B}}_{N}$ and $\breve{\boldsymbol{B}}_{N}$ should be a divisor of the number of segments used in the original bases $\invbreve{\boldsymbol{B}}$ and $\breve{\boldsymbol{B}}$. The reasoning behind is graphically illustrated in Figure~\ref{f_nested_bases}, that has been taken from the paper by \cite{Lee2013}. The two top plots depict two cubic B-spline bases of dimension $11$ and $7$, respectively. The squares and triangles denote the breakpoints (knots) that define each segment. As can be observed, the knots of the small basis correspond to a subset of the knots of the large basis. This implies that the space spanned by the small basis is a subset of the space spanned by the large one. This can be seen on the plot at the bottom, where both bases overlap.

Note that the use of nested bases reduces the number of coefficients associated to $f_{u,v}$ from $(L - 2)(P - 2)$ to $(L_N - 2)(P_N - 2)$. This reduction allows being generous with the number of B-spline basis functions used for the main effects and the smooth varying coefficient terms. Our experience suggest using (a) as many segments for $\invbreve{\boldsymbol{B}}$ and $\breve{\boldsymbol{B}}$ as number of rows and columns in the field, respectively; and (b) half the number of segments for the nested basis.

\section{Some results on mixed models and hat matrices\label{mixed_model_results}}	
For given values of the variance components ($\sigma^2$ and $\sigma_k^2$, $k = 1, \ldots, q$), BLUEs for $\boldsymbol{\beta}$ and BLUPs for $\boldsymbol{c}$ can be obtained as the solution to the linear system of equations \citep{Henderson1963}
\begin{equation}
\underbrace{
\begin{bmatrix}
\boldsymbol{X}^{t}\boldsymbol{R}^{-1}\boldsymbol{X} & \boldsymbol{X}^{t}\boldsymbol{R}^{-1}\boldsymbol{Z} \\
\boldsymbol{Z}^{t}\boldsymbol{R}^{-1}\boldsymbol{X} & \boldsymbol{G}^{-1} + \boldsymbol{Z}^{t}\boldsymbol{R}^{-1}\boldsymbol{Z}
\end{bmatrix}}_{\boldsymbol{C}}
\begin{bmatrix} 
\widehat{\boldsymbol{\beta}}\\
\widehat{\boldsymbol{c}}
\end{bmatrix}
=
\begin{bmatrix}
\boldsymbol{X}^{t}\boldsymbol{R}^{-1}\boldsymbol{y}\\
\boldsymbol{Z}^{t}\boldsymbol{R}^{-1}\boldsymbol{y}
\end{bmatrix}
\label{MX:linearsystem}
\end{equation}
which gives rise to closed-form expressions
\begin{align}
\widehat{\boldsymbol{\beta}} & = \left(\boldsymbol{X}^{t}\boldsymbol{V}^{-1}\boldsymbol{X}\right)^{-1}\boldsymbol{X}^{t}\boldsymbol{V}^{-1}\boldsymbol{y},\label{mm_fixedest}\\
\widehat{\boldsymbol{c}} & = \boldsymbol{G}\boldsymbol{Z}^{t}\boldsymbol{V}^{-1}(\boldsymbol{y} - \boldsymbol{X}\widehat{\boldsymbol{\beta}}) = \boldsymbol{G}\boldsymbol{Z}^{t}\boldsymbol{Q}\boldsymbol{y}\label{mm_randomest},
\end{align}
where $\boldsymbol{V}$ and $\boldsymbol{Q}$ have been defined in Section \ref{s_estproc}. As it will be seen, the last equivalence in (\ref{mm_randomest}) plays an important role in our approach, and it is simply obtained by substituting $\widehat{\boldsymbol{\beta}}$ by (\ref{mm_fixedest}). The previous expressions reveal that the hat matrices associated to the fixed, $\boldsymbol{\beta}$, and random, $\boldsymbol{c}$, effects are $\boldsymbol{H}_F = \boldsymbol{X}\left(\boldsymbol{X}^{t}\boldsymbol{V}^{-1}\boldsymbol{X}\right)^{-1}\boldsymbol{X}^{t}\boldsymbol{V}^{-1}$ and $\boldsymbol{H}_R = \boldsymbol{Z}\boldsymbol{G}\boldsymbol{Z}^{t}\boldsymbol{Q}$ respectively (as already discussed in Section \ref{eff_dim}).

It is worth remembering that in this paper we assume that $\boldsymbol{Z} = [\boldsymbol{Z}_1, \ldots, \boldsymbol{Z}_q]$, where each $\boldsymbol{Z}_k$ represents the design matrix associated to the $k$-th random factor $\boldsymbol{c}_k$, with $\boldsymbol{c} = \left(\boldsymbol{c}_1^{t},\ldots,\boldsymbol{c}_q^{t}\right)^{t}$, and that the variance-covariance  $\boldsymbol{G} = \bigoplus_{k = 1}^{q}\boldsymbol{G}_{k}$. As a consequence
\begin{align*}
\widehat{\boldsymbol{c}} = 
\begin{bmatrix} 
\widehat{\boldsymbol{c}}_1 \\ 
\widehat{\boldsymbol{c}}_2 \\ 
\vdots \\ 
\widehat{\boldsymbol{c}}_q
\end{bmatrix} 
& = 
\begin{bmatrix}
\boldsymbol{G}_1 & \boldsymbol{0} &  \cdots & \boldsymbol{0}\\
\boldsymbol{0} & \boldsymbol{G}_2 &  \cdots & \boldsymbol{0}\\
\vdots & \vdots &  \ddots & \vdots\\
\boldsymbol{0} & \boldsymbol{0} &  \cdots & \boldsymbol{G}_q\\
\end{bmatrix}
\begin{bmatrix}
\boldsymbol{Z}_1^{t}\\
\boldsymbol{Z}_2^{t}\\
\vdots\\
\boldsymbol{Z}_q^{t}\\
\end{bmatrix}
\boldsymbol{Q}\boldsymbol{y} \\
& = 
\begin{bmatrix} 
\boldsymbol{G}_1\boldsymbol{Z}_1^{t}\boldsymbol{Q}\boldsymbol{y}\\ 
\boldsymbol{G}_2\boldsymbol{Z}_2^{t}\boldsymbol{Q}\boldsymbol{y}\\ 
\vdots \\ 
\boldsymbol{G}_q\boldsymbol{Z}_q^{t}\boldsymbol{Q}\boldsymbol{y},
\end{bmatrix}. 
\end{align*}
This result implies that
\begin{align*}
\boldsymbol{H}_R\boldsymbol{y} & = \boldsymbol{Z}\widehat{\boldsymbol{c}} = \sum_{k = 1}^{q}\boldsymbol{Z}_k\widehat{\boldsymbol{c}}_k = \sum_{k = 1}^{q}\boldsymbol{Z}_k\boldsymbol{G}_k\boldsymbol{Z}_k^{t}\boldsymbol{Q}\boldsymbol{y} = \sum_{k = 1}^{q}\boldsymbol{H}_k\boldsymbol{y},
\end{align*} 
where $\boldsymbol{H}_k = \boldsymbol{Z}_k\boldsymbol{G}_k\boldsymbol{Z}_k^{t}\boldsymbol{Q}$. Accordingly, the hat matrix associated to the random part of model (\ref{mm_equation}) can be decomposed as a sum of independent hat matrices, each related to a specific random factor in the model, i.e., 
\[
\boldsymbol{H}_R = \sum_{k = 1}^{q}\boldsymbol{H}_k.
\]
Finally, it can also be shown \cite[see, e.g., eqn. (9d) and Appendix 1 in ][]{Johnson1995} that
\begin{equation}
\widehat{\boldsymbol{\varepsilon}} = \boldsymbol{y} - \boldsymbol{X}\widehat{\boldsymbol\beta} + \boldsymbol{Z}\widehat{\boldsymbol{c}} = \boldsymbol{R}\boldsymbol{Q}\boldsymbol{y},
\label{hat_residuals}
\end{equation}
and the residuals' hat matrix is thus $\boldsymbol{H}_{\boldsymbol{\varepsilon}} = \boldsymbol{R}\boldsymbol{Q}$.
\section{Equivalence between effective dimensions definitions\label{ed_equivalence}}
This section presents the equivalence between the definition given by \cite{Cui2010} of the effective dimension associated to a model's component and $\mbox{ED}_k$. Specifically, \cite{Cui2010} define the effective dimension of a random factor $\boldsymbol{c}_k$ as
\begin{align}
\mbox{ED}\left(\boldsymbol{Z}_k\right) & = \lim_{\varsigma \rightarrow + \infty}\mbox{trace}\left(\boldsymbol{Z}_k\boldsymbol{G}_k\boldsymbol{Z}_k^{t}\left(\boldsymbol{V} + \varsigma\boldsymbol{X}\boldsymbol{W}\boldsymbol{X}^{t}\right)^{+}\right)\label{ED_Cui_1}\\
& = \mbox{trace}\left(\boldsymbol{Z}_k\boldsymbol{G}_k\boldsymbol{Z}_k^{t}\left[\left(\boldsymbol{I}_n - \boldsymbol{P}_{\boldsymbol{X}}\right)\boldsymbol{V}\left(\boldsymbol{I}_n - \boldsymbol{P}_{\boldsymbol{X}}\right)\right]^{+}\right),
\label{ED_Cui_2}
\end{align}
where $\boldsymbol{\Gamma}^{+}$ denotes the Moore-Penrose pseudoinverse of $\boldsymbol{\Gamma}$, $\boldsymbol{W}$ is a positive definite matrix, $\varsigma$ is a positive scalar, and $\boldsymbol{P}_{\boldsymbol{X}} = \boldsymbol{X}\left(\boldsymbol{X}^{t}\boldsymbol{X}\right)^{-1}$. The expression given above defines the effective dimension of a model's component as the trace of the ratio of that ``component's modeled variance matrix'' ($\boldsymbol{Z}_k\boldsymbol{G}_k\boldsymbol{Z}_k^{t}$) to ``total variance matrix'' ($\boldsymbol{V} + \varsigma\boldsymbol{X}\boldsymbol{W}\boldsymbol{X}^{t}$). Note that this definition treats $\boldsymbol{\beta}$ as a random vector with variance-covariance $\varsigma\boldsymbol{W}$. However, as pointed out by the authors, a fixed effect can be viewed as the limiting case of a random effect for which the variance-covariance matrix goes to infinity (i.e., when  $\varsigma \rightarrow + \infty$). Similarly to $\mbox{ED}\left(\boldsymbol{Z}_k\right)$, \cite{Cui2010} define the effective dimension for the error term as
\begin{align}
\mbox{ED}\left(\boldsymbol{\varepsilon}\right) & = \lim_{\varsigma \rightarrow + \infty}\mbox{trace}\left(\boldsymbol{R}\left(\boldsymbol{V} + \varsigma\boldsymbol{X}\boldsymbol{W}\boldsymbol{X}^{t}\right)^{+}\right)\label{ED_Cui_sigma}\\
& = \mbox{trace}\left(\boldsymbol{R}\left[\left(\boldsymbol{I}_n - \boldsymbol{P}_{\boldsymbol{X}}\right)\boldsymbol{V}\left(\boldsymbol{I}_n - \boldsymbol{P}_{\boldsymbol{X}}\right)\right]^{+}\right)\nonumber,
\end{align}
and show that
\begin{align*}
n & = \mbox{ED}(\boldsymbol{X}) + \sum_{k = 1}^{q}\mbox{ED}(\boldsymbol{Z}_k) + \mbox{ED}\left(\boldsymbol{\varepsilon}\right)\\
  & = \mbox{rank}\left(\boldsymbol{X}\right) + \sum_{k = 1}^{q}\mbox{ED}(\boldsymbol{Z}_k) + \mbox{ED}\left(\boldsymbol{\varepsilon}\right).
\end{align*}
The definition given in \cite{Cui2010} thus partitions the number of observations $n$ into independent effective dimensions for the model's components and error. In author's words, this result jointly with (\ref{ED_Cui_1}) and (\ref{ED_Cui_sigma}), suggests interpreting the effective dimension of a model's component $\mbox{ED}\left(\boldsymbol{Z}_k\right)$ as the fraction of response variation attributed to that individual effect, and the same applies to the error term. 

To show that $\mbox{ED}_k= \mbox{ED}\left(\boldsymbol{Z}_k\right)$, we use results derived in the paper by \cite{Hoog1990}. Given that \cite[see identity (1) in ][]{Hoog1990}
\[
\boldsymbol{Q} = \boldsymbol{V}^{-1} - \boldsymbol{V^{-1}}\boldsymbol{X}\left(\boldsymbol{X^{t}V^{-1}X}\right)^{-1}\boldsymbol{X}^{t}\boldsymbol{V^{-1}} = \left[\left(\boldsymbol{I}_n - \boldsymbol{P}_{\boldsymbol{X}}\right)\boldsymbol{V}\left(\boldsymbol{I}_n - \boldsymbol{P}_{\boldsymbol{X}}\right)\right]^{+},
\]  
we have
\begin{align*}
\mbox{ED}_k & = \mbox{trace}\left(\boldsymbol{Z}_k^{t}\boldsymbol{Q}\boldsymbol{Z}_k\boldsymbol{G}\right)\\
						& = \mbox{trace}\left(\boldsymbol{Z}_k\boldsymbol{G}\boldsymbol{Z}_k^{t}\boldsymbol{Q}\right)\\
						& = \mbox{trace}\left(\boldsymbol{Z}_k\boldsymbol{G}\boldsymbol{Z}_k^{t}\left[\left(\boldsymbol{I}_n - \boldsymbol{P}_{\boldsymbol{X}}\right)\boldsymbol{V}\left(\boldsymbol{I}_n - \boldsymbol{P}_{\boldsymbol{X}}\right)\right]^{+}\right)\\
						& = \mbox{ED}\left(\boldsymbol{Z}_k\right).
\end{align*}
We would like to note that the equivalence between (\ref{ED_Cui_1}) and (\ref{ED_Cui_2}) can also be proved using results of \cite{Hoog1990}. Furthermore, we also have that
\begin{align*}
\mbox{ED}_{\boldsymbol{\varepsilon}} & = \mbox{trace}\left(\boldsymbol{R}\boldsymbol{Q}\right)\\
																		 & = \mbox{trace}\left(\boldsymbol{R}\left[\left(\boldsymbol{I}_n - \boldsymbol{P}_{\boldsymbol{X}}\right)\boldsymbol{V}\left(\boldsymbol{I}_n - \boldsymbol{P}_{\boldsymbol{X}}\right)\right]^{+}\right)\\
						& = \mbox{ED}\left(\boldsymbol{\varepsilon}\right).
\end{align*}
\section{\texttt{SpATS} package\label{s_software}}
This section contains a brief description of the developed \texttt{R}-package associated to this paper. The package can be freely downloaded from \url{https://cran.r-project.org/package=SpATS}, where a more detailed depiction of it use can be found. The main function of the package is \texttt{SpATS()}, which fits the spatial model presented in Section \ref{s_trials}. Numerical and graphical summaries of the fitted spatial model can be obtained, as usual in \texttt{R}, by using \texttt{summary.SpATS()}, \texttt{variogram.SpATS()}, \texttt{predict.SpATS()} and \texttt{plot.SpATS()}. In the implementation of the package, the sparse structure of the design matrix associated with the genotype has been taken into account, which, in combination with the estimation procedure presented in Section \ref{s_estproc} and the possible use of nested B-spline bases, makes the package computational efficient, allowing the analysis of very large datasets. 

By way of example, we present here the syntax for the Australian wheat trial example discussed in the paper by \cite{Gilmour1997}. The aim of this trial was the evaluation of advance breeding lines and commercial varieties. The trial consisted of $107$ varieties, which were sown in three replicates, each replicate being a complete block. Each block comprised $5$ columns and $22$ rows, yielding a total of $330$ ($5\times 22\times 3$) plots on the field. To meet the $110$ plots per replicate, from the $107$ varieties, three were sown twice in each of these. For more details about the trial, we refer the readers to the cited paper. On the basis of the results shown in \cite{Gilmour1997}, the following statistical model was assumed
\begin{equation*}
\boldsymbol{y} = \boldsymbol{X}_g\boldsymbol{\beta}_g + f\left(\boldsymbol{u}, \boldsymbol{v}\right) + \boldsymbol{Z}_r\boldsymbol{c}_r + \boldsymbol{Z}_c\boldsymbol{c}_c + \boldsymbol{\varepsilon},
\label{tpspline_model_gil}
\end{equation*}
where $\boldsymbol{\beta}_g$ is a ($106 \times 1$) vector of fixed variety (genetic) effects, and $\boldsymbol{X}_g$ is the corresponding ($330 \times 106$) design matrix. Note that the dimension of the genetic effect is $m_g - 1$ (where $m_g = 107$) since the intercept is included in $f\left(\boldsymbol{u}, \boldsymbol{v}\right)$. Here, $\boldsymbol{c}_r \sim N\left(\boldsymbol{0}, \sigma_r^2 \boldsymbol{I}_{22}\right)$ and $\boldsymbol{c}_c \sim N\left(\boldsymbol{0}, \sigma_c^2 \boldsymbol{I}_{15}\right)$ are vectors of row and column random effects respectively, and $\boldsymbol{\varepsilon} \sim N\left(\boldsymbol{0},\sigma^2\boldsymbol{I}_{330}\right)$.

The dataset can be found in the \texttt{R}-package \texttt{agridat}, under the name \texttt{gilmour.serpentine}. Here there is a brief summary of the data
\begin{verbatim}
> library(agridat)
> GS <- gilmour.serpentine
> summary(GS)
     col          row       rep                gen          yield      
 Min.   : 1   Min.   : 1.0   R1:110   TINCURRIN   :  6   Min.   :194.0  
 1st Qu.: 4   1st Qu.: 6.0   R2:110   VF655       :  6   1st Qu.:469.0  
 Median : 8   Median :11.5   R3:110   WW1477      :  6   Median :617.5  
 Mean   : 8   Mean   :11.5            (WWH*MM)*WR*:  3   Mean   :591.8  
 3rd Qu.:12   3rd Qu.:17.0            (WqKPWmH*3Ag:  3   3rd Qu.:713.5  
 Max.   :15   Max.   :22.0            AMERY       :  3   Max.   :925.0  
                                      (Other)     :303
\end{verbatim}
The dataset contains the column and row positions (\texttt{col} and \texttt{row} variables), the block (variable \texttt{rep}), the variety (\texttt{gen}) and the yield (\texttt{yield}). In order to incorporate in the model the random factors of rows and columns, we need first to create the corresponding factor variables, that we denote as \texttt{col\_f} and \texttt{row\_f}, and we then fit the model
\begin{verbatim}
>  GS$col_f = factor(GS$col)
>  GS$row_f = factor(GS$row)

> fit.SpATS <- SpATS(response = "yield", genotype = "gen", genotype.as.random = FALSE,
+ spatial = ~ PSANOVA(col, row, nseg = c(16,20), degree = 3, nest.div = 2), 
+ fixed = NULL, random = ~ row_f + col_f, 
+ data = GS, control =  list(tolerance = 1e-03, monitoring = 1))

Timings:
SpATS 0.38 seconds
All process 0.57 seconds
\end{verbatim}
Through \texttt{response} and \texttt{genotype} arguments, users specify the name of the variables in the dataset that contains, respectively, the response variable (phenotype) of interest and the genotype or variety. The genotype can be included in the model either as fixed (default) or random (\texttt{genotype.as.random = TRUE}). For modeling the spatial trend, argument \texttt{spatial}, we consider $16$ segments (\texttt{nseg}) for the column position and $20$ for the row. This, jointly with the fact we use cubic B-splines, \texttt{degree = 3}, gives rise to B-spline bases of dimension $P = 16 + 3 = 19$ and $L = 20 + 3 = 23$ for the columns and rows, respectively. By specifying the argument \texttt{nest.div = 2}, we indicate the use of nested bases, with half the number of segments of the original ones (see \ref{s_p2mix}). The fixed and random effects to be included in the model are indicated in \texttt{fixed} and \texttt{random}, and argument \texttt{control} allows to modify some default parameters that control the fitting process. For instance, the tolerance for the convergence criterion for the variance components can be altered using this argument, as well as the maximum number of iterations. Under this representation, the model has a total of $322$ coefficients, but it took less than $1$ seconds to be fitted. A numerical summary of the fitted model can be obtained by calling the function \texttt{summary()}. By indicating the argument \texttt{which = "all"}, we obtain both the estimates of the variance components and the effective dimensions
\begin{verbatim}
> summary(fit.SpATS, which = "all")
[...]

Variance components:
                   Variance            SD     log10(lambda)
row_f             4.397e+02     2.097e+01           0.67320
col_f             4.442e+03     6.665e+01          -0.33128
f(col)            1.245e+04     1.116e+02          -0.77895
f(row)            7.240e+01     8.509e+00           1.45657
f(col):row        7.847e+02     2.801e+01           0.42159
col:f(row)        6.490e-06     2.548e-03           8.50408
f(col):f(row)     2.530e+03     5.030e+01          -0.08684
                                                           
Residual          2.072e+03     4.552e+01                  

Dimensions:
                  Effective     Model     Nominal     Ratio     Type
gen                   106.0       106         106      1.00        F
Intercept               1.0         1           1      1.00        F
row_f                  12.6        22          21      0.60        R
col_f                  10.3        15          14      0.74        R
col                     1.0         1           1      1.00        S
row                     1.0         1           1      1.00        S
row:col                 1.0         1           1      1.00        S
f(col)                  2.3        17          17      0.14        S
f(row)                  1.0        21          21      0.05        S
f(col):row              2.6        17          17      0.15        S
col:f(row)              0.0        21          21      0.00        S
f(col):f(row)           7.5        99          99      0.08        S
                                                                    
Total                 146.3       322         320      0.46         
Residual              183.7                                         
Nobs                    330                                         

Type codes: F 'Fixed'    R 'Random'    S 'Smooth/Semiparametric'
\end{verbatim}
In this example, there are seven variance components, five associated to the spatial trend, one associated to the row random effects, one to the column random effects, and the residual variance $\sigma^2$. The column \texttt{log10(lambda)} shows the logarithm of base $10$ of the smoothing parameters, i.e., the ratio between the residual variance and the variance component. As far as the dimensions is concerned, for each component in the model (either fixed, random or spatial), the function returns (a) the \texttt{effective} dimension or effective degrees of freedom, (b) the \texttt{model} dimension, i.e., the number of parameters to be estimated, (c) the \texttt{nominal} dimension, which, for the random components is the model dimension minus one, lost due to the constraint of zero-mean imposed to them; and (d) the \texttt{ratio} between the \texttt{effective} and the \texttt{nominal} dimension. It is worth remembering that, if the variety had been included as random, the \texttt{ratio} for the variety would have provided an estimate of the so-called generalized heritability as proposed by \cite{Oakey2006}. If we focus on the random effects for the rows and columns, we have that the effective dimensions are, respectively, about the $60\%$ and the $74\%$ of the nominal dimensions. As discussed in Section \ref{mm_psplines}, for the spatial trend we have, in total, $8$ components (excluding the intercept). The linear effects for the rows and the columns (\texttt{row} and \texttt{column}), as well as the linear interaction (\texttt{row:col}), represent the fixed or unpenalized part of the tensor-product P-spline. The remaining five components, i.e., the main effects (\texttt{f(row)} and \texttt{f(col)}), the smooth varying coefficient terms (\texttt{f(col):row} and \texttt{row:f(col)}); and the smooth-by-smooth interaction component (\texttt{f(col):f(row)}) correspond to the penalized or random part, and have been extensively discussed in Sections \ref{s_splines_2D} and \ref{eff_dim}. On the basis of the effective dimensions associated to each of these five components, we may inferred that most of the trend has been captured by the main effect and the smooth varying coefficient term along the column position, but also by the smooth-by-smooth interaction term, for which we have an effective dimension of $7.6$. 

\begin{figure}
    \begin{center}
    \includegraphics[width=12cm]{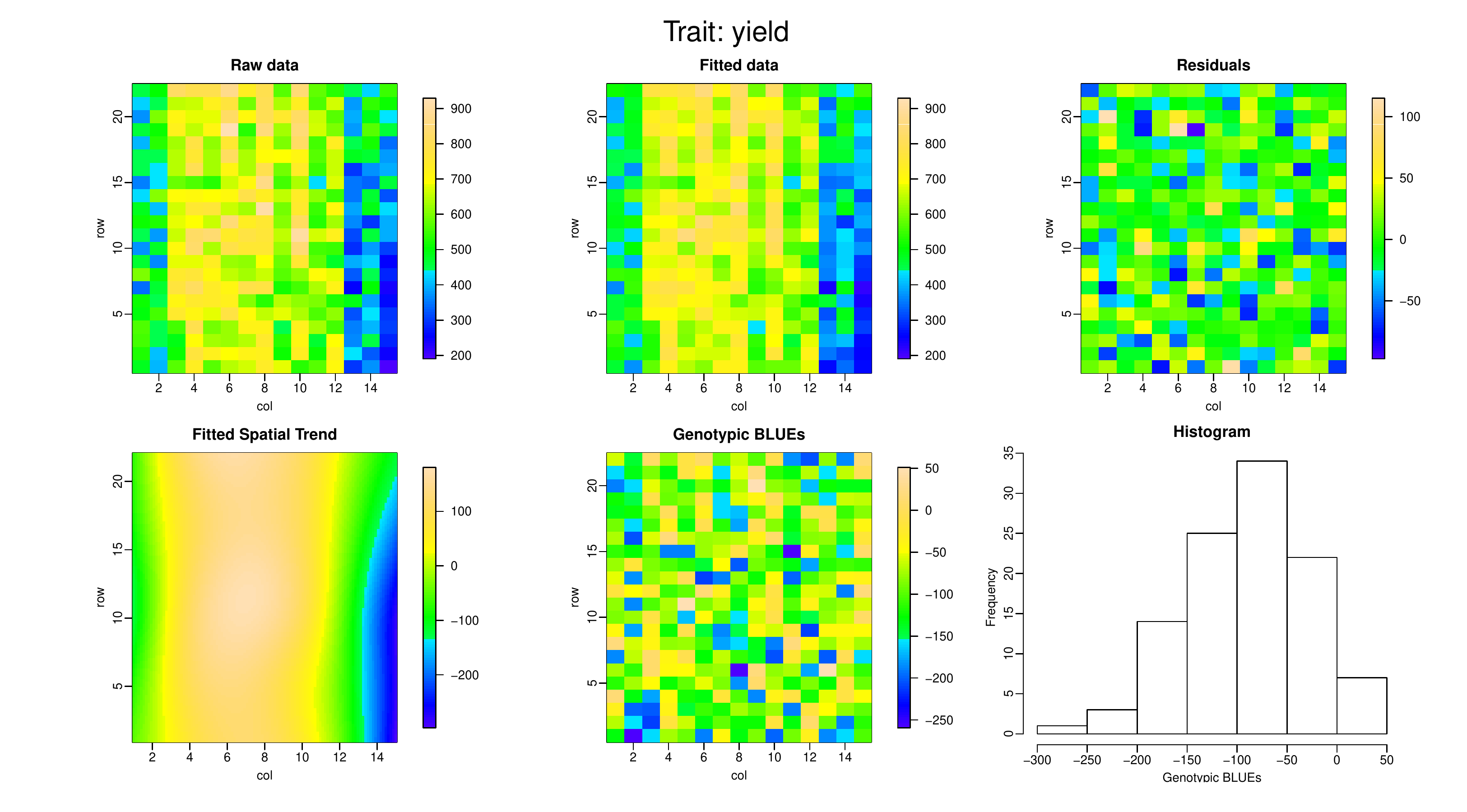}
    \end{center}
    \caption{Graphical results provided by the \texttt{SpATS} package for the Australian wheat trial.}
		\label{SpATS_gresult_Gilmour}
  \end{figure}

\begin{figure}
    \begin{center}
    \includegraphics[width=5cm]{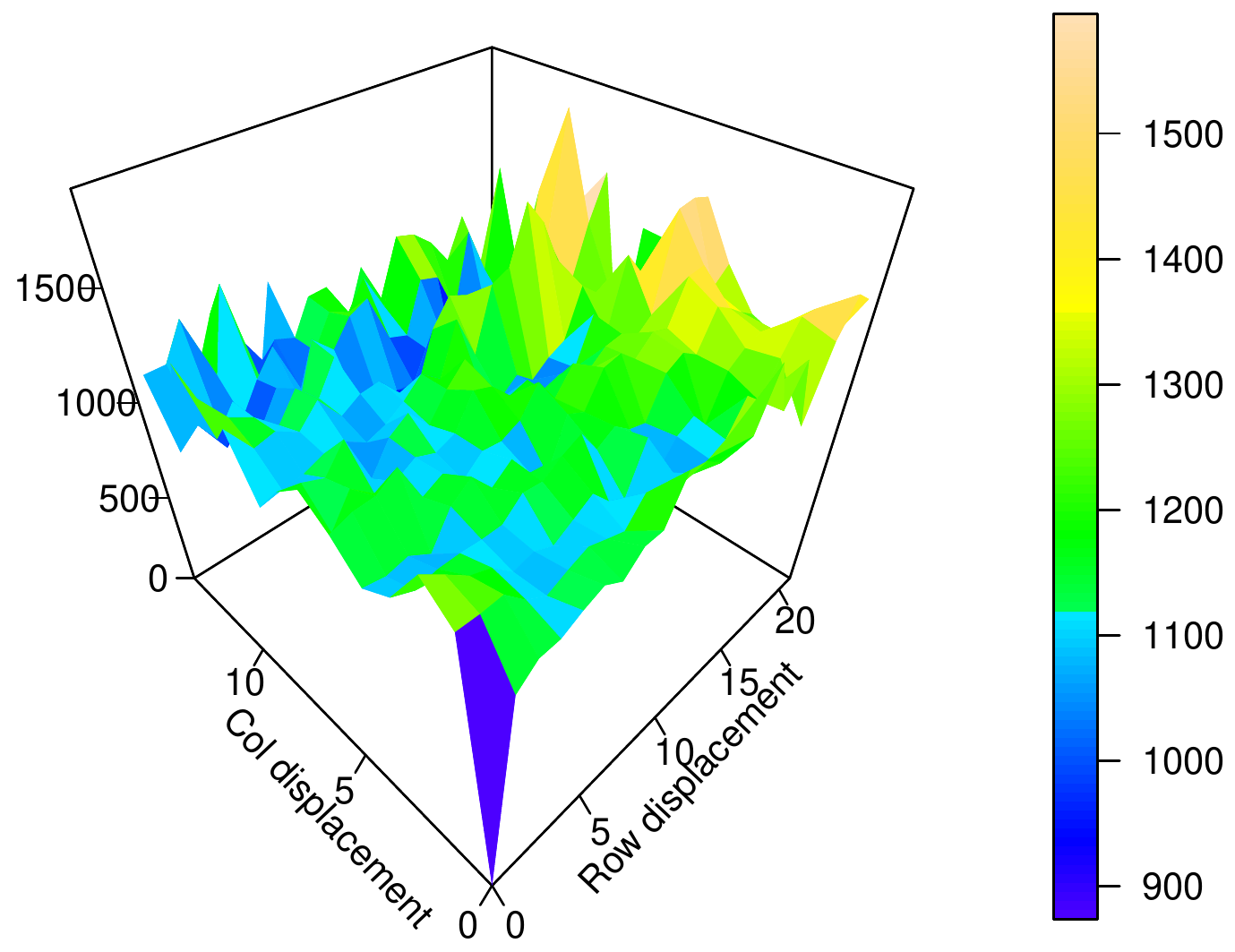}
    \end{center}
    \caption{Sample variogram of the residuals provided by the \texttt{SpATS} package for the Australian wheat trial.}
		\label{SpATS_variogram_Gilmour}
  \end{figure}

To complement those numerical results, the \texttt{SpATS} package furnishes different graphical results that can be used to further explore the fitted model. Specifically, the sample variogram can be obtained using the function \texttt{variogram()}, which can also be plotted; and the function \texttt{plot()} depicts six different graphics: the raw data, the fitted data, the residuals, the estimated spatial trend (excluding the intercept), the genotypic BLUEs (or BLUPs) and their histogram. Except for the histogram, the plots are depicted in terms of the spatial coordinates (e.g., the rows and columns of the field).
\begin{verbatim}
> plot(fit.SpATS)

> plot(variogram(fit.SpATS))
\end{verbatim}
The result of the above code is shown in Figures \ref{SpATS_gresult_Gilmour} and \ref{SpATS_variogram_Gilmour}. The spatial plots of the residuals and the genotypic BLUEs ($\widehat{\boldsymbol{\beta}}_g$) do not suggest the presence of any extra spatial pattern that should have been taken into account. A similar conclusion can be drawn from the sample variogram of the residuals shown in Figure \ref{SpATS_variogram_Gilmour}. Finally, note that the fitted spatial trend takes values between $-300$ and $200$, whereas the residuals vary between $-100$ and $100$. This result highlights what could have been expected based on the raw data, that spatial (plot-to-plot) variation is larger than random (plot-to-plot) variation.

\bibliographystyle{chicago}
\bibliography{spat_splines}
\end{document}